\documentclass[reprint, aip, pof, amsmath, amssymb]{revtex4-1}

\usepackage{graphicx}
\usepackage{dcolumn}
\usepackage{bm}
\usepackage{wrapfig}

\usepackage[utf8]{inputenc}
\usepackage[T1]{fontenc}
\usepackage{mathptmx}
\usepackage{etoolbox}
\usepackage{tikz}
\usetikzlibrary{shapes.geometric, arrows.meta, positioning}

\tikzset{
  block/.style = {rectangle, draw, thick, fill=none, text width=10em, align=center, rounded corners, minimum height=4em}, arrow/.style = {-{Stealth}, thick},
}

\tikzset{
  wide block/.style = {rectangle, draw, thick, fill=none, text width=21.8em, align=center, rounded corners, minimum height=4em},
}

\draft 

\begin{document}


\title[Lagrangian entropic lattice Boltzmann method]{Lagrangian entropic lattice Boltzmann method for Courant-free supersonic compressible flow simulation} 



\author{Woohyuk Noh}
 \email{alexandremargar123@gmail.com}
 \affiliation{School of Mathematics and Computing, Yonsei University, Seoul 03722, Korea}
\author{Changhoon Lee}%
 \email{clee@yonsei.ac.kr}
 \affiliation{School of Mathematics and Computing, Yonsei University, Seoul 03722, Korea}
 \affiliation{Department of Mechanical Engineering, Yonsei University, Seoul 03722, Korea}

\date{\today}

\begin{abstract}
This work presents the Lagrangian entropic lattice Boltzmann method (LELBM), a novel framework for supersonic lattice Boltzmann method that surpasses traditional restrictions on velocity, temperature, viscosity, and memory, resulting in stable and efficient Courant‐free LBM formulation, while still maintaining explicit time stepping. A regularized central moment collision method coupled with a double-distribution function (DDF) that considers internal degrees of freedom enabled the precise control of shear viscosity, bulk viscosity, and thermal conductivity. Lagrangian acoustic stencils (LAS), a generalized multispeed shifted lattice construction method, is introduced to adapt to arbitrary local velocity and temperature. Additionally, entropic population reconstruction (EPR) has been used to obtain entropy-satisfying post-collision population while achieving both positivity and conservation of moments. Memory requirements are drastically reduced from a polynomial order of temperature to a constant by employing moment streaming (MS) algorithm. The unified LELBM framework is validated through comprehensive benchmark problems. In one dimension, the model is evaluated through Sod’s shock tube, Lax problem, and Shu–Osher wave. In two dimensions, 2D Riemann problem, double Mach reflection, oblique shock, supersonic flow past a circular cylinder, and supersonic flow past a NACA0012 airfoil have been tested to assess the robustness and accuracy of the proposed algorithm.
\end{abstract}


\maketitle 

\section{\label{sec:intro}Introduction}
The lattice Boltzmann method (LBM), developed in the late 1980s\cite{McNamara1988}, is based on lattice gas automata (LGA) from the 1970s\cite{Hardy1973, Frisch1987, Frisch1986}. Unlike LGA, LBM streams particle populations to simulate the discrete Boltzmann equation. The conventional lattice Boltzmann method is characterized by three processes: integration, collision, and streaming\cite{Kruger2017}. The LBM algorithm is simple and easy to implement, since all processes except streaming are local, which is very favorable for the highly parallelized computation, demanded for modern CFD solvers. However, several limitations hinder its practical application. These include stability issues in velocity, temperature, and viscosity\cite{Chao2004, Sterling1996, Nie2008_Gal}. Additionally, because traditional LBM stores all populations, its memory demand is significantly higher than conventional continuum methods\cite{Matyka2021}.

The essence of the lattice Boltzmann method lies in the Bhatnagar-Gross-Krook (BGK) collision model\cite{Bhatnagar1954}, which simplified the nonlinear interaction of particles to a linear decay rate of non-equilibrium dynamics\cite{Higuera1989}. The most traditional collision model is the so-called single relaxation time model (SRT), which first constructs the equilibrium distribution using Hermite expansion\cite{Shan1998} and conducts a collision based on the relaxation time $\tau$ to the raw population to adjust the viscosity. After the introduction of the multiple relaxation time model (MRT) \cite{Lallemand2000, DHumires2002, Asinari2008}, the collision process can be broadly classified into two categories: population-based and moment-based. The population-based method explicitly performs collision on the population, while enhancing the stability by either regularization \cite{Latt2006, Malaspinas2015, Coreixas2017, Jacob2018, Feng2019} or entropic stabilization\cite{Frapolli2015, Frapolli2016, Frapolli2017, Frapolli2020, Hosseini2023, Hosseini2024}. The moment-based method conducts collision in moment space and conducts back-transformation to reconstruct the population that satisfies the post-collision moments, while also relaxing higher-order moments to equilibrium. Developed from the traditional raw moment method, the advanced method conducts collisions in different moment space such as central moments \cite{Geier2006_thesis, Geier2006, Premnath2011, DeRosis2019}, or cumulants\cite{Geier2015, Geier2017_1, Geier2017_2}.

Considering that the mean velocity of the gas particles is proportional to the square root of temperature \cite{Salinas2001}, higher-order velocity stencils are required to simulate high-temperature flow. The first multispeed lattice was introduced by Alexander et al.\cite{Alexander1992} to simulate thermohydrodynamic flows using the lattice Boltzmann method. Although originally developed for subsonic thermal flows, it was later extended to handle supersonic flows\cite{Shan2006, Chikatamarla2009, Chen2010}. However, this approach consequently increased the number of velocity stencils even for the case of supersonic flow under low temperature. To address this issue, Sun\cite{Sun1998, Sun2000, Sun2003, Sun2004} proposed the shift of velocity stencils by bulk velocity and reconstructing stencils in a relative frame. But this method has long been forgotten until Frapolli\cite{Frapolli2017} revisiting the shifted lattice for their supersonic entropic LBM framework.

The earliest energy models in the LBM involved the direct integration of second-order moments\cite{Succi2018}. However, this approach was limited to monatomic gases and lacked the flexibility in adjusting key thermodynamic properties, such as thermal conductivity and specific heat. From the Chapman-Enskog expansion, it becomes clear that using a single collision time is insufficient to fully capture the fluid properties described by the Navier-Stokes-Fourier (NSF) equations, particularly because third-order moments are closely linked to heat flux\cite{Teixeira1992, McNamara1995}. To address this limitation, He et al.\cite{He1998} introduced the double-distribution function (DDF) model, which incorporates a second distribution function for internal energy to solve the energy equation alongside the kinetic equations. Still, this model required finite differencing to capture the effect of viscous heating. Developing from the internal energy model, Guo\cite{Guo2007} proposed a modified approach about the total energy, enabling the accurate modeling of viscous coupling without finite differentiation. In parallel, Nie\cite{Nie2008_therm} introduced the DDF method for the internal degree of freedom, which is now the common approach for modern supersonic LBM models. Besides the complete LBM framework for the energy equation, hybrid LBM (HLBM)\cite{Feng2019, Mezrhab2004, Li2014, Nee2020} instead adopted the traditional finite-volume method for energy, while still using LBM for the mass and momentum equations.

Modern supersonic lattice Boltzmann methods can broadly be classified into four distinct approaches. While the hybrid recursive regularization LBM (HRR-LBM) solves the energy equation using a traditional continuum method, this method also attempts to improve the stability of LBM formulation through regularization process \cite{Feng2019, Guo2020, Renard2021, Guo2024}. While HRR-LBM constructs populations for cell-centered velocity stencils, semi-Lagrangian LBM (SLLBM) reconstructs a population on a compact, non-cell-conforming velocity stencils for each lattice, to enhance stability under compressible flow conditions \cite{Kraemer2017, Wilde2020, Wilde2021_1, Wilde2021_2, Spelten2024}. As a more "physical" approach, the entropic LBM (ELBM) \cite{Frapolli2015, Frapolli2016, Frapolli2017} dynamically adjusts the collision frequencies to achieve local maximum entropy in each cell in accordance with the Boltzmann $H$-theorem\cite{Boltzmann2003}. Besides the traditional Hermite polynomial-based population reconstruction approach, numerical equilibria LBM (NE-LBM) constructs populations through the application of Lagrange multipliers to satisfy the maximum entropy principle \cite{Latt2020, Coreixas2020, Thyagarajan2023}. This approach results in an exponential-shaped distribution function that preserves both positivity and conservation of properties while also enhancing the numerical stability.

As an even more sophisticated approach, the particle on demand method combined with the discrete unified gas kinetic scheme (PonD/DUGKS) and entropic stabilization has recently been developed \cite{Kallikounis2022, Kallikounis2024}. The characteristic of PonD is its shift from an absolute to a relative reference frame, while the moments are also scaled to an optimal local temperature. Consequently, the populations are constructed through Hermite expansion in that shift-scaled frame, then mapped back to the absolute frame using Lagrange polynomials. DUGKS \cite{Guo2013, Guo2015, Guo2021}, in turn, employs a finite-volume formulation that explicitly computes population fluxes across cell interfaces, improving spatial resolution. By combining with entropic stabilization, the PonD/DUGKS framework proved its robustness under extreme conditions, such as Mach 30 astrophysical jets.

Despite numerous unique formulations proposed over the past decades to simulate supersonic flows using the lattice Boltzmann method, however, to the best of the author's knowledge, no universal numerical scheme has been yet developed that incorporates all benefits of previous methods. This paper presents a novel formulation, the Lagrangian entropic lattice Boltzmann method (LELBM), which unifies multiple approaches, including the philosophy of moment-matching of MRT model, the double distribution function (DDF), shifted multispeed lattice, and entropic formulations for stability. Additionally, this work adopted the moment expression of memory storage, inspired by the work of Valero-Lara\cite{ValeroLara2017}, to reduce the memory demand bottleneck. The paper is structured into six main sections. Section~\ref{sec:lelbm} introduces the LELBM formulation, providing a detailed theoretical framework that bridges various techniques developed in previous research and provides algorithmic flowchart for better understanding in practical application. Section~\ref{sec:bc} introduces the boundary condition techniques applied for the evaluation of the present model. Section~\ref{sec:verifications} verifies the theory of the present framework, from fluid properties to computational cost. Section~\ref{sec:results} evaluates both the accuracy and robustness of the present model on various canonical to extreme compressible flow problems. Finally, Section~\ref{sec:conclusion} concludes the paper by summarizing the key findings and contributions, while also listing the required improvements for practical implementations.

\section{\label{sec:lelbm}Lagrangian Entropic Lattice Boltzmann Method}
\subsection{\label{sec:theory}Theoretical Background}

Lattice Boltzmann method is an automaton-based method that, rather than solving partial differential equations about macroscopic variables as in traditional continuum methods, solves discretized Boltzmann equation in a less "math-oriented" approach. The Boltzmann equation is written as below:
\begin{align}
    \frac{\partial f}{\partial t} + \xi_\alpha \frac{\partial f}{\partial x_\alpha} + \frac{F_\alpha}{\rho} \frac{\partial f}{\partial \xi_\alpha} = \Omega(f). \label{eq:IIA1}
\end{align}
Here, $f$ is the continuous distribution function of the particle in $2D+1$ dimensions, defined in both $D$ dimensions of position and momentum, with an additional dimension from time. $\bm{\xi}$ denotes particle velocity vector, and $\bm{F}$ is the external forces. $\Omega$ is the collision kernel that results in macroscopic phenomena such as hydrodynamic stress or heat conduction\cite{Grad1949}. With BGK collision model\cite{Bhatnagar1954}, we can write the collision kernel in a simplified form of
\begin{align}
    \Omega(f) = -\frac{1}{\tau}(f - f^{\mathrm{eq}}), \label{eq:IIA2}
\end{align}
where $\tau$ is the collision time that determines the decay rate of non-equilibrium contribution on distribution function. Here, $f^{\mathrm{eq}}$ is the equilibrium distribution, which is known to follow the Boltzmann distribution.
\begin{align}
   f^{\mathrm{eq}}(\rho, \bm{\xi}, \bm{u}, RT) = \frac{\rho}{(2\pi RT)^{D/2}} \exp\left( -\frac{|\bm{\xi} - \bm{u}|^2}{2RT} \right). \label{eq:IIA3}
\end{align}
From the above distribution $\rho$ is the macroscopic mass density of fluid in position of interest, $\bm{\xi}$ is the microscopic velocity of particles, $\bm{u}$ for the macroscopic bulk velocity, and $RT$ is the temperature with gas constant. Now, if we conduct temporal discretization on Eq. \eqref{eq:IIA1} in a total derivative form and separating the equilibrium and non-equilibrium distributions from the raw distribution function $f$, we obtain
\begin{align}
    &f(\bm{x} + \bm{\xi} \, \Delta t, \bm{\xi}, t + \Delta t) \nonumber \\
    &= f^{\mathrm{eq}}(\bm{x}, \bm{\xi}, t) + \left(1 - \omega\right) f^{\mathrm{neq}}(\bm{x}, \bm{\xi}, t) + \mathcal{O}(\Delta t^2). \label{eq:IIA4}
\end{align}
where $f^{\mathrm{neq}}$ is the non-equilibrium distribution function which is defined as
\begin{align}
    f^{\mathrm{neq}} = f - f^{\mathrm{eq}}. \label{eq:IIA5}
\end{align}

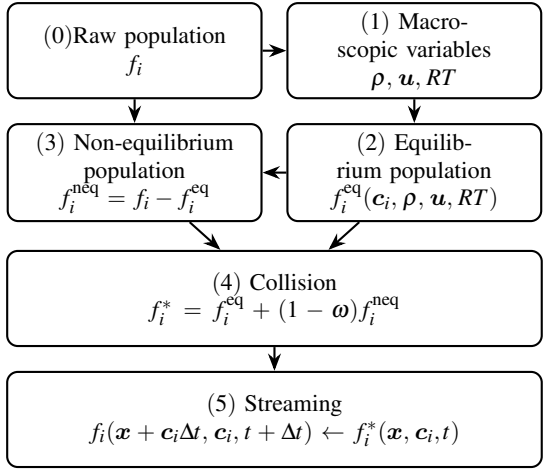
\begin{figure}
  \centering
  \begin{tikzpicture}[node distance=1em]
    \node[block] (raw_pop) {$(0)$Raw population\\$f_i$};
    \node[block, right=of raw_pop] (mac_var) {$(1)$ Macroscopic variables\\ $\rho,\, \bm{u},\, RT$};
    \node[block, below=of raw_pop] (nequil) {$(3)$ Non-equilibrium population\\$f_i^\mathrm{neq}=f_i-f_i^\mathrm{eq}$};
    \node[block, below=of mac_var] (equil) {$(2)$ Equilibrium population\\$f_i^\mathrm{eq}(\bm{c}_i,\,\rho,\,\bm{u},\,RT)$};
    \path (nequil.east) -- (equil.west) coordinate[midway] (midpoint);
    \node[wide block, below=of midpoint, yshift=-2.25em] (collision) {$(4)$ Collision\\$f_i^*=f_i^\mathrm{eq}+(1-\omega)f_i^\mathrm{neq}$};
    \node[wide block, below=of collision] (strm) {$(5)$ Streaming\\$f_i(\bm{x}+\bm{c}_i\Delta t,\,\bm{c}_i,\,t+\Delta t)\leftarrow f_i^*(\bm{x},\,\bm{c}_i,t)$};

    \draw[arrow] (raw_pop) -- (mac_var);
    \draw[arrow] (raw_pop) -- (nequil);
    \draw[arrow] (mac_var) -- (equil);
    \draw[arrow] (equil) -- (nequil);
    \draw[arrow] (nequil) -- (collision);
    \draw[arrow] (equil) -- (collision);
    \draw[arrow] (collision) -- (strm);
  \end{tikzpicture}
  \caption{\label{fig:traditional_algorithm}Traditional LBM algorithm. The numbering indicate the cascaded operation procedure and arrow direction indicates the transfer of memory.}
\end{figure}

As explained in Appendix~\ref{sec:implicit}, $\omega$ in Eq. \eqref{eq:IIA4} is the implicit collision frequency arising from the implicit time stepping. Since $f$ is continuous both in position and momentum, the distribution is discretized according to the velocity stencils $\bm{c}_i$.
\begin{align}
    f_i (\bm{x} + \bm{c}_i \Delta t, \bm{c}_i, t + \Delta t ) = f_i^{\mathrm{eq}} (\bm{x}, \bm{c}_i, t) + (1 - \omega) f_i^{\mathrm{neq}} (\bm{x}, \bm{c}_i, t). \label{eq:IIA6}
\end{align}
Equation~\eqref{eq:IIA6} is the automaton algorithm in which the population $f_i$ defined by $\bm{x}$ and $\bm{c}_i$ is landing at position $\bm{x}+\bm{c}_i\Delta t$ after $\Delta t$. Since an LBM algorithm in general streams the population to exactly the cell center in each time step, the streaming algorithm shows a lattice-like network. The algorithm of the traditional method can be simply expressed as Figure~\ref{fig:traditional_algorithm}. See \cite{Kruger2017, Succi2018} for details on LBM algorithm.

In the fields of LBM, the distinction between the continuous distribution function $f$ and the population (or the discrete distribution function) $f_i$ has long been ambiguous. However, their dimensional units of $f$ and $f_i$ are  $[\rho U^{-D}]$ and $[\rho]$, respectively, where $[\rho]$ and $[U]$ are the dimensions of density and velocity, and $D$ is the number of translational DoF. Since LBM generally expects the population to be located at the lattice center, their relationship can be expressed as
\begin{align}
    f_i(\bm{x},\, \bm{c}_i) = \iiint_{\Delta C_i} f(\bm{x},\, \bm{\xi}) d\bm{\xi}, \label{eq:IIA7}
\end{align}
where $\Delta C_i$ is the translational volume of speed for the specified $i$th population we are interested in, which is centered at $\bm{c}_i$. For general uniform lattice,
\begin{align}
    \Delta C_i = \prod_{\alpha=1}^{D} \Delta c_{i\alpha}. \label{eq:IIA8}
\end{align}
Thus, the population $f_i$ has an approximation relation with the distribution function $f$ by
\begin{align}
    f_i(\bm{x},\, \bm{c}_i) \approx f(\bm{x},\, \bm{c}_i)\Delta C_i. \label{eq:IIA9}
\end{align}
Additionally, for the LELBM framework, the population itself follows the cell-averaging concept of the finite-volume method. The volume-integrated population $\langle f_i \rangle$ is expressed as
\begin{align}
    \langle f_i \rangle (\bm{x},\, \bm{c}_i) &= \iiint_{\Delta V}  \iiint_{\Delta C_i} f(\bm{x},\, \bm{\xi}) d\bm{\xi}  dV, \label{eq:IIA10}
\end{align}
or approximately
\begin{align}
    \langle f_i \rangle(\bm{x},\, \bm{c}_i) \approx f_i(\bm{x},\, \bm{c}_i) \Delta V \approx f(\bm{x},\, \bm{c}_i) \Delta C_i \Delta V. \label{eq:IIA11}
\end{align}
Here, $\Delta V$ denotes the physical volume of the cell at position $\bm{x}$. The dimensional unit of $\langle f_i \rangle$ is $[\rho L^D]$, corresponding to the mass fraction distributed throughout the velocity space with the mean particle velocity of $\bm{c}_i$ for each cell, within the control volume of $\Delta V$. For notational consistency with the traditional LBM framework, $\langle f_i \rangle$ is simply expressed as $f_i$ throughout the remainder of this paper. These dimensional formulations are unnecessary in the standard lattice unit setting, where $\Delta C_i = 1$, $\Delta x = 1$, $\Delta V = 1$, $\Delta t = 1$, which is also adopted in this work. Although the rigorous definitions introduced in this section are not required for the current uniform lattice implementation, this background knowledge is necessary for understanding key concepts in supersonic LBM models.

Additionally, the definition of moment is required for the present framework. The raw moment of an arbitrary continuous distribution function $f$ can be calculated as
\begin{align}
    M_{\alpha \beta \gamma} = \iiint_{-\infty}^{\infty} \xi_x^{\alpha} \xi_y^{\beta} \xi_z^{\gamma} f d\bm{\xi}. \label{eq:IIA12}
\end{align}
Defining $\xi_x^{\alpha} \xi_y^{\beta} \xi_z^{\gamma}$ as the moment basis function $T_N(\bm{\xi})$, we can define arbitrary $N$th-order raw moments in the form of
\begin{align}
    M_N = \iiint_{-\infty}^{\infty} T_N(\bm{\xi}) f d\bm{\xi}. \label{eq:IIA13}
\end{align}

Central moments are evaluated in a similar approach, but with a shifted population velocity of $\bm{\xi} - \bm{u}$.
\begin{align}
    \widetilde{M}_N = \iiint_{-\infty}^{\infty} T_N(\bm{\xi} - \bm{u}) f d\bm{\xi}. \label{eq:IIA14}
\end{align}

Moments of an arbitrary discrete distribution function can be obtained by taking a summation instead. $T_N$ is now the discrete moment basis function, which is in matrix form.
\begin{subequations}
\begin{align}
    M_N &= \sum_i T_N(\bm{c}_i) f_i, \label{eq:IIA15a} \\
    \widetilde{M}_N &= \sum_i T_N(\bm{c}_i - \bm{u}) f_i. \label{eq:IIA15b}
\end{align}
\end{subequations}
The dimensions of both continuous and discrete moments are $[\rho U^N]$. However, since LELBM conducts all operations in volume-integrated form, the normalization and scaling processes are involved throughout the algorithm.

\subsection{\label{sec:collision}Regularized Central Moment Collision}

With the introduction of the multiple relaxation time (MRT) model \cite{Lallemand2000, DHumires2002, Asinari2008}, moment-based collision generally follows a three-step process consisting of moment transformation, moment collision, and inverse transformation:
\begin{subequations}
\begin{align}
    \bm{M} &= \bm{T} \bm{f}, \label{eq:IIB1}, \\
    \bm{M}^* &= \bm{M}^{\mathrm{eq}} + (\bm{I} - \bm{\Omega}) \bm{M}^{\mathrm{neq}}, \label{eq:IIB2} \\
    \bm{f}^* &= \bm{T}^{-1} \bm{M}^*. \label{eq:IIB3}
\end{align}
\end{subequations}
Here, $\bm{f}$ and $\bm{M}$ are the vector of population and moments. $\bm{T}$ is the moment basis matrix, $\bm{I}$ is the identity matrix, and $\bm{\Omega}$ is the matrix of collision frequencies. Traditionally, MRT models have been applied mainly to athermal flows with fixed $DdQ3^d$ lattices. This was due to the sporadic emergence of a negative post-collision population $\bm{f}^*$ arising from direct inversion of the moment basis matrix $\bm{T}$ in higher-order lattice. As a result, simulations on multispeed lattices often led to nonphysical solutions.

However, as proved in Appendices~\ref{sec:chapman} and \ref{sec:implicit}, assigning distinct collision frequencies to a different order of moments allows independent control of fluid properties. When combined with a positivity-preserving reconstruction scheme, moment-based collision models can be extended to simulate thermal flows with flexible adjustment of thermohydrodynamic properties. This section provides the procedure for the moment-based collision model tailored for thermal LBM. The corresponding reconstruction strategy is addressed in Section~\ref{sec:epr}.

Founded upon the concept of cascaded digital lattice automata \cite{Geier2006_thesis}, central moment method performs collisions on moments evaluated in a relative reference frame \cite{Premnath2011, DeRosis2019}. This approach effectively decoupled the momentum and energy in their moments, and was found to enhance stability even in a high Reynolds number flow. The regularized central moment collision method extends the athermal framework of the central moment collision model to thermal flows.

The LELBM algorithm operates with two distribution populations: fluons ($f_i$) and phonons ($g_i$). In short, the fluons and phonons are related to the dynamics of translational DoF $D$ and internal DoF $K$, respectively (see Appendix~\ref{sec:polyatomic} for details). The raw and central moments of these populations are defined as follows:
\begin{subequations}
\begin{align}
    \{ M_N,\, G_N \} &= \sum_i T_N(\bm{c}_i) \{ f_i,\, g_i \}, \label{eq:IIB4a} \\
    \{ \widetilde{M}_N,\, \widetilde{G}_N \} &= \sum_i T_N(\bm{c}_i - \bm{u}) \{ f_i,\, g_i \}. \label{eq:IIB4b}
\end{align}
\end{subequations}
However, in the LELBM algorithm, such moment transformation process is neglected, since the raw moments are explicitly stored instead of populations, with implicit application of moment transformation through the moment-streaming algorithm (explained in Section~\ref{sec:ms}). The full set of moments stored through the moment streaming algorithm are:
\begin{align}
    \{ \bm{M},\, \bm{G} \} =
    \left\{
    \begin{array}{l}
    \begin{bmatrix}
    M_0,\, M_{\alpha},\, \cdots,\, M_{\alpha \beta},\, \cdots,\, M_{D \gamma},\, \cdots
    \end{bmatrix}^T, \\[6pt]
    \begin{bmatrix}
    G_0,\, G_{\gamma},\, \cdots
    \end{bmatrix}^T
    \end{array}
    \right\}. \label{eq:IIB5}
\end{align}
Refer to Appendix~\ref{sec:polyatomic} for description of each moment component.

An arbitrary moment $M_N$ is composed of equilibrium $M_N^{\mathrm{eq}}$ and non-equilibrium part $M_N^{\mathrm{neq}}$. Since the collision is carried out in central moment space, a transformation from raw to central moments is necessary. To perform transformation, the macroscopic variables-density $\rho$, velocity $u_{\alpha}$, and temperature $RT$ are first computed as:
\begin{subequations}
\begin{align}
    \rho &= \frac{M_0}{\Delta V}, \label{eq:IIB6a} \\
    u_{\alpha} &= \frac{M_{\alpha}}{\rho \Delta V}, \label{eq:IIb6b} \\
    RT &= \frac{1}{D + K} \left( \frac{M_{\alpha \alpha} + G_0}{\rho \Delta V} - u_{\alpha}^2 \right). \label{eq:IIB6c}
\end{align}
\end{subequations}

Since the equilibrium distribution is analytically Boltzmann, the expression for the corresponding equilibrium raw and central moments can also be analytically obtained.
\begin{subequations}
\begin{align}
    M_{\alpha \beta}^{\mathrm{eq}} &= \rho (u_{\alpha} u_{\beta} + RT \delta_{\alpha \beta}) \Delta V, \label{eq:IIB7a} \\
    M_{D \gamma}^{\mathrm{eq}} &= \rho (|\bm{u}|^2 + (D + 2) RT) u_{\gamma} \Delta V, \label{eq:IIB7b} \\
    G_0^{\mathrm{eq}} &= \rho KRT \Delta V, \label{eq:IIB7c} \\
    G_{\gamma}^{\mathrm{eq}} &= \rho KRT u_{\gamma} \Delta V, \label{eq:IIB7d} \\
    \widetilde{M}_0^{\mathrm{eq}} &= \rho \Delta V, \label{eq:IIB7e} \\
    \widetilde{M}_{\alpha}^{\mathrm{eq}} &= 0, \label{eq:IIB7f} \\
    \widetilde{M}_{\alpha \beta}^{\mathrm{eq}} &= \rho RT \delta_{\alpha \beta} \Delta V, \label{eq:IIB7g} \\
    \widetilde{M}_{D \gamma}^{\mathrm{eq}} &= 0, \label{eq:IIB7h} \\
    \widetilde{G}_0^{\mathrm{eq}} &= \rho KRT \Delta V, \label{eq:IIB7i} \\
    \widetilde{G}_{\gamma}^{\mathrm{eq}} &= 0. \label{eq:IIB7j}
\end{align}
\end{subequations}

Central moments can be expressed as the combination of raw moments. By expanding the polynomials of the moment basis function defined in Eq. \eqref{eq:IIB4b}, the transformation becomes
\begin{subequations}
\begin{align}
    \widetilde{M}_{\alpha \beta} &= \sum_i (c_{i\alpha} - u_{\alpha})(c_{i\beta} - u_{\beta}) f_i \notag \\
    &= M_{\alpha \beta} - u_{\alpha} M_{\beta} - u_{\beta} M_{\alpha} + u_{\alpha} u_{\beta} M_0, \label{eq:IIB8a} \\
    \widetilde{M}_{D \gamma} &= \sum_i (c_{i\alpha} - u_{\alpha})^2 (c_{i\gamma} - u_{\gamma}) f_i = M_{D \gamma} - 2 u_{\alpha} M_{\alpha \gamma} \notag \\
    & - u_{\gamma} M_{\alpha \alpha} + u_{\alpha}^2 M_{\gamma} + 2 u_{\alpha} u_{\gamma} M_{\alpha} - u_{\alpha}^2 u_{\gamma} M_0, \label{eq:IIB8b} \\
    \widetilde{G}_0 &= \sum_i g_i = G_0, \label{eq:IIB8c} \\
    \widetilde{G}_{\gamma} &= \sum_i (c_{i\gamma} - u_{\gamma}) g_i = G_{\gamma} - u_{\gamma} G_0. \label{eq:IIB8d}
\end{align}
\end{subequations}

Given that the moments can be decomposed into equilibrium and non-equilibrium components, the non-equilibrium central moments can systematically be derived by applying Eqs. \eqref{eq:IIB7a}--\eqref{eq:IIB7j} into the central moment transformation Eqs. \eqref{eq:IIB8a}--\eqref{eq:IIB8d} in a cascaded manner.
\begin{subequations}
\begin{align}
    \widetilde{M}_{\alpha \beta}^{\mathrm{neq}} &= M_{\alpha \beta}^{\mathrm{neq}} = M_{\alpha \beta} - \rho \left( u_{\alpha} u_{\beta} + RT \delta_{\alpha \beta} \right) \Delta V, \label{eq:IIB9a} \\
    \widetilde{M}_{D \gamma}^{\mathrm{neq}} &= M_{D \gamma} - 2 u_{\alpha} M_{\alpha \gamma}^{\mathrm{neq}} - u_{\gamma} M_{\alpha \alpha}^{\mathrm{neq}} \notag \\
    & - \rho \left( |\bm{u}|^2 + (D+2)RT \right) u_{\gamma} \Delta V, \label{eq:IIB9b} \\
    \widetilde{G}_0^{\mathrm{neq}} &= G_0^{\mathrm{neq}} = G_0 - \rho KRT \Delta V, \label{eq:IIB9c} \\
    \widetilde{G}_{\gamma}^{\mathrm{neq}} &= G_{\gamma} - u_{\gamma} G_0. \label{eq:IIB9d}
\end{align}
\end{subequations}

The second-order non-equilibrium central moment $\widetilde{M}_{\alpha\beta}^{\mathrm{neq}}$ consists of both shear and bulk stress components. Then the bulk stress contribution can be expressed as
\begin{align}
    \widetilde{M}_{\alpha\beta}^{\mathrm{bulk}} = -\frac{1}{D} \widetilde{G}_0^{\mathrm{neq}} \delta_{\alpha\beta}. \label{eq:IIB10}
\end{align}

The shear stress component is obtained by subtracting the bulk stress part:
\begin{align}
    \widetilde{M}_{\alpha\beta}^{\mathrm{shear}} = \widetilde{M}_{\alpha\beta}^{\mathrm{neq}} - \widetilde{M}_{\alpha\beta}^{\mathrm{bulk}}. \label{eq:IIB11}
\end{align}

The LELBM algorithm also regularizes the heat flux using contracted third-order non-equilibrium central moments. The fluon contribution to the conductive heat flux is analytically
\begin{align}
    \widetilde{M}_{D\gamma}^{\mathrm{neq}} = -\rho (D+2) R T \tau_h \partial_{\gamma} RT \Delta V. \label{eq:IIB12}
\end{align}
For each translational dimension, the individual contributions are given by
\begin{align}
    \widetilde{M}_{\alpha\alpha|\gamma}^{\mathrm{neq}} = - (1 + 2 \delta_{\alpha\gamma}) \rho R T \tau_h \partial_{\gamma}RT \Delta V. \label{eq:IIB13}
\end{align}
Note that the repeated index $\alpha\alpha$ in Eq. \eqref{eq:IIB13} does not imply Einstein summation. Each component is placed as an additional constraint in each translation dimension during the population reconstruction process. The contribution of each translational dimension can be expressed as
\begin{align}
    \widetilde{M}_{\alpha\alpha|\gamma}^{\mathrm{neq}} = \frac{1 + 2 \delta_{\alpha\gamma}}{D + 2} \widetilde{M}_{D\gamma}^{\mathrm{neq}}. \label{eq:IIB14}
\end{align}
For example, for the $D$-dimensional case
\begin{subequations}
\begin{align}
    \widetilde{M}_{xxx}^{\mathrm{neq}} &= \frac{3}{D+2} \widetilde{M}_{D\gamma}^{\mathrm{neq}}, \label{eq:IIB15a} \\
    \widetilde{M}_{yyx}^{\mathrm{neq}} = \widetilde{M}_{zzx}^{\mathrm{neq}} = \cdots &= \frac{1}{D+2} \widetilde{M}_{D\gamma}^{\mathrm{neq}}. \label{eq:IIB15b}
\end{align}
\end{subequations}

After all non-equilibrium moments are computed, the collision step is applied to each order of moments:
\begin{subequations}
\begin{align}
    \widetilde{M}_{0}^{*} &= \rho \Delta V, \label{eq:IIB16a} \\
    \widetilde{M}_{\alpha}^{*} &= 0, \label{eq:IIB16b} \\
    \widetilde{M}_{\alpha\beta}^{*} &= \rho R T \delta_{\alpha\beta} \Delta V + (1 - \omega_f) \widetilde{M}_{\alpha\beta}^{\mathrm{shear}} + (1 - \omega_b) \widetilde{M}_{\alpha\beta}^{\mathrm{bulk}}, \label{eq:IIB16c} \\
    \widetilde{M}_{\alpha\alpha|\gamma}^{*} &= (1 - \omega_h) \widetilde{M}_{\alpha\alpha|\gamma}^{\mathrm{neq}}, \label{eq:IIB16d} \\
    \widetilde{G}_{0}^{*} &= \rho K R T \Delta V + (1 - \omega_b) \widetilde{G}_0^{\mathrm{neq}}, \label{eq:IIB16e} \\
    \widetilde{G}_{\gamma}^{*} &= (1 - \omega_h) \widetilde{G}_{\gamma}^{\mathrm{neq}}. \label{eq:IIB16f}
\end{align}
\end{subequations}
Here, $\omega_f$ and $\omega_b$ denote the collision frequencies for shear and bulk viscosity, and $\omega_h$ for heat conduction. To ensure the Gaussian shape of the post-collision phonon distribution function, additional second-order $G$ moments are imposed as
\begin{align}
    \widetilde{G}_{\alpha\beta}^{*} = \rho K R T^{2} \delta_{\alpha\beta} \Delta V. \label{eq:IIB17}
\end{align}

For convenience during the population reconstruction step, all moments involved in the collision process are normalized by
\begin{align}
    \{ \bm{M}^{\prime},\, \bm{G}^{\prime} \} = \left\{ \tfrac{1}{\rho \Delta V} \bm{M},\, \tfrac{1}{\rho K R T \Delta V} \bm{G} \right\}. \label{eq:IIB18}
\end{align}

Since all relevant post-collision moments can be exactly recovered from the application of an appropriate population reconstruction technique, the resulting dynamics remain identical regardless of whether the collision has applied to raw or central moments, under properly imposed collision. However, in the present model, the central moment formulation has been adopted for the sake of simplicity and improving numerical precision. This approach is especially advantageous in high-speed flows, where using a relative frame of reference effectively removes bulk velocity contributions, which can reach several orders of magnitude larger than the non-equilibrium terms, and could cause significant precision loss from floating point limitations.

It has been observed that the population reconstruction technique proposed in Section~\ref{sec:epr} tends to fail at shock interfaces, due to excessively high non-equilibrium moment contribution. To mediate this issue, the Knudsen limiter has been applied to artificially equilibrate non-equilibrium components toward the continuum regime by adjusting $\omega$ toward $1$. The implementation of the limiter is explained in the Appendix~\ref{sec:knudsen}. Additionally, the bulk stress collision frequency $\omega_b$ is adjusted to $1$ for general problems throughout this paper, to suppress the acoustic effects responsible for oscillatory artifacts in the solution field.

\subsection{\label{sec:las}Lagrangian Acoustic Stencils}

One of the most fundamental stability limitations of the LBM has long been expected to be associated with the achievable Mach number. This constraint has very often been confused with the Courant number limitation, typically seen in explicit finite-difference methods, where the Courant number, defined below, is recommended to remain below unity for numerical stability.
\begin{align}
    \mathrm{Co} = (|u| + c_s)\tfrac{\Delta t}{\Delta x} \label{eq:IIC1}
\end{align}

As defined in Eq.~\eqref{eq:IIC1}, the Courant number represents the ratio between the distance information can travel in each timestep and the size of the maximum spatial discretization. In Lagrangian streaming of LBM, this interpretation does not transfer directly. Historically, confusion has arisen because, in the standard $DdQ3^d$ lattice, populations move only to adjacent cells (or nodes) in each timestep, resembling an explicit forward‐time scheme. However, in multispeed lattices, the information can be transferred across multiple cells in a single timestep, due to the large particle velocity. Furthermore, additional confusion arises from the distinction between the population velocity $\bm{c}_i$ and the bulk velocity $\bm{u}$.  

To resolve decades-old confusion, this research introduces three new metrics for evaluating velocity stencils: the conformation number $n$, the lattice Courant number $\mathrm{Co_{lat}}$, and the interlattice Courant number $\mathrm{Co_{int}}$. If we expect a closely isotropic relative particle velocity $\bm{c}_i - \bm{u}$ spanned over the radius of $R_{\mathrm{ac}}$, the conformation number is defined as
\begin{align}
  n = \frac{||\overline{\bm{c}}-\bm{u}|-R_{\mathrm{ac}}|}{\sqrt{RT}}. \label{eq:IIC2}
\end{align}
$\overline{\bm{c}}$ denotes the mean value of the particle velocity $\bm{c}_i$. The denominator $\sqrt{RT}$ corresponds to the thermal speed, which is proportional to the speed of sound, and also the standard deviation of the Boltzmann distribution. This number indicates how well the constructed velocity stencils cover the distribution function. To ensure accurate construction of discretized distribution function $f_i$, $n>1$ is recommended. This condition conversely implies that the high temperature simulation is capable of using LBM under the construction of appropriate velocity stencils that satisfy $n>1$. Even without this concept, multispeed lattice developed over the past decades resulted in the satisfaction of this conformation condition \cite{Frapolli2017, Watari2006, Frapolli2014}.

Now, the lattice Courant number defined by
\begin{align}
  \mathrm{Co_{lat}} = \max_i | \bm{c}_i | \frac{\Delta t}{\Delta V^{1/D}}  \label{eq:IIC3}
\end{align}
describe the ratio between the maximum distance population propagating in each timestep and the generalized distance $\Delta V^{1/D}$. In standard $DdQ3^d$ lattices, $\mathrm{Co_{lat}}=\sqrt{D}$, and greater for multispeed lattice. This is analogous to the classical Courant number.

Similarly, the interlattice Courant number is defined as
\begin{align}
  \mathrm{Co_{int}} = |\overline{\bm{c}} - \bm{u}| \frac{\Delta t}{\Delta V^{1/D}},
  \label{eq:IIC4}
\end{align}
quantifying the deviation of the macroscopic flow speed from the mean particle speed. For both stability and accuracy, $\mathrm{Co_{int}}$ should remain below some value proportional to $\sqrt{RT}\tfrac{\Delta t}{\Delta x}$, which bridges the connection with the conformation number $n$. Even without the concept of $\mathrm{Co_{int}}$, earlier studies developed a shifted lattice \cite{Sun1998, Sun2000, Sun2003, Sun2004, Frapolli2017} to enable simulation under high bulk velocity, which consequently resulted in a reduction of interlattice Courant number. The confusion between $n$, $\mathrm{Co_{lat}}$ and $\mathrm{Co_{int}}$ with the classical Courant number $\mathrm{Co}$ led to the historical misconception that LBM requires $\mathrm{Co}\ll1$ or similarly $\mathrm{Ma}\ll1$ for stability.

\begin{figure}
    \includegraphics[width=\columnwidth]{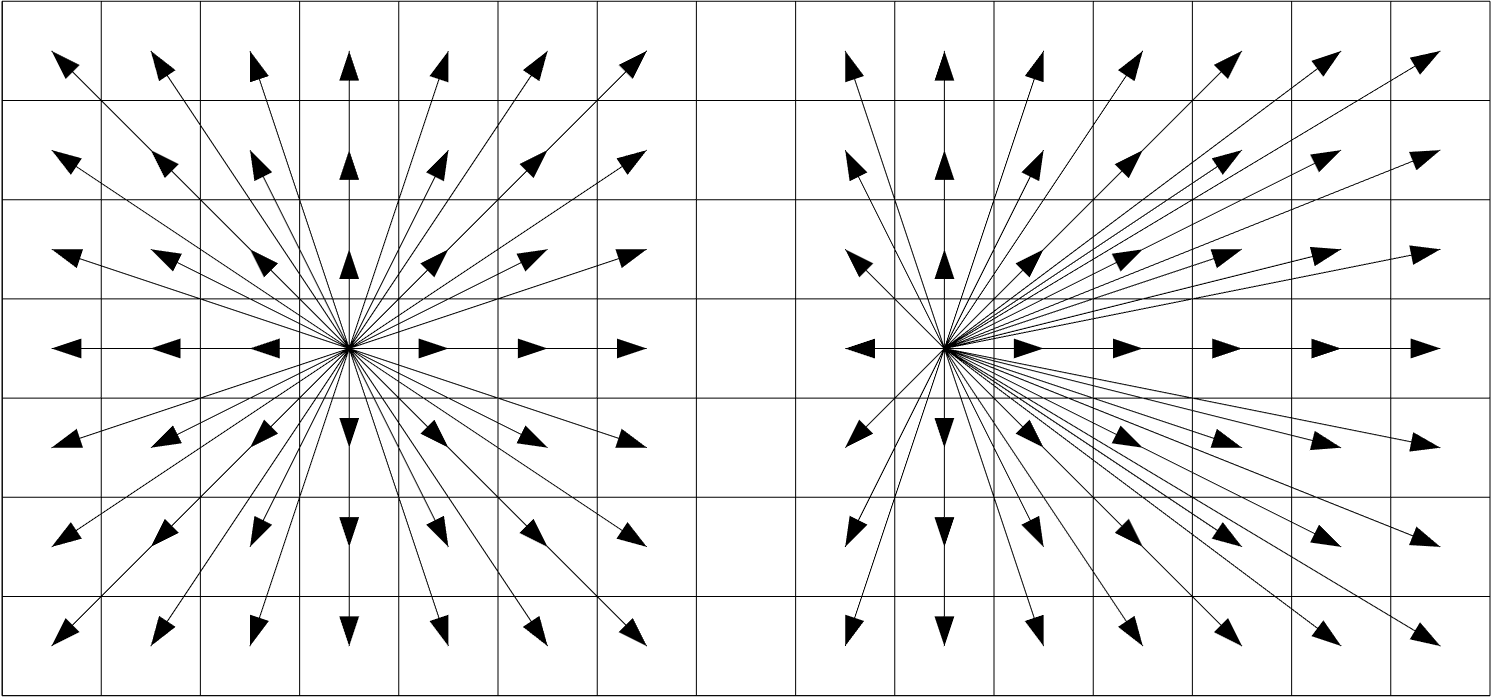}
    \caption{\label{fig:2} $D2Q49$ velocity stencils. The static stencil is shown on the left, and the shifted stencil with $U = 2$ is on the right.}
\end{figure}

Progressively, LBM community has adopted advanced velocity stencils to simulate supersonic flows while empirically satisfying both conformation number and the interlattice Courant number conditions. For instance, Frapolli \cite{Frapolli2017} employed a $DdQ7^d$ lattice with a shift of $U=2$ to simulate supersonic flow with entropic stabilization (Fig.~\ref{fig:2}). However, hardware memory limits soon restricted further stencil extension. Additionally, the globally uniform, statically designed velocity stencils that account for maximum anticipated velocity and temperature across the entire domain reduced the flexibility and increased the likelihood of reconstruction failures in local extremes.

To address this issue, Coreixas \cite{Coreixas2020} proposed a locally adaptive velocity stencil that shifts and extends the lattice based on the bulk velocity and the temperature of the cell, and using a Grad‐type reconstruction to recover the missing populations. However, from the author’s point of view, this recovery process is unnecessary when considering the Lagrangian nature of particle motion.

This section presents the novel velocity stencil construction technique, namely the Lagrangian acoustic stencils (LAS), which unifies the concept of a multispeed and shifted lattice. A key feature of LAS is the dynamic reconstruction of velocity stencils in each cell and time step, adapting to local flow conditions. This method is designed to minimize memory usage, enable local refinement for low-temperature regimes, and ensure high conformity of populations for smooth solution transitions. When combined with the population reconstruction technique and the moment streaming algorithms (Section~\ref{sec:epr} and \ref{sec:ms}), the resulting LELBM formulation enables the unbounded extension of velocity stencils to match local flow conditions, without the constraints of memory limitation.

\begin{figure}
    \includegraphics[width=0.75\columnwidth]{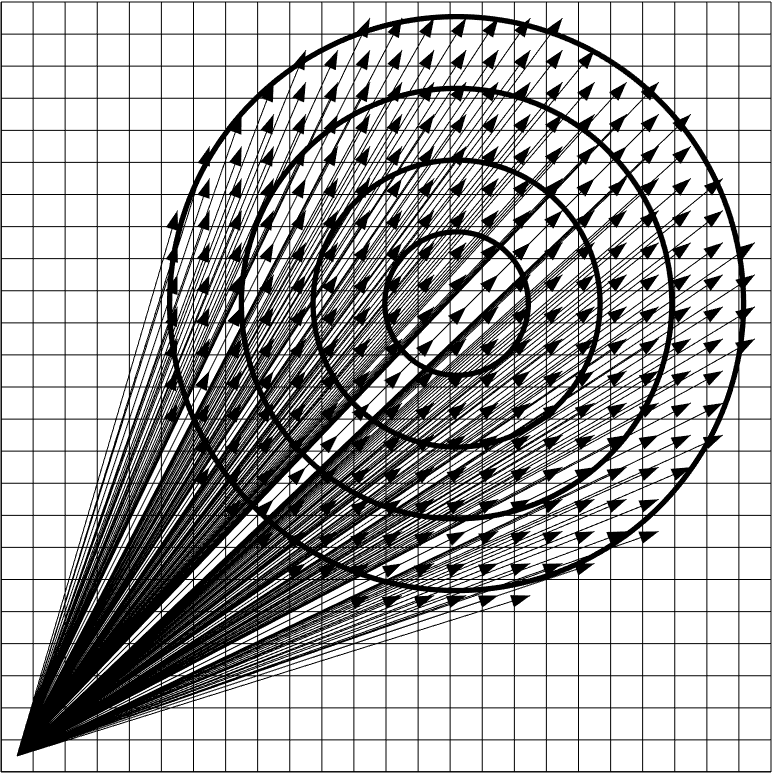}
    \caption{\label{fig:3} Velocity stencils constructed by LAS for $RT = 5$, $u_x = 13.7$, and $u_y = 14.1$, with $\mathrm{Co_{lat}}=28.7$. Each concentric circles indicate $R_{\mathrm{ac}}=n\sqrt{RT} \Delta t$, centered at $(x,\,y)=(u,\,v)\Delta t$, for $n=1,\,2,\,3,$ and $4$.}
\end{figure}

From the Chapman-Enskog expansion, we can expect the distribution function to be approximately in equilibrium under a small Knudsen number. For a gas of density $\rho$ in equilibrium with bulk velocity $\bm{u}$ and temperature $RT$, the distribution function follows the Boltzmann distribution as shown in Eq.~\eqref{eq:IIA3}. Statistically, this is equivalent to a multidimensional Gaussian distribution centered at $\bm{u}$, with a standard deviation of $\sqrt{RT}$ and scaled by $\rho$. Physically, the propagation of gas particles is acoustic and is governed by the local speed of sound, which is also proportional to $\sqrt{RT}$. Based on this insight, LAS constructs local velocity stencils that satisfy:
\begin{align}
    |c_i - [\bm{u}]|\Delta t \leq R_{\mathrm{ac}} = \max \left( \left[ n \sqrt{R T} \Delta t \right],\, 2 \Delta x \right). \label{eq:IIC6}
\end{align}
Here, $R_{\mathrm{ac}}$ is the acoustic radius, which defines the maximum distance of population that streams in the relative frame during each time step. The rounding operator $[ \cdot ]$ ensures an integer stencil radius. The conforming number $n$ controls how well the discrete stencil conforms to the continuous Gaussian distribution. Eq. \eqref{eq:IIC6} ensures that all velocity stencils are centered on the bulk velocity and span a region large enough to capture the significant portion of the distribution. Following the empirical rule of statistics\cite{Grafarend2006}, this work adopts $n=4$, which captures 99.99\% of a standard Gaussian distribution. The minimum radius $R_{\mathrm{ac}} \geq 2 \Delta x$ ensures that at least five stencil points align along any direction, sufficient for reconstructing up to third-order moments, while still maintaining the isotropy. This formulation ensures that the conformation number remains $n>1$, while $\mathrm{Co_{int}}$ is maintained below 0.5. This explains the seemingly Courant-free nature of the LELBM framework. Using LAS formulation, the resulting lattice Courant number is approximately
\begin{align}
    \mathrm{Co_{lat}}\approx (|\bm{u}|+R_{\mathrm{ac}}) \frac{\Delta t}{\Delta V^{1/D}}. \label{eq:IIC7}
\end{align}
In the remainder of the paper, $\mathrm{Co_{lat}}$ is estimated using Eq. \eqref{eq:IIC7}. Since the lower bound of the acoustic radius $R_{\mathrm{ac}}$ is $2\Delta x$, $\mathrm{Co_{lat}}$ is in minimum of 2 for the present formulation, and without the upper limit. Figure~\ref{fig:3} shows example LAS velocity stencils for $RT = 5$, $u_x = 13.7$, and $u_y = 14.1$, which contains a total of 284 velocity stencils in a 2D space. $\mathrm{Co_{lat}}=28.7$ for this case.

Asymptotically, for high temperature, the number of velocity stencils $q$ produced by LAS scales as
\begin{align}
  q \sim \left(n\sqrt{RT}\right)^D. \label{eq:IIC8}
\end{align}

\begin{figure}
    \includegraphics[width=0.5\columnwidth]{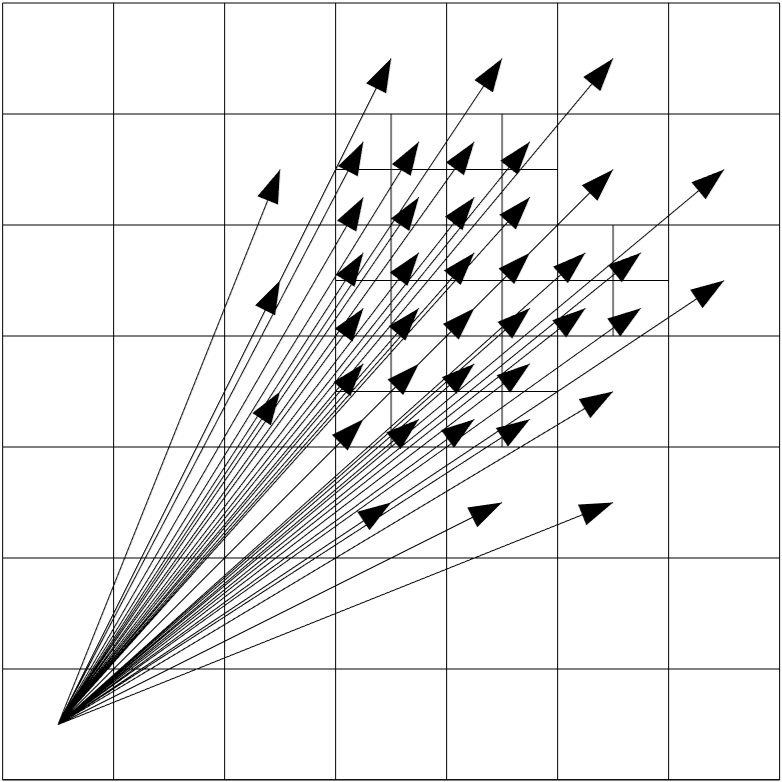}
    \caption{\label{fig:4} Refined velocity stencils for $RT = 0.2$, $u_x = 3.7$, and $u_y = 4.1$.}
\end{figure}

However, it has been found that EPR fails when the local temperature becomes $RT \lessapprox 0.4$ in uniform $\Delta C_i$. This is because the discrete velocity resolution becomes insufficient to capture the narrow Gaussian peak. To address this issue, the present work implemented internal refinement of velocity stencils, defined by the refinement number of
\begin{align}
    N_r = \left[ \frac{\Delta x}{3 \sqrt{RT} \Delta t} \right]. \label{eq:IIC9}
\end{align}
$N_r$ is the maximum refinement level and the bracket operator $[\cdot]$ denotes the rounding to the nearest integer. For each level of refinement, the stencils are refined by two if
\begin{align}
    |\bm{c}_i - [\bm{u}]|\Delta t \leq 3 \times  0.5^{\mathrm{Lvl}} \Delta x. \label{eq:IIC10}
\end{align}
Here, $\mathrm{Lvl}$ is the refinement level. This refinement process also factors the velocity volume by
\begin{align}
    \Delta C_i^{\mathrm{Lvl+1}} = 0.5^D \times \Delta C_i^{\mathrm{Lvl}}  \label{eq:IIC11}
\end{align}
The resulting refined stencils accurately reconstruct populations at moderately low temperature. Figure~\ref{fig:4} shows the refined stencils for $RT = 0.2$, $u_x = 3.7$, and $u_y = 4.1$, which contains a total of 41 velocity stencils.

Unfortunately, internal velocity refinement is not the panacea for low-temperature stability. Even with local refinements, EPR tends to fail around a temperature of $RT \lessapprox 0.15$ and introduce significant numerical viscosity from the insufficient resolution. The viscosity effects are discussed in Section~\ref{sec:shear_wave} and \ref{sec:thermal_wave}.

\subsection{\label{sec:epr}Entropic Population Reconstruction}

The essence of the moment-based collision method is the reconstruction of the post-collision population $f_i^*$ corresponding to $\bm{c}_i$, while accurately preserving the post-collision moment $M_N^*$. The most traditional method of inverse transformation involves the direct inversion of the moment basis matrix $\bm{T}$, as described in Eq.~\eqref{eq:IIB3}. However, this method requires $\bm{T}$ to be a square matrix of size $q \times q$, where $q$ is the number of populations, and equally the number of moment constraints. This constraint leads to high memory demand, increases computational cost, and more critically results in an overconstrained system that almost certainly produces nonphysical solutions, even when the post-collision moments are conserved. This limitation has confined the MRT method to mainly $DdQ3^d$ lattices.

The stability issues regarding the population reconstruction technique can broadly be classified into two categories: the failure of positivity preservation and the failure of reconstruction. The first category has been extensively studied over the past several decades \cite{Sterling1996, Brownlee2007}, and the positivity-preserving reconstruction techniques, which are consistent with the Boltzmann $H$-theorem\cite{Karlin1998, Karlin1998_Equil, Ansumali2000, Ansumali2002, Boghosian2003, Brownlee2007}, have been successfully developed to mitigate this issue. The failure of reconstruction is caused by the insufficient velocity stencils itself, typically related to the the dissatisfaction of conditions described in Section~\ref{sec:las}.

In 2020, Latt et al. introduced numerical equilibria (NE) to construct equilibrium populations for population-based collision\cite{Latt2020, Coreixas2020, Thyagarajan2023}, in which the theoretical foundation derived from the discrete velocity model (DVM) was developed many decades ago \cite{Broadwell1964, Mieussens2001, Mieussens2000, Dubroca2001}. DVM is the conceptual ancestor of the lattice Boltzmann method, which solves the Boltzmann equation in a large set of velocity stencils at each time step \cite{Tallec1997, Charrier2004, Baranger2014}. The essence of DVM involves the collision process from the exponential form of the equilibrium distribution function that can be derived from the maximum entropy principle (MEP)\cite{Karlin1998, Karlin1998_Equil, Press2013, upanovi2018}. Since the exponential form with a real exponent is strictly positive, this method inherently preserves positivity. Conceptually, this problem is equivalent to a truncated Hamburger moment problem using the maximum entropy principle \cite{Shohat1943, Akhiezer1965, Krein1977, Mead1984, Borwein1991}. Given that the populations in entropic LBM (ELBM) on a $DdQ3^d$ lattice arise from the same exponential form as in numerical equilibria, NE-LBM and ELBM share the same underlying foundation.

NE-LBM is the only method that successfully adopted moment-based collision for multispeed lattice. Since LELBM operates in a relative reference frame, the present method generalizes population reconstruction process to arbitrary velocity stencils and normalized central moments, extending prior absolute‐frame raw moment-based collision methods. For clarity and ease of reference, this generalized method is referred to as entropic population reconstruction (EPR) throughout this paper.

First, the $H$ functional is defined as
\begin{align}
    H = \iiint_{-\infty}^{\infty} f \ln \left(f/\rho\right) d\bm{\xi}. \label{eq:IID1}
\end{align}
By the Boltzmann $H$ theorem\cite{Boltzmann2003}, $H$ is always decreasing over time and is related to Gibbs entropy\cite{upanovi2018} by
\begin{align}
    S=-k_B H, \label{eq:IID2}
\end{align}
where $k_B$ is the Boltzmann constant. Now, the volume-integrated $H$ can be expressed as
\begin{align}
    \langle H \rangle = \iiint_{\Delta V} \iiint_{-\infty}^{\infty} f \ln \left(f/\rho\right) d\bm{\xi} dV. \label{eq:IID3}
\end{align}

From the previous definition of the LELBM population described in Eqs.~\eqref{eq:IIA7}--\eqref{eq:IIA11} and the normalization process in Eq. \eqref{eq:IIB18}, the normalized entropy functional can be expressed by the discrete population of fluons and phonons.
\begin{align}
    \left\{\langle H_f \rangle',\,\langle H_g \rangle'\right\} \approx \sum_i \left\{f_i' \ln  \left(f_i'/\Delta C_i\right),\, g_i' \ln  \left(g_i'/\Delta C_i\right)\right\}, \label{eq:IID4}
\end{align}
where $f_i'$ and $g_i'$ are the normalized populations. The result of Eq. \eqref{eq:IID4} corresponds to the form of the Shannon entropy\cite{Shannon1948_1, Shannon1948_2}, weighted by $\Delta C_i$. Since the entropy must be maximized under local moment constraints, the entropy-satisfying populations can be obtained by applying Lagrange multipliers to minimize $\left\{\langle H_f \rangle',\, \langle H_g \rangle' \right\}$.
\begin{align}
    \frac{\partial \langle H \rangle'}{\partial f_i'} + \frac{\partial}{\partial f_i'} \sum_j \tilde{\lambda}_{f,j} \widetilde{M}_j' = 0 \label{eq:IID5}, \\
    \sum_i T_j(\tilde{c}_i) f_i' = \widetilde{M}_j', \label{eq:IID6}
\end{align}
where $\tilde{\lambda}_{f,j}$ is the Lagrange multiplier for $f_i'$, corresponding to each $j$th moment constraint. $\widetilde{\bm{c}}_i$ is the relative velocity defined as $\widetilde{\bm{c}}_i = \bm{c}_i - \bm{u}$, used for the central moment basis function. The same approach applies to the phonon. By relating Eqs. \eqref{eq:IID4}--\eqref{eq:IID6}, the general form of the population is derived:
\begin{align}
    f_i' = \Delta C_i \exp\left[-\left(1 + \sum_j \tilde{\lambda}_{f,j} T_j(\tilde{c}_i) \right)\right]. \label{eq:IID7}
\end{align}
Now, using Newton-Raphson iteration \cite{Latt2020}, Eqs. \eqref{eq:IID5} and \eqref{eq:IID6} can be used to solve Eq. \eqref{eq:IID7}.
\begin{align}
    \sum_k T_i(\tilde{c}_k) \sum_j \Delta\tilde{\lambda}_{f,j} f_k^{\prime(n)} T_j(\tilde{c}_k) &= \sum_k T_i(\tilde{c}_k) f_k^{\prime(n)} - \tilde{M}_i', \label{eq:IID8} \\
    \tilde{\lambda}^{(n+1)}_{f,j} &= \tilde{\lambda}^{(n)}_{f,j} + \Delta\tilde{\lambda}^{(n)}, \label{eq:IID9}
\end{align}
where $f_k^{\prime(n)}$ is the population at $n$th iteration and $\Delta\tilde{\lambda}^{(n)}_{f,j}$ is the change of $\tilde{\lambda}_{f,j}$ in each iteration. In matrix form, Eq. \eqref{eq:IID8} can be expressed as
\begin{align}
    \widetilde{\bm{A}}_f  \delta\widetilde{\bm{\Lambda}}_f = \delta\widetilde{\bm{M}}, \label{eq:IID10}
\end{align}
where $\bm{A}_f$, $\delta\bm{\Lambda}_f$, and $\delta\bm{M}$ are matrix and vectors defined by
\begin{subequations}
\begin{align}
    \widetilde{A}_{f,ij} &= \sum_k T_i(\tilde{\bm{c}}_k) f_k'^{(n)} T_j(\tilde{\bm{c}}_k), \label{eq:IID11a} \\
    \delta\widetilde{\Lambda}_{f,j} &= \Delta\tilde{\lambda}^{(n)}_{f,j}, \label{eq:IID11b} \\
    \delta \widetilde{M}_i &= \sum_k T_i(\tilde{\bm{c}}_k) f_k'^{(n)} - \widetilde{M}_i'. \label{eq:IID11c}
\end{align}
\end{subequations}
$\widetilde{\bm{A}}_f$ is a matrix of size $N_f \times N_f$, where $N_f$ is the number of relevant fluon moments under consideration. As we can see, the matrix defined as Eq. \eqref{eq:IID11a} must be recomputed in every iteration, which is the source of the major computational overhead. Theoretically, the computational complexity of each iteration is dominated by the process of constructing $\widetilde{\bm{A}}_f$ and $\widetilde{\bm{A}}_g$, which is, for a $q$ number of velocity stencils, the cost becomes $\mathcal{O}(qN^2_f)$ and $\mathcal{O}(qN^2_g)$, respectively. In addition, the direct inversion of matrix costs $\mathcal{O}(N^3_f)$ and $\mathcal{O}(N^3_g)$. Noticing that $\{N_f,\,N_g\} \ll q$ due to the large number of velocity stencils generated by LAS, we expect that direct inversion of the matrix does not contribute significantly to the computational cost. For both fluons and phonons, two independent iterations must be carried out simultaneously for the polyatomic fluid, using $\{\widetilde{\bm{M}}^{*\prime},\, \widetilde{\bm{G}}^{*\prime}\}$ as constraint.

The Lagrange multiplier matrix $\widetilde{\bm{A}}$ is dense, but small, symmetric, and positive definite. Therefore, the Cholesky decomposition \cite{Golub2013} has been applied to directly solve the matrix efficiently. The residual of moments are set to $10^{-12}$ for simulations throughout this paper.

Preconditioning can help reduce the number of iterations. By approximating the distribution function to be Boltzmann and abiding the discretization definition Eq. \eqref{eq:IIA11}, we make a good approximation of Eq. \eqref{eq:IID7} to be
\begin{align}
    f_i^{\prime} &= \Delta C_i \exp\left[-\left(1 + \sum_j \tilde{\lambda}_j T_j(\tilde{\bm{c}}_i)\right)\right] \notag \\
    &\approx \frac{\Delta C_i}{(2\pi RT)^{D/2}} \exp\left(-\frac{|\tilde{\bm{c}}_i|^2}{2RT}\right). \label{eq:IID12}
\end{align}
Then we can approximate $\tilde{\lambda}$ as
\begin{align}
    &\left\{\tilde{\lambda}_0,\,\tilde{\lambda}_\alpha,\,\tilde{\lambda}_{\alpha\beta},\,\tilde{\lambda}_{\alpha\alpha|\gamma}\right\} = \left\{-1 + \tfrac{D}{2} \ln(2\pi RT),\,0,\,\tfrac{1}{2RT} \delta_{\alpha\beta},\,0 \right\}. \label{eq:IID13}
\end{align}
This approximation applies to both $f_i^{\prime}$ and $g_i^{\prime}$. With preconditioning, the convergence is generally achieved in less than five iterations, regardless of local cell conditions. Even faster convergence is achievable by explicitly storing the Lagrange multipliers for each cell as a precondition for the next timestep, with increased memory demand.

\subsection{\label{sec:ms}Moment Streaming}

Another challenge in the development of the multispeed LBM lies in their intensive memory demands. For example, the conventional population-based approach is limited to the $D3Q343$ lattice, due to hardware limitation, while remaining stable only up to a Mach number of $\mathrm{Ma} \approx 2$ \cite{Frapolli2017}. This configuration requires the storage of at least 686 populations per cell to simulate the thermohydrodynamics of compressible polyatomic gases. Such an enormous memory demand becomes a significant restriction, making the adoption of a higher-order lattice impractical.

To overcome this limitation, the LELBM framework implemented a moment streaming (MS) algorithm, inspired by the idea of Valero-Lara \cite{ValeroLara2017}. This method fundamentally reduces memory demand by storing only the necessary moments instead of entire populations.

According to Grad’s moment closure theory\cite{Grad1949}, the minimal set of moments required to reconstruct the NSF equations in three-dimensional translational space includes: one zeroth-order moment $M_0$ for mass conservation, three first-order moments $M_\alpha$ for momentum, six second-order moments $M_{\alpha\beta}$ for stress and energy, and three contracted third-order moments $M_{D\gamma}$ for heat flux. The derivation of the NSF equations using this closure is explained in details in Appendix~\ref{sec:chapman}. While the classical multispeed LBM typically discretizes hundreds of populations, the moment closure implies that the essential continuum physics can accurately be represented with only 13 moments with 4 additional components required for DDF models: $G_0$ and $G_\gamma$.

In the classical LBM framework, the streaming step involves the transferring of post-collision populations to their designated cell at each timestep:
\begin{align}
    f_i(\bm{x} + \bm{c}_i \Delta t,\, \bm{c}_i,\, t + \Delta t) \leftarrow f_i^*(\bm{x},\, \bm{c}_i,\, t). \label{eq:IIE1}
\end{align}
Because the Lagrangian streaming process is collisionless, it is intrinsically stable, regardless of Courant number. This explains the Courant-free nature of LBM formulation, under seemingly high lattice Courant number. The only requirement for stability in LBM is the successful and accurate population reconstruction process.

Besides, for moment-based collision, the moment transformation is typically performed in the next timestep using the stored pre-collision populations. The moment streaming algorithm combines the processes of streaming and moment transformation into a single operation. Instead of streaming and storing the full sets of distribution functions, only the fraction of raw moments $\{\delta M_{N,i},\, \delta G_{N,i}\}$ are streamed accordingly
\begin{subequations}
\begin{align}
    &\{\delta M_{N,i},\, \delta G_{N,i}\}(\bm{x},\, t) \notag \\
    &= T_N(\bm{c}_i) \left\{ \rho \Delta V f_i^{*\prime},\, \rho K RT \Delta V g_i^{*\prime} \right\}(\bm{x},\, \bm{c}_i,\, t), \label{eq:IIE3a} \\
    &\{M_N,\, G_N\}(\bm{x} + \bm{c}_i \Delta t,\, t + \Delta t) \notag \\
    &=\{M_N,\, G_N\}(\bm{x} + \bm{c}_i \Delta t,\, t + \Delta t) + \{\delta M_{N,i},\, \delta G_{N,i}\}(\bm{x},\, t). \label{eq:IIE3b}
\end{align}
\end{subequations}
This procedure ensures that only the relevant moments are stored, while the moment transformation process is implicitly applied during the streaming process.

The memory requirement for this approach in a general $D$-dimensional simulation is reduced to a fixed quantity of $\tfrac{1}{2}(D^2 + 5D + 2)$ memory units, with additional $D + 1$ memory units for polyatomic fluids. This represents a substantial compression compared to the conventional streaming method, where the number of populations stored scales as $\mathcal{O}(RT^{D/2})$. The extensive discussion about the maximum and the minimum available temperature is provided in the thesis of Frapolli\cite{Frapolli2017}. For example, the $D3Q343$ lattice required the minimum of $686$ memory units for polyatomic fluid, with a maximum temperature of $RT \approx 1.63$\cite{Frapolli2017}. By combining LAS with the MS algorithm, LELBM can simulate compressible flows without an upper temperature limit, while requiring only $13+4$ memory units for three-dimensional problems, effectively transcending the classical hardware limitations faced by many previous supersonic multispeed LBM models.

\subsection{\label{sec:imple}Implementations}

\begin{figure}
  \centering
  \begin{tikzpicture}[node distance=1em]
    \node[block] (raw_moments) {$(0)$Raw moments\\$\{\bm{M},\,\bm{G}\}(\bm{x},\,t)$};
    \node[block, below=of raw_moments] (col_mod) {$(1)$ Collision module\\$\{\widetilde{\bm{M}}^{*\prime},\,\widetilde{\bm{G}}^{*\prime}\},\,\{\rho,\,\bm{u},\,RT\}$};
    \node[block, right=of col_mod] (las_mod) {$(2)$ LAS module\\$\bm{c}_i=\bm{c}_i(\bm{u},\,RT)$};
    
    \path (col_mod.east) -- (las_mod.west) coordinate[midway] (midpoint);
    \node[wide block, below=of midpoint, yshift=-2.25em] (epr_mod) {$(3)$ EPR module\\$\left\{f_i^{*\prime},\,g_i^{*\prime}\right\}=\left\{f_i^{*\prime}(\bm{c}_i,\,\widetilde{\bm{M}}^{*\prime}),\,g_i^{*\prime}(\bm{c}_i,\,\widetilde{\bm{G}}^{*\prime})\right\}$};
    
    \node[wide block, below=of epr_mod] (ms_mod) {$(4)$ MS module\\$\{\bm{M},\,\bm{G}\}(\bm{x}+\bm{c}_i\Delta t,\,t+\Delta t)\mathrel{+}=\{\delta \bm{M}^*_i,\, \delta \bm{G}^*_i \}(\bm{x},\,t)$};

    \draw[arrow] (raw_moments) -- (col_mod);
    \draw[arrow] (col_mod) -- (las_mod);
    \draw[arrow] (col_mod) -- (epr_mod);
    \draw[arrow] (las_mod) -- (epr_mod);
    \draw[arrow] (epr_mod) -- (ms_mod);
    \draw[arrow] (col_mod.west) -- ++(-1em,0) |- (ms_mod.west);
    \draw[arrow] (las_mod.east) -- ++(1em,0) |- (ms_mod.east);
  \end{tikzpicture}
  \caption{\label{fig:LELBM_algorithm}Modularized LELBM flowchart. Numbering indicate the sequence of operations and arrows indicate the transfer of data for next process.}
\end{figure}
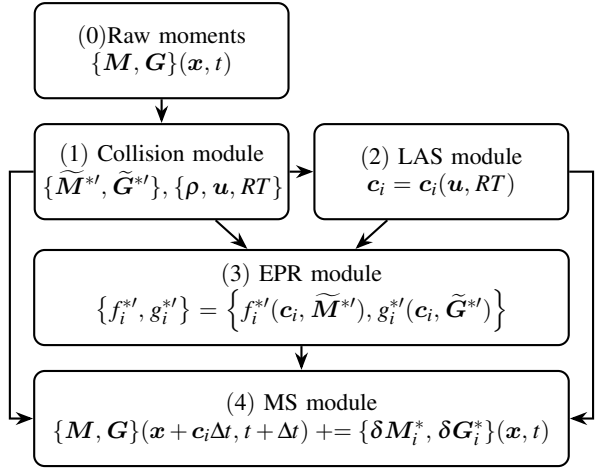

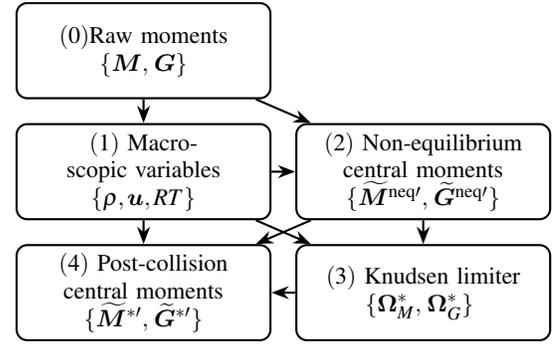
\begin{figure}
  \centering
  \begin{tikzpicture}[node distance=1em]
    \node[block] (raw_moments) {$(0)$Raw moments\\$\{\bm{M},\,\bm{G}\}$};
    \node[block, below=of raw_moments] (mac_var) {$(1)$ Macroscopic variables\\$\{\rho,\bm{u},RT\}$};
    \node[block, right=of mac_var] (neq_cent_mmt) {$(2)$ Non-equilibrium central moments\\$\{\widetilde{\bm{M}}^{\mathrm{neq}\prime},\,\widetilde{\bm{G}}^{\mathrm{neq}\prime}\}$};
    \node[block, below=of neq_cent_mmt] (knud_lim) {$(3)$ Knudsen limiter\\
    $\hspace{-0.3em}\{\bm{\Omega}_M^*,\,\bm{\Omega}_G^* \}$};
    \node[block, below=of mac_var] (col_fin) {$(4)$ Post‐collision central moments\\
    $\{\widetilde{\bm{M}}^{*\prime},\,\widetilde{\bm{G}}^{*\prime}\}$};

    \draw[arrow] (raw_moments) -- (mac_var);
    \draw[arrow] (raw_moments) -- (neq_cent_mmt);
    \draw[arrow] (mac_var) -- (neq_cent_mmt);
    \draw[arrow] (mac_var) -- (knud_lim);
    \draw[arrow] (mac_var) -- (col_fin);
    \draw[arrow] (neq_cent_mmt) -- (knud_lim);
    \draw[arrow] (neq_cent_mmt) -- (col_fin);
    \draw[arrow] (knud_lim) -- (col_fin);
  \end{tikzpicture}
  \caption{\label{fig:collision_module}Collision module flowchart.}
\end{figure}

The Lagrangian entropic lattice Boltzmann method fully utilizes its inherent Lagrangian formulation in the relative frame of reference to achieve a Courant‐free, explicitly time‐stepping algorithm. Although the theoretical foundations are profound, the resulting implementation is remarkably straightforward and "almost local", making it exceptionally well suited to large‐scale parallel computing. The LELBM formulation can be decomposed into a sequence of modular processes. The complete single‐cell operation is shown in Figure~\ref{fig:LELBM_algorithm}. This section explains a simplified algorithmic workflow for each modules:

\textbf{Collision Module}: As shown in Figure~\ref{fig:collision_module}, the raw moments are used to extract macroscopic variables $\{\rho,\,\bm{u},\,RT\}$ and obtain normalized post-collision central moments: $\{\widetilde{\bm{M}}^{*\prime},\,\widetilde{\bm{G}}^{*\prime}\}$. Knudsen limiter explained in Appendix~\ref{sec:knudsen} is applied during the collision process if the local Knudsen number is found to be excessive.

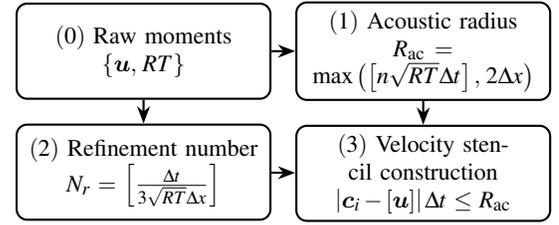
\begin{figure}
  \centering
  \begin{tikzpicture}[node distance=1em]
    \node[block] (mac_var) {$(0)$ Raw moments\\$\{\bm{u},\,RT\}$};
    \node[block, right=of mac_var] (acous_rad) {$(1)$ Acoustic radius\\$R_{\mathrm{ac}} = \max \left( \left[ n \sqrt{RT} \Delta t \right],\, 2 \Delta x \right)$};
    \node[block, below=of mac_var] (ref_num) {$(2)$ Refinement number\\$N_r = \left[ \tfrac{\Delta t}{3 \sqrt{RT} \Delta x} \right]$};
    \node[block, below=of acous_rad] (vel_stenc) {$(3)$ Velocity stencil construction\\$\left|\bm{c}_i - [\bm{u}] \right|\Delta t \leq R_{\mathrm{ac}}$};

    \draw[arrow] (mac_var) -- (acous_rad);
    \draw[arrow] (mac_var) -- (ref_num);
    \draw[arrow] (acous_rad) -- (vel_stenc);
    \draw[arrow] (ref_num) -- (vel_stenc);
  \end{tikzpicture}
  \caption{\label{fig:las_module}LAS module flowchart.}
\end{figure}

\textbf{Lagrangian Acoustic Stencils (LAS) Module}: The velocity and temperature obtained from the collision module is transferred to the LAS module (Figure~\ref{fig:las_module}), which constructs an optimized set of discrete velocity stencils $\bm{c}_i$, optimally tailored for local hydroacoustic properties. To reduce computational cost in the upcoming population reconstruction step, one may reduce the conforming number $n$, at the expense of minor oscillatory numerical artifacts.

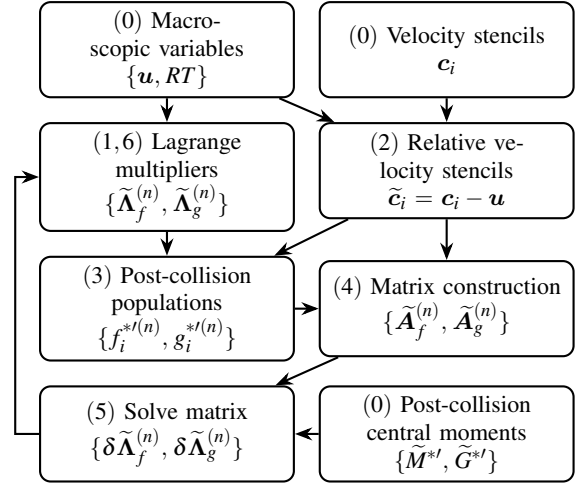
\begin{figure}
  \centering
  \begin{tikzpicture}[node distance=1em]
    \node[block] (mac_var) {$(0)$ Macroscopic variables\\$\{\bm{u},\,RT\}$};
    \node[block, right=of mac_var] (vel_stenc) {$(0)$ Velocity stencils\\$\bm{c}_i$};
    \node[block, below=of mac_var] (lag_mul) {$(1, 6)$ Lagrange multipliers\\
    $\{\widetilde{\bm{\Lambda}}^{(n)}_f,\,\widetilde{\bm{\Lambda}}^{(n)}_g\}$};
    \node[block, below=of vel_stenc] (rel_vel) {$(2)$ Relative velocity stencils\\$\widetilde{\bm{c}}_i=\bm{c}_i-\bm{u}$};
    \node[block, below=of lag_mul] (pops) {$(3)$ Post-collision populations\\
    $\{f_i^{*\prime(n)},\,g_i^{*\prime(n)}\}$};
    \node[block,right=of pops] (mtrx) {$(4)$ Matrix construction\\$\{\widetilde{\bm{A}}_f^{(n)},\,\widetilde{\bm{A}}_g^{(n)}\}$};
    \node[block, below=of pops] (slv_mtrx) {$(5)$ Solve matrix\\
    $\{\delta\widetilde{\bm{\Lambda}}_f^{(n)},\,\delta\widetilde{\bm{\Lambda}}_g^{(n)}\}$};
    \node[block, right=of slv_mtrx] (post_col_mmt) {$(0)$ Post-collision central moments\\$\{\widetilde{M}^{*\prime},\,\widetilde{G}^{*\prime}\}$};

    \draw[arrow] (mac_var) -- (lag_mul);
    \draw[arrow] (mac_var) -- (rel_vel);
    \draw[arrow] (vel_stenc) -- (rel_vel);
    \draw[arrow] (rel_vel) -- (mtrx);
    \draw[arrow] (rel_vel) -- (pops);
    \draw[arrow] (lag_mul) -- (pops);
    \draw[arrow] (pops) -- (mtrx);
    \draw[arrow] (mtrx) -- (slv_mtrx);
    \draw[arrow] (post_col_mmt) -- (slv_mtrx);
    \draw[arrow] (slv_mtrx.west) -- ++(-1em,0) |- (lag_mul.west);
  \end{tikzpicture}
  \caption{\label{fig:epr_module}EPR module flowchart.}
\end{figure}

\textbf{Entropic Population Reconstruction (EPR) Module}: From normalized post‐collision central moments $\{\widetilde{\bm{M}}^{*\prime},\,\widetilde{\bm{G}}^{*\prime}\}$ and Lagrangian acoustic stencils $\bm{c}_i$, the EPR module reconstructs the normalized post‐collision population iteratively, according to the procedure in Figure~\ref{fig:epr_module}.

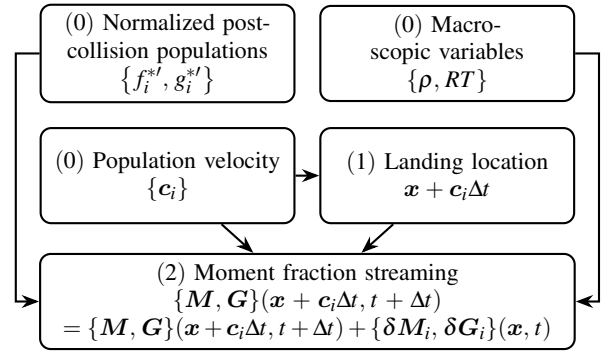
\begin{figure}
  \centering
  \begin{tikzpicture}[node distance=1em]
    \node[block] (pops) {$(0)$ Normalized post-collision populations\\$\left\{f_i^{*\prime},\,g_i^{*\prime}\right\}$};
    \node[block, right=of pops] (mac) {$(0)$ Macroscopic variables\\$\{\rho,\, RT\}$};
    \node[block, below=of pops] (pop_vel) {$(0)$ Population velocity\\$\{\bm{c}_i\}$};
    \node[block, right=of pop_vel] (landing) {$(1)$ Landing location\\$\bm{x}+\bm{c}_i\Delta t$};

    \path (pop_vel.east) -- (landing.west) coordinate[midway] (midpoint);
    \node[wide block, below=of midpoint, yshift=-2.25em] (frac_mmt) {$(2)$ Moment fraction streaming\\$\{\bm{M},\, \bm{G}\}(\bm{x} + \bm{c}_i \Delta t,\, t + \Delta t)$\\$=\{\bm{M},\, \bm{G}\}(\bm{x} + \bm{c}_i \Delta t,\, t + \Delta t) + \{\delta \bm{M}_i,\, \delta \bm{G}_i\}(\bm{x},\, t)$};

    \draw[arrow] (pop_vel) -- (landing);
    \draw[arrow] (pop_vel) -- (frac_mmt);
    \draw[arrow] (landing) -- (frac_mmt);
    \draw[arrow] (pops.west) -- ++(-1em,0) |- (frac_mmt.west);
    \draw[arrow] (mac.east) -- ++(1em,0) |- (frac_mmt.east);
  \end{tikzpicture}
  \caption{\label{fig:ms_module}MS module flowchart.}
\end{figure}

\textbf{Moment Streaming (MS) Module}: Finally, the reconstructed fractional moments are streamed along their respective positions during the streaming step. This step is the only non‑local operation in the LELBM framework. In practical CUDA implementations, the \texttt{atomicAdd} function has been used to ensure data consistency.

\subsection{\label{sec:compare}Comparison with Other Supersonic LBM Models}

\begin{table*}
  \caption{\label{tab:framework_compare}Comparison of key characteristics of supersonic LBM frameworks.}
  \begin{ruledtabular}
  \begin{tabular}{l *{10}{c}}
    Framework   & Energy  & Flux   & Lattice  & Streaming & Frame     & Collision & Population & Stabilization & Memory   \\
                & model   & recon. & type     &           &           &           & recon.     &               &          \\
    \hline
    HRR-LBM\cite{Feng2019, Guo2020, Renard2021, Guo2024}     & Hybrid  & —      & Static   & V–V\footnote{Vertices}       & Absolute  & $\bm{f}$                & Hermite    & RR            & $\bm{f},\,T$\footnote{Temperature} \\
    SLLBM\cite{Kraemer2017, Wilde2020, Wilde2021_1, Wilde2021_2, Spelten2024}       & DDF     & —      & Static   & C–I\footnote{Intracell}       & Absolute  & $\bm{f},\,\bm{g}$       & Hermite    & —             & $\bm{f},\,\bm{g}$ \\
    ELBM\cite{Frapolli2015, Frapolli2016, Frapolli2017, Frapolli2020, Hosseini2023, Hosseini2024}        & DDF     & —      & Static   & C–C\footnote{Cell-center}       & Absolute  & $\bm{f},\,\bm{g}$       & Hermite    & Entropic      & $\bm{f},\,\bm{g}$ \\
    NE-LBM\cite{Latt2020, Coreixas2020, Thyagarajan2023}      & DDF     & —      & Static   & C–C       & Absolute  & $\bm{M},\,\bm{G}$       & NE         & Entropic, Kn. & $\bm{f},\,\bm{g}$ \\
    PonD/DUGKS\cite{Kallikounis2022, Kallikounis2024}  & DDF     & Yes    & Adaptive & C–C       & Relative  & $\widetilde{\bm{f}},\,\widetilde{\bm{g}}$ & Hermite    & Entropic      & $\bm{f},\,\bm{g}$ \\
    LELBM       & DDF     & —      & Adaptive & C–C/I     & Relative  & $\widetilde{\bm{M}},\,\widetilde{\bm{G}}$ & NE         & Entropic, Kn. & $\bm{M},\,\bm{G}$ \\
  \end{tabular}
  \end{ruledtabular}
\end{table*}

The Lagrangian Entropic Lattice Boltzmann Method combined many advanced techniques to fully utilize its Lagrangian formulation while strictly abiding the physical principles to maintain both robustness and stability at high Courant numbers. In this section, the LELBM framework is compared with several contemporary supersonic LBM variants. Table~\ref{tab:framework_compare} summarizes key features of LELBM with hybrid recursive regularized LBM (HRR‐LBM) \cite{Feng2019, Guo2020, Renard2021, Guo2024}, semi‐Lagrangian LBM (SLLBM) \cite{Kraemer2017, Wilde2020, Wilde2021_1, Wilde2021_2, Spelten2024}, entropic LBM (ELBM) \cite{Frapolli2015, Frapolli2016, Frapolli2017, Frapolli2020, Hosseini2023, Hosseini2024}, numerical equilibria (NE‐LBM) \cite{Latt2020, Coreixas2020, Thyagarajan2023}, and particle on demand/discrete unified gas kinetics scheme (PonD/DUGKS) \cite{Kallikounis2022, Kallikounis2024} method.

The characteristic of HRR-LBM is the stabilization technique that filters out higher-order ghost moments through recursive regularizations \cite{Dellar2003}. Although the term "regularization" might imply the connection with regularized central moment collision (Section~\ref{sec:collision}), the term has been used by definition of decoupling and regularizing the non-equilibrium moments to enforce exact fluid properties. LELBM rather embraces the higher-order ghosts within its entropic stabilization framework. Additionally, HRR‐LBM employs finite‐volume formulation to solve energy equation on each cell. In contrast, like other thermal lattice Boltzmann methods, LELBM adopted a DDF approach to represent energy contribution from internal DoF.

As another approach, SLLBM streams the distribution functions from cell centers to intracell Gauss-Lobatto-Chebyshev nodes \cite{Wilde2020, Wilde2021_1, Wilde2021_2}, and perform numerical integration on populations (or distribution function) to improve stability. Although the present framework uses first‐order integration over the velocity space at each lattice site, this idea is analogous to velocity‐volume integration in present model. Additionally, this intracell streaming unintentionally shares a similar concept with the internal velocity refinement strategy in LAS. However, node‐based integration of SLLBM sacrifices moment conservation \cite{Wilde2020} by neglecting velocity-space resolution and relying solely on spatial quadrature at interlattice nodes, unlike LELBM unconditionally maintaining moment conservation. The terminology "Lagrangian" mainly derived from the characteristic of moment streaming in high Courant number flow has little connection with quadrature integration as in SLLBM.

Comparing ELBM, NE‐LBM, and LELBM, we can see that since all of them employ a maximum entropy principle to stabilize post‐collision populations, all methods diverge in their treatment of equilibrium and collision. ELBM begins with an exponential‐form equilibrium derived via Lagrange multipliers for a $DdQ3^d$ lattice \cite{Ansumali2003, Frapolli2020, Hosseini2023}. However, lacking an analytical expression for multispeed lattices, the method eventually reverted to a Hermite polynomial expansion. It then enforced the entropy principle to the post‐collision populations, while retaining a traditional population‐level collision, and introduced quasi-populations to correct thermal conductivity.

Unlike ELBM, NE-based methods maintained the exponential form for multispeed lattice, and applied iterative solvers to acquire correct Lagrange multipliers. Early NE‐LBM formulations used an entropy‐based reconstruction solely to obtain the equilibrium populations and applied collision directly at the population level \cite{Latt2020}. The recent NE‐LBM model extended this entropic formulation by performing collisions on the raw moments and then reconstructing the post‐collision populations from the updated moment set. This moment-based collision effectively removes the reliance on quasi-populations.

The key distinction between LELBM and NE-LBM lies in the use of adaptive velocity stencils, population reconstruction within a relative reference frame, and subsequent collision on central moments. In contrast, both the ELBM and the NE-LBM frameworks operate on a fixed lattice. Although prior research on NE-LBM mentioned a potential extension to adaptive lattices \cite{Latt2020, Coreixas2020}, further discussion on this topic remains limited.

Finally, NE‐LBM introduced the Knudsen number-based viscosity enhancement near discontinuities as an explicit stabilization \cite{Latt2020, Coreixas2020, Thyagarajan2023}. Although the calculation of the Knudsen number and its treatment differ slightly, this concept is also shared with LELBM. However, ELBM did not apply such explicit enhancement but solely relied on entropic control on collision frequency, yet yielding similar effect to turbulent viscosity, as proposed in the Smagorinsky model \cite{Hosseini2023}.

PonD/DUGKS is an integrated model of PonD, DUGKS, and ELBM. In essence, PonD is the method that shifts between the absolute and relative reference frame, while DUGKS is the scheme that applied the population flux evaluation on cell-faces, just like in the finite-volume method \cite{Guo2013, Guo2015, Guo2021}. PonD and LELBM share some similarities in their concepts, especially in their relative reference frame. However, the key distinction arises again from the collision method. Since PonD/DUGKS is developed upon ELBM, the fundamental differences are identical. Additionally, PonD constructs populations based on thermally scaled moments, and the scaling reversion process introduces an interpolation error that slightly violates moment conservation \cite{Kallikounis2022}, while LELBM perfectly satisfies. The principal limitation of LELBM lies in the low‐temperature (or equally a small $\Delta t$) regime where PonD/DUGKS remains stable.

The definite advantage of LELBM lies in its memory compression. Conventionally, LBM formulations require to store all populations, which is an impractical burden in high‐temperature regimes. As an attempt to reduce the memory demand, prior works therefore implemented lattice pruning. As an example, Frapolli\cite{Frapolli2017} and Latt\cite{Latt2020} reduced the $D3Q343$ lattice to $D3Q39$ for their supersonic LBM model. PonD approach of thermal scaling and implementation of small timestep size also coincides with this idea, since the smaller timestep  is equivalent to the lower temperature in the LBM algorithm. Although PonD formulation enabled the dynamic population compression between $DdQ3^d$ and $DdQ4^d$, regardless of local temperature, this subsequently required a very low Courant number ($\mathrm{Co}\ll1$, as defined by Eq.~\eqref{eq:IIC1}) to remain stable \cite{Kallikounis2024}. However, LELBM demands only a fixed number of data, regardless of local flow conditions, by implementing the moment representation \cite{ValeroLara2017}. Although each cell generates a large number of velocity stencils during its calculation, these are only temporarily allocated, and released once the single-cell operation finishes. Such memory compression enabled the adoption of very large, compact velocity stencils and allowed LELBM to maintain stability at high Courant numbers.

\section{\label{sec:bc}Boundary Conditions}

In the lattice Boltzmann method, the computational domain is generally partitioned by explicitly designating the fluid and solid cells. Within fluid cells, populations propagate to neighboring cells via collisionless streaming, and solid cells remain inactive and act solely as impermeable obstacles that shape the flow field.

Fluid boundaries, such as inlets and outlets, are implemented by prescribing target values or gradients of macroscopic variables (density, momentum, energy) on a designated "boundary cells", then completely reconstructing the populations that satisfy boundary conditions, which are to be streamed to nearby cells inside the domain. Any populations that land on fluid boundary lattices are generally discarded. Solid boundaries, however, enforce boundary conditions by modifying boundary-touching populations to satisfy boundary conditions, and are re‑streamed in a reflective manner.

Since the populations are primarily serving as the medium of moment transfer in the present model, the conventional population-based boundary condition treatments are inapplicable. Consequently, this work introduces a novel boundary condition scheme tailored for moment‑based LBM. The following section explains how the fluid and solid boundaries are applied for test problems in this paper.

\begin{figure}
    \includegraphics[width=0.7\columnwidth]{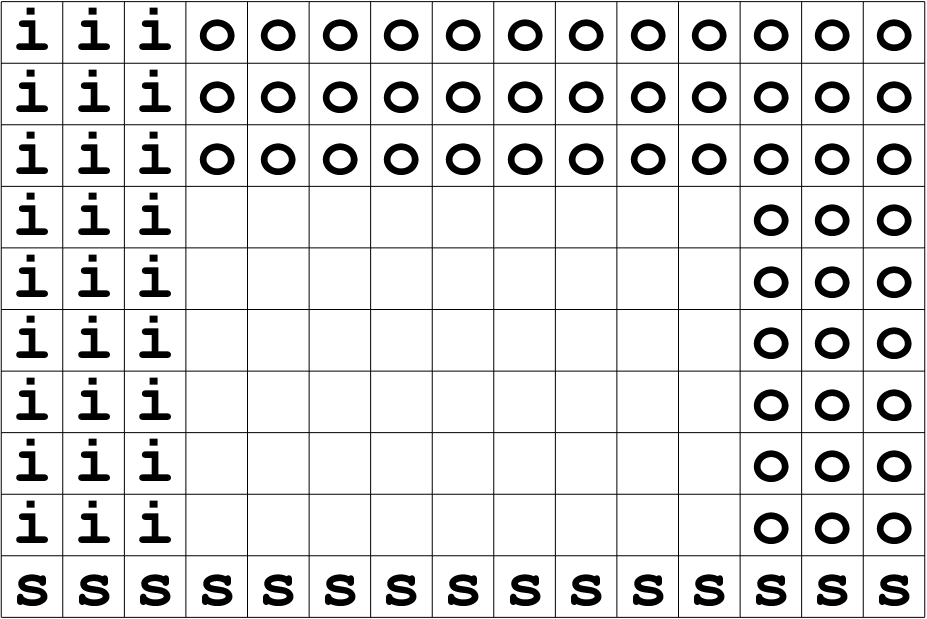}
    \caption{\label{fig:5} Cell region with mixed boundary conditions.}
\end{figure}

The general boundary conditions in fluid simulations include the inlet, outlet, and solid boundaries. Figure~\ref{fig:5} shows the common domain setup of flow simulation with all three types of boundary conditions. Cells labeled ($\bm{\mathtt{i}}$) indicate inlet cells, ($\bm{\mathtt{o}}$) for outlet cells, and ($\bm{\mathtt{s}}$) mark solid cells. The unlabeled cells correspond to the interior fluid domain, where the general LBM calculations are carried out. Since populations in a multispeed model can stream across multiple cells in a single timestep, several layers of fluid boundary cells must be padded around the simulation domain.

\subsection{\label{sec:fluidbc}Fluid Boundary Condition}
For the inlet boundary condition, macroscopic variables such as density $\rho$, velocity $\bm{u}$, and temperature $RT$ are prescribed according to the Dirichlet boundary condition. In the case of Neumann boundary conditions, a first-order extrapolation is applied to the adjacent fluid cell within the simulation domain for the relevant moments. Non-equilibrium moments are typically neglected at the inlet but are taken into account at the outlet. After defining the macroscopic properties, each order of equilibrium moments is reconstructed as described in the Section~\ref{sec:collision}. The extracted non-equilibrium moments are also assembled with reconstructed moments for outlet boundary condition. The remaining steps follow standard fluid cell operations. The treatment of the fluid boundary condition is performed simultaneously with other standard cell operations.

\subsection{\label{sec:solidbc}Solid Boundary Condition}

The earliest approach on implementing the solid boundary condition in the LBM framework is the bounce-back method \cite{Ladd1994_1, Ladd1994_2, Ladd2001, Wagner2002}, originally developed from the concept of Lees–Edwards boundary condition in particle simulations used to impose a momentum gradient within periodic domains \cite{Lees1972}. Despite the simplicity in its implementation, the classic bounce-back method is limited to the "legoland" approximation on curved geometry. To overcome this issue, a variety of interpolation-based bounce-back methods have been developed. This group of interpolation-based methods stream the off-lattice population obtained from the geometric interpolation, modify according to the boundary conditions, then bounce back to the designated cell center \cite{Chen1996, Mei1999, Bouzidi2001, Lallemand2003, Dorschner2015, Dorschner2017, Sanjeevi2018}. With similar concept to the finite-volume method, these bounce-back methods assume the location of boundary to be located at the surface of the cell.

Alternatively, there exist node-based boundary conditions similar to finite-difference methods, in which the state properties are geometrically defined on each node \cite{Noble1995, Inamuro1995, Zou1997, Guo2002, Malaspinas2011}. This method resolves missing populations by solving a linear system that accounts for boundary conditions and had shown a higher order of accuracy than the traditional bounce-back type. However, these methods suffer from an underdetermination of the problem near the geometric corner and mass leakage, resulting practical implementations being outdated for modern-day LBM applications \cite{Xu2022}.

The solid boundary treatments discussed so far are limited to the classic $DdQ3^{d}$ lattice structure. Only a handful of studies have extended solid boundary condition treatments to higher-order velocity stencils \cite{Frapolli2014, Lee2018, Klass2021}. Following the same underlying principle as the bounce-back method, these approaches reconstruct the reflected population from multiple layers of inner solid cells that satisfies both the boundary condition and mass conservation. Frapolli et al. \cite{Frapolli2016, Frapolli2017} further generalized this framework by adopting the interpolation method to implement solid boundaries beyond 'legoland' geometries. However, the implementations are limited to only the quadrature-based population framework, with uniform velocity stencils. Consequently, there exists a demand for a new solid boundary condition method for moment-based LBM with arbitrary velocity stencils. This section introduces the extended bounce-back method for general solid boundary condition with velocity Dirichlet boundary condition.

In the Lees–Edwards boundary condition~\cite{Lees1972}, a uniform shear rate is enforced by imposing a tangential velocity of $\pm\bm{u}_w$ at the upper and lower boundaries, respectively. Any particle crossing the upper boundary with velocity $\bm{u}$ reenters through the lower boundary with a shifted velocity of
\begin{equation}
  \bm{u}' = \bm{u} - 2\,\bm{u}_w \label{eq:IIIB1}
\end{equation}
and vice versa.

Based on this idea, the bounce–back method for a wall moving at a velocity $\bm{u}_w$ modifies the bulk velocity of populations that contact the solid boundary in a similar way. Consider a population with a bulk velocity of $\bm{u}$, impinging on the wall. The bulk velocity of the population after being bounced by the moving wall can be obtained by first calculating the relative velocity in the frame of the moving wall, then applying no–slip reflection, and finally, returning to the absolute frame, as shown by \eqref{eq:IIIB2}--\eqref{eq:IIIB4}
\begin{subequations}
\begin{align}
  \bm{u}_{\mathrm{rel}} &= \bm{u} - \bm{u}_w, \label{eq:IIIB2} \\
  \bm{u}_{\mathrm{rel}}' &= -\bm{u}_{\rm rel}, \label{eq:IIIB3} \\
    \bm{u}'&= \bm{u}_{\mathrm{rel}}' + \bm{u}_w= -(\bm{u} - \bm{u}_w) + \bm{u}_w = -\bm{u} + 2\,\bm{u}_w. \label{eq:IIIB4}
\end{align}
\end{subequations}
Thus, upon reflection on a wall moving with $\bm{u}_w$, each population $f_i(\bm{c}_i,\,\bm{u})$ is modified to $f_i(-\bm{c}_i,\,-\bm{u} + 2\bm{u}_w)$ \cite{Wagner2002}.

\begin{figure}
  \centering
  \includegraphics[width=\columnwidth]{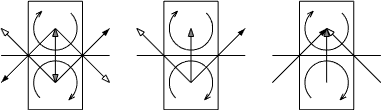}
  \caption{\label{fig:6}Equivalent bounce-back algorithm. Left: streaming trajectories of populations from both the real and image lattices. Middle: populations entering the simulation domain from the image lattice. Right: localized periodic boundary condition applied to surfaces of the real lattice not in contact with solid boundaries. The upper lattice represents the real lattice, while the lower lattice corresponds to the image lattice. Arrows of the same color indicate equivalent populations and velocity stencils.}
\end{figure}

These operations are linear and admit a geometric interpretation via an "image lattice" analogy. Consider two adjacent real and image lattices in contact by a solid boundary moving at $\bm{u}_w$, with the image lattice having antisymmetric momentum with the corresponding real lattice. In the wall reference frame, the relative bulk velocity of the real lattice is equal to Eq. \eqref{eq:IIIB2}. Since the image lattice has antisymmetric momentum, the relative bulk velocity of the image lattice is Eq. \eqref{eq:IIIB3}. Then returning back to the laboratory frame results Eq. \eqref{eq:IIIB4}. Finally, imposing periodicity along the wall-normal surfaces of the cell yields the same result as the traditional bounce-back method. Figure~\ref{fig:6} is the equivalent image lattice diagram that recovers the bounce–back method.

\begin{figure}
  \centering
  \includegraphics[width=0.8\columnwidth]{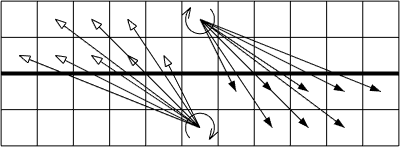}
  \caption{\label{fig:7}Trajectory of a population passing through the solid boundary (black arrows) and the corresponding streams from the image lattice (white arrow).}
\end{figure}

Developed upon on this analogy, the extended bounce–back method extended the concept of image lattice from the boundary‐adjacent cells to the entire simulation domain, designed especially for multispeed lattice. Figure~\ref{fig:7} shows how a real population that passes the solid boundary (black arrows) is compensated by its corresponding population entering from the image lattice (white arrow). Each image population is constructed to retain the bulk velocity of $-\bm{u} + 2\bm{u}_w,$ in agreement with the traditional bounce–back method. Similarly, antisymmetric populations are designated to the population velocity of $-\bm{c}_i$. Unlike the traditional bounce-back method, the extended multispeed formulation disregards the localized periodicity of the boundary-adjacent lattice, since the distance each population stream in each timestep is significantly greater than the size of single lattice.

However, this raises a mass conservation issue. Simple bounce–back in a $DdQ3^d$ lattice inherently conserves the mass, but in a multispeed formulation, the sum of the population streamed from the image lattice is not necessarily equal to the sum of the real population absorbed at the boundary. Moreover, storing entire mass accumulation in each image lattice is algorithmically complicated, especially for complex geometries. To address this, the back-streaming approach is introduced to identify the mass accumulation in each image lattice in a simpler algorithm.

First, during the streaming process of each cell, any population that encounters a solid boundary is specularly reflected until it finishes its streaming of $\bm{c}_i\Delta t$. Then the populations that have touched $n$th solid boundary are accumulated to the corresponding mass imbalance. This procedure can be written as
\begin{equation}
  \Delta m_{n}\bigl(\bm{x}\bigr)\mathrel{+}=f_{i}\bigl(\bm{x},\,\bm{c}_{i}\bigr)\bigr|_{n}, \label{eq:IIIB8}
\end{equation}
where $f_{i}(\bm{x},\,\bm{c}_{i})\bigr|_{n}$ indicates the population that contacted the $n$th solid boundary, and $\Delta m_{n}(\bm{x})$ is the corresponding accumulated mass.

Once all fluid cell operations are complete, each site with $\Delta m_{n}(\bm{x})>0$ reconstructs a normalized population that satisfies the boundary condition. This follows the moment decomposition and reassembly procedure explained in Section~\ref{sec:collision}. First, one extracts the non‐equilibrium central moments from the full raw moments from the real lattice site $\bm{x}$.
\begin{equation}
  \widetilde{\bm{M}}^{\mathrm{neq}'}\bigl(\bm{x}\bigr)\twoheadleftarrow\bm{M}\bigl(\bm{x}\bigr). \label{eq:IIIB9}
\end{equation}
Then one reassembles the post‐collision moments by combining the equilibrium moments that satisfy both the boundary condition and the post-collision non‐equilibrium moments:
\begin{equation}
    \widetilde{\bm{M}}^{*\prime}(\bm{x})\bigr|_{n}=\widetilde{\bm{M}}^{\mathrm{eq}\prime}(\bm{x},\,\bm{u}_{n},\,RT)+(\bm{I}-\bm{\Omega})\widetilde{\bm{M}}^{\mathrm{neq}\prime}(\bm{x}). \label{eq:IIIB10}
\end{equation}
For an adiabatic wall, the temperature $RT$ is unchanged, and $\bm{u}_{n}$ is the velocity of $n$th boundary condition, evaluated as
\begin{equation}
  \bm{u}_{n}(\bm{x})=\bm{u}(\bm{x})-2\bm{u}_{w}\bigr|_{n}. \label{eq:IIIB11}
\end{equation}
Moreover, if the wall allows slip in a given direction $\bm{e}_i$, that velocity component remains unchanged by the boundary treatment.
\begin{equation}
  \bm{u}_{n}(\bm{x})\cdot\bm{e}_i= \bm{u}(\bm{x})\cdot\bm{e}_i. \label{eq:IIIB12}
\end{equation}
With this formulation, the combination of slip and no-slip boundary conditions can be imposed in each direction.

After constructing $\widetilde{\bm{M}}^{*\prime}\bigr|_{n}$, the new boundary‐satisfying populations are reconstructed
\begin{equation}
  f_{i}'\bigl(\bm{x},\,\bm{c}_{i}\bigr)\bigr|_{n} = f_{i}'\bigl(\bm{x},\,\bm{c}_{i},\,\widetilde{\bm{M}}^{*\prime}(\bm{x})\bigr|_{n}\bigr). \label{eq:IIIB13}
\end{equation}
These reconstructed populations are then streamed again using the antisymmetric velocity stencil $\bm{c}_{i}\bigr|_{n}=-\bm{c}_{i}$. In the same way as the streaming of real populations, the image populations are specularly reflected at the solid boundary, and the sum of population that last touches the $n$th boundary is recorded:
\begin{equation}
    \Delta m_{n}'(\bm{x})\mathrel{+} = f_{i}'\bigl(\bm{x},\,\bm{c}_{i}\bigr|_{n}\bigr)\bigr|_{n,n}. \label{eq:IIIB14}
\end{equation}
$f_{i}'\bigl(\bm{x},\,\bm{c}_{i}\bigr|_{n}\bigr)\bigr|_{n,n}$ indicates the normalized population reconstructed for the $n$th boundary condition in the last contact with the $n$th boundary during the streaming. Finally, to enforce mass conservation, outgoing populations are scaled to ensure mass conservation and allocated to the designated real lattice.
\begin{equation}
    f_{i}(\bm{x}^*,\,\bm{c}_{i}^*)=\frac{\Delta m_{n}(\bm{x})}{\Delta m_{n}'(\bm{x})}f_{i}'(\bm{x},\,\bm{c}_{i}\bigr|_{n})\bigr|_{n,n}. \label{eq:IIIB15}
\end{equation}
$\bm{c}_{i}^*$ is the post-specular-reflection population velocity, by which the sign of each component changes according to the normal direction of the wall in each reflection. $\bm{x}^*$ is the position of the final landing position inside the simulation domain.

\begin{figure}
  \centering
  \includegraphics[width=0.8\columnwidth]{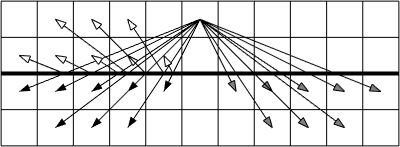}
  \caption{\label{fig:8}Schematic of the extended bounce-back method. Black arrows indicate the intended streaming directions without reflection, white arrows represent the actual population trajectories following specular reflections at the boundary, and gray arrows depict intermediate stencils that satisfied boundary conditions before the application of anti-symmetry transformation.}
\end{figure}

This procedure guarantees exact enforcement of the moving‐wall boundary condition while ensuring mass conservation. In addition, the collision and population reconstruction steps are not necessarily limited to the formulations discussed in Section~\ref{sec:lelbm}. Any collision model and population reconstruction technique that can correctly reconstruct $\bm{M}^*\bigr|_{n}$ and $\bm{f}^*\bigr|_{n}$ is applicable. Although this method can formally impose a temperature‐specified boundary condition, in general, it does not produce accurate solutions. Hence, the extended bounce‐back method is restricted to adiabatic walls. Figure~\ref{fig:8} illustrates the schematics of the extended bounce‐back method.

\section{\label{sec:verifications}Verifications of Theory}
In thermoviscous simulations, four key fluid properties must be accurately imposed: shear viscosity $\mu_s$, bulk viscosity $\mu_b$, thermal conductivity $\kappa$, and the specific heat ratio $\gamma$. This section assesses the accuracy of the imposed physical property of the present model through several benchmark problems and a simple "hot-box advection" problem to evaluate the computational complexity.

\subsection{\label{sec:shear_wave}Shear Wave}

\begin{figure}
  \centering
  \includegraphics[width=\columnwidth]{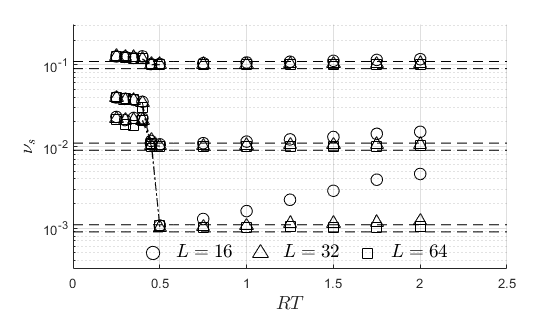}
  \caption{\label{fig:VA1}Measured kinematic viscosity under variation of temperature $RT$ and resolution $L$. Dashed lines indicate $\pm 10 \%$ margin for target values.}
\end{figure}

The shear viscosity is verified by the decay of a longitudinal shear wave imposed in a 2D fully periodic domain. The initial conditions follow those of references\cite{Coreixas2020, Thyagarajan2023}:
\begin{align}
  (\rho, u_x, u_y, RT) &= \left(\rho_0, \ u_0, \ u_y' \cos(2\pi x/L), \ RT_0 \right). \label{eq:VA1}
\end{align}
where $L$ denotes the size of the simulation domain. Unlike references \cite{Coreixas2020, Thyagarajan2023}, which varied the Mach number in the $x$ direction, the problem has been solved with a fixed velocity of $u_0 = 0.5$. This is typically due to the irrelevance of bulk velocity on accuracy of the present model, yet is intended to maximize numerical diffusion by ensuring that the center of distribution is consistently landing between the cells in every timestep. This approach provides the evaluation of the resulting shear viscosity of the model, in the worst possible flow condition. Simulations are carried out over a range of lattice temperatures, $RT_0 \in [0.25,\,2.0]$, and in resolutions of $L = 16$, $32$, and $64$, allowing us to assess the accuracy of the measured shear viscosity for both temperature and the grid resolution.  

The analytical solution for the initial condition of Eq. \eqref{eq:VA1} was given by \cite{Kovasznay1953}
\begin{align}
  u_y(x,t) = u_y' \cos\left(k\,(x - u_{0}\,t)\right)\exp\left(-\nu_s\,k^2\,t\right), \label{eq:VA2}
\end{align}
where $\nu_s$ is the kinematic shear viscosity and the wavenumber $k$ is defined as
\begin{align}
  k = \frac{2\pi}{L}. \label{eq:VA3}
\end{align}
Rearranging Eq. \eqref{eq:VA2} yields a direct estimate of the viscosity:
\begin{align}
  \nu_s = -\frac{1}{k^2\,t}\ln\frac{\max_x|u_y(x,t)|}{u_y'}. \label{eq:VA4}
\end{align}
where $\max_x$ indicate the maximum value in the $x$ direction. In this study, the kinematic viscosity is measured at the timestep when the Fourier number
\begin{align}
  \mathrm{Fo} = \nu_s k^2t \label{eq:VA5}
\end{align}
reaches unity. All simulations are conducted at $\Pr = 1$ with internal degrees of freedom set to $K=10^6$, to minimize thermal effects. The bulk stress collision frequency is set to $\omega_b = 1.0$ to fully damp the acoustic modes. The results for target viscosity of $\nu_s = 10^{-1},\,10^{-2},\,10^{-3}$ are shown in Figure~\ref{fig:VA1}, in which a prominent change in accuracy is observed near $RT \approx 0.45$, corresponding to the activation threshold of the internal velocity stencil refinement of $RT\lessapprox0.44$. The resulting numerical viscosity from the refined population velocity is measured to be $\nu_{\mathrm{num}}\approx0.02$. As $RT$ further increases, the numerical diffusion increases due to the extended steaming distance per timestep. Hence, for viscosity sensitive problems, such as turbulence, it is advisable to maintain a low lattice temperature of around $RT\approx0.5$.

\subsection{\label{sec:thermal_wave}Thermal Wave}

\begin{figure}
  \centering
  \includegraphics[width=\columnwidth]{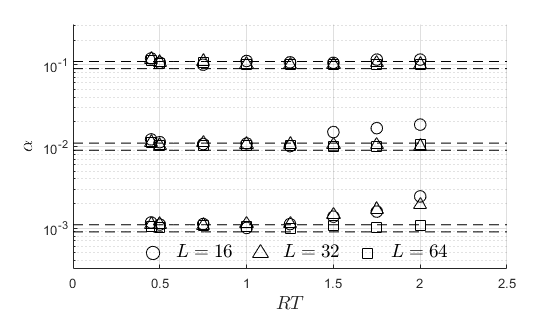}
  \caption{\label{fig:VB1}Measured thermal diffusivity under variation of lattice temperature $RT$ and resolution $L$. Dashed lines indicate $\pm 10\%$ margins around the target values.}
\end{figure}

Following the methodology used in Section~\ref{sec:shear_wave} and relevant references \cite{Coreixas2020, Thyagarajan2023}, the thermal diffusivity $\alpha$ is evaluated using the thermal wave problem. Since the thermal conductivity is defined as $\kappa = \rho C_p \alpha$, the accurate imposition of $\alpha$ corresponds to the accurate $\kappa$. In a 2D periodic domain, the following initial conditions are imposed:
\begin{align}
  (\rho,\, u_x,\, u_y,\, RT) &= \left(\rho_0 + \rho' \cos(2\pi x/L),\, u_0,\ 0.0,\ \rho_0 RT_0/\rho \right). \label{eq:VB1}
\end{align}
These conditions yield an analytic solution of
\begin{align}
  \rho(x,t) = \rho_0 + \rho' \cos\left(\tfrac{2\pi}{L}(x - u_0 t)\right)\exp(-\alpha k^2 t). \label{eq:VB2}
\end{align}
From Eq. \eqref{eq:VB2}, the thermal diffusivity can be obtained as
\begin{align}
  \alpha = -\frac{1}{k^2 t} \ln\left( \frac{\max_x |\rho(x,t) - \rho_0|}{\rho'} \right). \label{eq:VB3}
\end{align}

In this study, the parameters are set to $\rho_0 = 1.0$, $\rho' = 0.01$, and $u_0 = 0.5$, with $\omega_b = 1.0$, again with same reason as in Section~\ref{sec:shear_wave}. However, the internal degrees of freedom have been set to $K = 0$, to maximize the thermal contribution. Simulations are performed over a range of lattice temperatures, $RT_0 \in [0.25,\,2.0]$, and spatial resolutions of $L = 16$, $32$, and $64$. The thermal diffusivity is measured at the timestep when the Fourier number reaches unity. The results are shown in Figure~\ref{fig:VB1}. The thermal diffusivity has been marked from $RT = 0.45$, due to the same reason reported in Section~\ref{sec:shear_wave}. The internal refinement approach resulted in excessive numerical diffusion such that the evaluation of the decay rate is infeasible. Optimal accuracy has been observed in the range of $RT = 0.5$ to $1.0$. Combined with the finding in Section~\ref{sec:shear_wave}, it can be concluded that a lattice temperature of $RT = 0.5$ remains a recommendable choice for thermoviscous problems.

\subsection{\label{acoustic_plane}Acoustic Plane Wave}
To assess the bulk viscosity, a one-dimensional acoustic plane wave problem has been analyzed. Following the work of Viggen \cite{Viggen2010,Viggen2014}, an oscillating point mass source at the origin is defined by
\begin{align}
  j(t) &= \dot{\rho}\sin(\omega_0 t), \label{eq:VC1}
\end{align}
where $\dot{\rho}$ is the mass injection rate and $\omega_0$ is the oscillation frequency. The oscillatory analytical solution of the pressure profile is \cite{Kinsler2000}
\begin{align}
  p'(x,t) &= \frac{c_s^2\dot{\rho}}{2\omega_0}\sin(\omega_0 t)\exp(-\alpha_x x), \label{eq:VC2}
\end{align}
where $c_s$ is the speed of sound $c_s=\sqrt{\gamma RT}$ for ideal gas and $\alpha_x$ is the spatial absorption coefficient defined as
\begin{align}
  \alpha_x \approx \frac{\omega_0^2}{2\,c_s^3}\left[\left(2-\tfrac{2}{D}\right)\nu_s + \nu_b + (\gamma-1)\alpha\right]. \label{eq:VC3}
\end{align}
Here, $\nu_s$ and $\nu_b$ are the kinematic shear and the bulk viscosity and $\alpha$ is the thermal diffusivity. The bulk viscosity can then be derived from the exponential decay rate between the successive peaks:
\begin{align}
  \nu_b &= -\frac{2c_s^3}{\omega_0^2\Delta x}\ln\left[\tfrac{p'(x+\Delta x,t)}{p'(x,t)}\right] -\left[\left(2-\tfrac{2}{D}\right)\nu_s + (\gamma-1)\alpha\right]. \label{eq:VC4}
\end{align}

In an application to the LBM framework, Viggen\cite{Viggen2010,Viggen2014} modified the Boltzmann equation to include a mass source by
\begin{align}
  \partial_t f + \xi_\alpha\,\partial_{\alpha} f = -\frac{1}{\tau}f^{\mathrm{neq}} + j. \label{eq:VC5}
\end{align}
Since the source is implemented at the population level, the corresponding central moment source terms are defined as
\begin{subequations}
\begin{align}
    &\left\{\widetilde{S}_0,\, \widetilde{S}_\alpha,\, \widetilde{S}_{\alpha\beta},\, \widetilde{S}_{D\gamma} \right\} = \dot{\rho}\sin(\omega_0 t)\left\{1,\, 0,\, RT\delta_{\alpha\beta},\, 0 \right\}, \label{eq:VC6}\\
    &\left\{\widetilde{Q}_0,\, \widetilde{Q}_\alpha,\, \widetilde{Q}_{\alpha\beta}\right\} = \dot{\rho}KRT\sin(\omega_0 t)\left\{1,\,0 ,\,RT\delta_{\alpha\beta}\right\}, \label{eq:VC7}
\end{align}
\end{subequations}
which are the central moment source required for the population reconstruction technique. $\widetilde{\bm{S}}$ is the source term for fluons and $\widetilde{\bm{Q}}$ for phonons. The post‐collision moments are then updated as
\begin{align}
  &\left\{\widetilde{\bm{M}}^{*}, \widetilde{\bm{G}}^{*} \right\} = \left\{\widetilde{\bm{M}}^{\mathrm{eq}}, \widetilde{\bm{G}}^{\mathrm{eq}} \right\} \notag \\
  &+ \left\{\left(\bm{I}-\bm{\Omega}_M \right)\widetilde{\bm{M}}^{\mathrm{neq}}, \left(\bm{I}-\bm{\Omega}_G \right)\widetilde{\bm{G}}^{\mathrm{neq}} \right\} + \Delta t\left\{\bm{\widetilde{S}}, \bm{\widetilde{Q}}\right\}. \label{eq:VC8}
\end{align}
The remainder of the algorithm follows the standard LELBM procedure.

Simulations are performed over a lattice temperature range of $RT\in\{0.5,\,2.0\}$ and frequencies of $\omega_0=2\pi/16$, $2\pi/32$, and $2\pi/64$ to evaluate both the temperature and the frequency effect on bulk viscosity. To make simulations closely adiabatic, the amplitude of the sinusoidal mass source has been prescribed to be $\dot{\rho}=10^{-6}$. Unlike Sections~\ref{sec:shear_wave} and \ref{sec:thermal_wave}, the low‐temperature regime of $RT<0.5$ is neglected, since the stencil refinement activated below $RT\approx0.44$ has already been shown to introduce significant numerical diffusion. To measure the bulk viscosity, $\nu_s$ and $\alpha$ are fixed at $10^{-4}$, and a calculation was performed for the 1D LELBM formulation. Furthermore, the internal degrees of freedom of $K=4$ was imposed to set $\gamma=1.4$.

\begin{figure}
  \centering
  \includegraphics[width=\columnwidth]{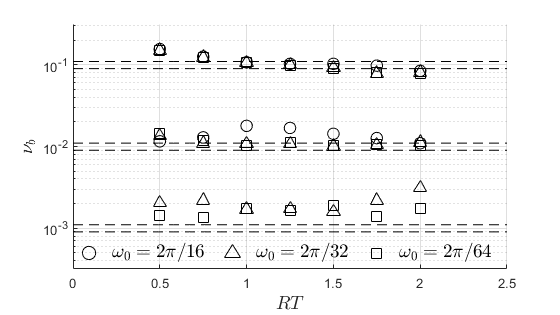}
  \caption{\label{fig:VC1}Measured kinematic bulk viscosity under variation of lattice temperature $RT$ and frequency $\omega_0$. Dashed lines indicate $\pm10\%$ margins around the target values.}
\end{figure}

Figure~\ref{fig:VC1} presents the bulk viscosity measured in a range of configurations. Unlike the shear and thermal cases, the bulk viscosity estimates exhibit large errors in all temperature regimes. In particular, the solver was unable to accurately impose acoustic absorption, for $\omega_0=2\pi/16$ with a target $\nu_b=10^{-3}$, making $\nu_b$ indeterminable. It can be observed that the accuracy is achieved only under low frequency with higher bulk viscosity. Although no strong temperature dependence is observed, $RT\approx1.0$ is typically recommended for acoustic problems, where errors remain within $\pm10\%$ under moderate conditions. Since the acoustic modes are already damped by $\omega_b=1.0$ for stabilization throughout the rest of the paper, further modification to improve the precision of the bulk viscosity is beyond the scope of this paper.

\subsection{\label{sec:acoustic_pulse}Acoustic Pulse}

\begin{figure}
  \centering
  \includegraphics[width=\columnwidth]{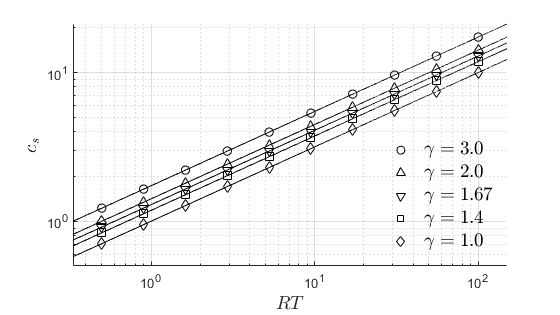}
  \caption{\label{fig:VD1}Measured sound speeds (markers) versus the theory (solid lines) for the indicated values of $\gamma=3.0$ to $1.0$.}
\end{figure}

The speed of sound is determined by the one‐dimensional acoustic pulse. An initially quiescent medium of $\rho_0=1.0$ and $u_x=0.0$ is perturbed in the center by $\Delta\rho=10^{-6}$, and the peak of the pressure pulse $x_{\rm peak}(t)$ is tracked to measure the speed of sound by
\begin{align}
  c_s = \frac{x_{\mathrm{peak}}(t)}{t}. \label{eq:VD1}
\end{align}
To evaluate precision under varying lattice temperature and specific heat ratio, a test has been carried out for a temperature range from $RT=0.5$ to $RT=100$. The specific heat ratio has been adjusted by prescribing the internal degrees of freedom by $K = 0,\, 1,\, 2,\, 4,$ and $10^6$-corresponding to the specific heat ratios of $\gamma = 3.0,\, 2.0,\, 1.67,\, 1.4,$ and $1.0$.

Figure~\ref{fig:VD1} compares the measured speeds (markers) with the predictions of the ideal gas model $c_s=\sqrt{\gamma RT}$ (solid lines). Excellent agreement is observed in all temperature ranges, even up to $RT=100$, validating the capability of capturing acoustic propagation over a wide temperature range using the present model.

\subsection{\label{sec:hot_box}Hot‐Box Advection}

To assess computational cost, the "hot-box advection" problem has been tested. The problem consists of a fully periodic domain initialized with uniform conditions $\{\rho,\,\mathbf{u},\,RT\}$. Here, the density has been set to unity for simplicity, and the bulk velocity has been arbitrarily set to $(u_x,u_y)=(10,\,10)$. Simulations were performed on an NVIDIA RTX 4070 GPU in the domains of $N=N_x\times N_y = 256^2,\,512^2,\,1024^2,\,2048^2,$ and $4096^2$ (and correspondingly in 1D), over the temperatures of $RT = 0.5,\,1,\,2,\,4,\,8,\,16,$ and $32$.

\begin{figure}
  \centering
  \includegraphics[width=\columnwidth]{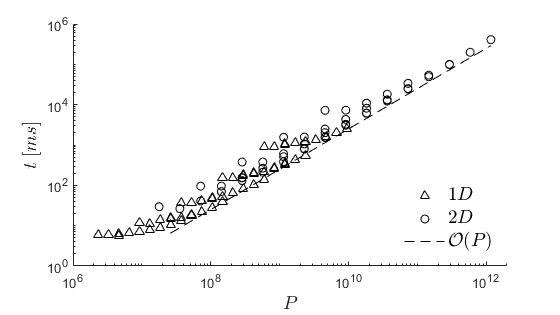}
  \caption{\label{fig:VE1}Average computational time $t$ measured from hot box problem in milliseconds, versus the problem size $P$. Triangular and circular markers indicate 1D and 2D setup each. Dashed line indicate the theoretical asymptotic scaling, proportional to $P$.}
\end{figure}

As derived from Sections~\ref{sec:las} and \ref{sec:epr}, the problem size $P$ of the present model scales as
\begin{align}
  P=\left(n\sqrt{RT}\right)^D\left(N_M^2 + N_G^2\right)N. \label{eq:VE1}
\end{align}
Consequently, the computational time should follow this same asymptotic relationship.  Figure~\ref{fig:VE1} shows the measured average wall-clock time per timestep against the problem size $P$. The results show excellent agreement with the theoretical scaling. The deviations are the consequence of small $q$ of the low-temperature setup, where the computational cost of other operations begins to be significant. Typically, $N_x \times N_y=4096^2$ case took approximately 10.9~s at $RT=0.5$, 33.8~s at $RT=2.0$, and 408.5~s at $RT=32$, corresponding to the general initial condition, mild post‑shock, and strong post‑shock temperatures respectively.

\section{\label{sec:results}Results of Algorithm Assessment}
To assess the capabilities of the proposed LELBM formulation, comprehensive canonical and challenging test problems are presented. In one dimension, the problems involve Sod’s shock tube, the Lax problem, and Shu–Osher wave. Both the Sod’s shock tube problem and the Shu-Osher wave are simulated at multiple spatial resolutions to evaluate the discretization effect, while the Lax problem is evaluated under acoustic scaling of the initial conditions, which is equivalent to the variation of timestep size.

In two dimensions, first, the classical 2D Riemann problem and the double Mach reflection problem have been simulated. Each problem is solved with several spatial resolutions to observe the convergence behavior. Furthermore, to examine performance under high-Mach number flow, the oblique shock and the supersonic flow past a circular cylinder have been simulated. The oblique shock problem is known to be one of the few problems that has an analytical solution for a compressible ideal gas. To examine both the robustness and physicality of the LELBM formulation, a test has been carried out over a broad parameter space of $\mathrm{Ma}\in[2,\,20]$ and $\theta\in[1^\circ,\,40^\circ]$, and their post-shock properties are compared with analytical solutions. Given that the supersonic flow past a circular cylinder also has a semianalytic solution for the shock standoff distance ratio, the simulation has been conducted for a range of Mach number of $\mathrm{Ma}\in[1.6,\,10]$, over the variation of freestream temperature and the resolution to examine the error convergence, and compared with the theoretical prediction. Finally, supersonic flow past a NACA0012 airfoil at $\mathrm{Ma}=1.5$ has been simulated, and the resulting pressure coefficient has been compared with published references. Throughout all cases, the specific heat ratio is designated as $\gamma=1.4$, and the bulk collision frequency is set to $\omega_b=1.0$.

\subsection{\label{sec:sod}Sod’s Shock Tube}

\begin{figure}
  \centering
  \includegraphics[width=0.93\columnwidth]{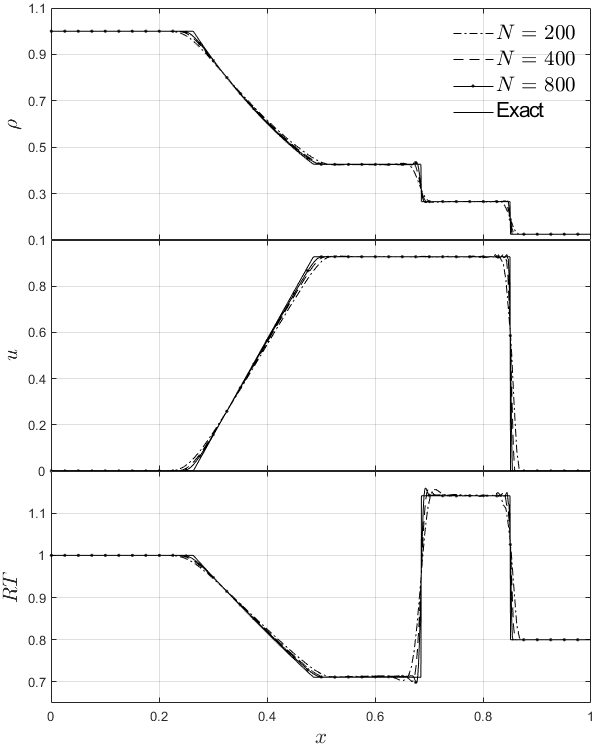}
  \caption{\label{fig:VIA1}Solution profiles of Sod's shock tube problem at $t=0.2$. Solid lines denote the exact solutions.}
\end{figure}

As a classical benchmark, the Sod's shock tube problem \cite{Sod1978} has been tested with initial conditions as the follows
\begin{align}
  (\rho,\, u_x,\, RT) 
  &= 
  \begin{cases}
    (1.0,\,0.0,\,1.0), & 0 \le x < 0.5,\\
    (0.125,\,0.0,\,0.8), & 0.5 \le x < 1.
  \end{cases}
  \label{eq:VIA1}
\end{align}
Figure~\ref{fig:VIA1} shows the results at $t=0.2$ for resolutions of $N = 200,\,400,$ and $800$, demonstrating excellent agreement with the exact solution.

\subsection{\label{sec:lax}Lax Problem}

To investigate the temperature effect on accuracy, Lax problem \cite{Lax1954} has been solved with initial conditions of
\begin{align}
  (\rho,\, u_x,\, RT)
  &=
  \begin{cases}
    (0.445,\,0.698,\,3.528/\rho), & 0 \le x < 0.5,\\
    (0.500,\,0.000,\,0.517/\rho), & 0.5 \le x < 1,
  \end{cases}
  \label{eq:VIB1}
\end{align}
which yields a maximum lattice temperature of $RT\approx7.93$. To compare the effect of temperature on accuracy, the acoustic scaling has been applied:
\begin{align}
  (\rho,\, u_x,\, RT)\longleftarrow\left(\rho,\,u_x/1.55,\,RT/1.55^2\right)
  \label{eq:VIB2}
\end{align}
The scaling approach reduces the maximum temperature to $RT\approx3.30$, and is also equivalent to adjusting the timestep size to $\Delta t = 1/1.55$, with the unchanged initial conditions. The lattice Courant numbers of unscaled and scaled cases are 12.5 and 8, respectively, throughout the entire simulation. The calculation has been conducted for $t=0.14$ in physical units, with the discretizations done with $N=1000$.

\begin{figure}
  \centering
  \includegraphics[width=0.93\columnwidth]{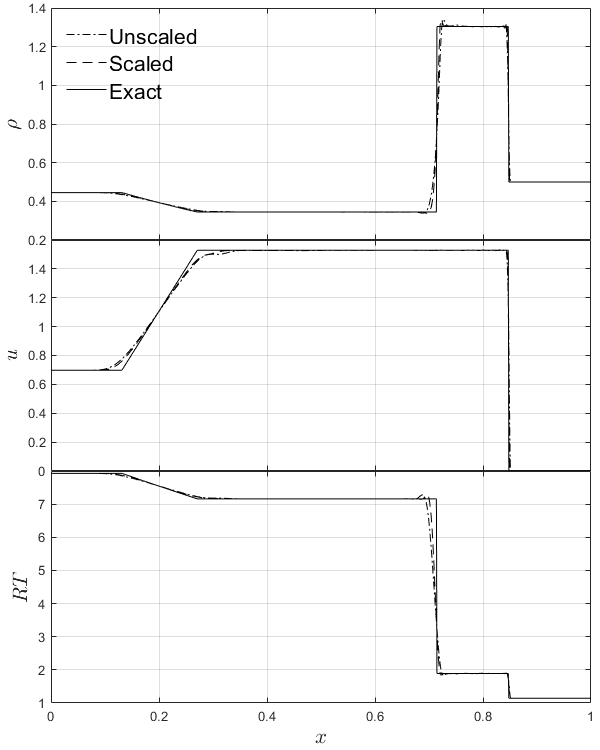}
  \caption{\label{fig:VIB1}Solution profiles of Lax problem at $t=0.14$. Solid lines denote the exact solution. The result from scaled setup have been unscaled for direct comparison.}
\end{figure}

Figure~\ref{fig:VIB1} compares the high- and low-temperature solutions with the analytical solution. Although the unscaled case exhibits marginally increased numerical diffusion, the differences are negligible, which demonstrates the robust accuracy of LELBM, even at high lattice temperatures.

\subsection{\label{sec:shu}Shu-Osher Wave}

\begin{figure}
  \centering
  \includegraphics[width=\columnwidth]{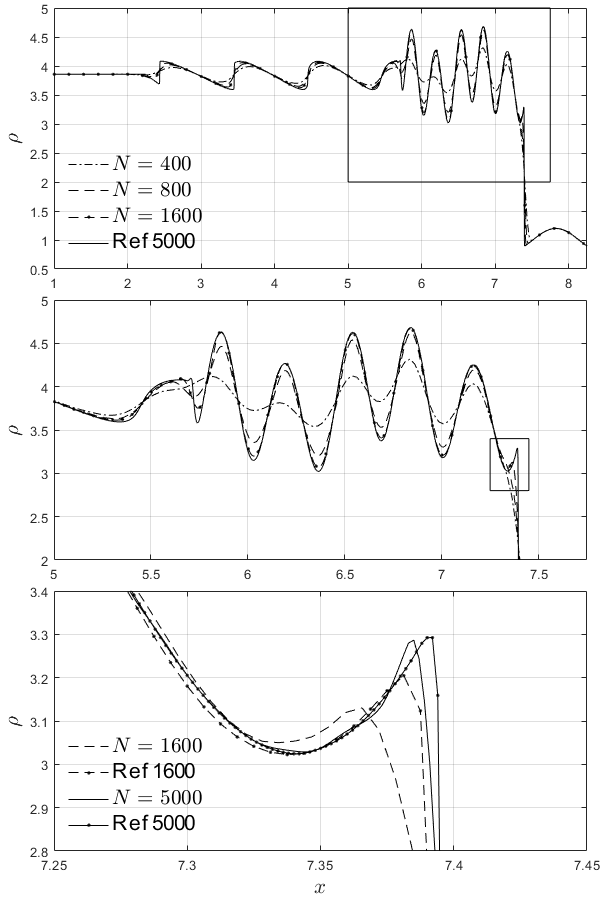}
  \caption{\label{fig:VIC1}Density at $t=0.18$ for Shu-Osher wave: full domain (top), perturbed region (middle), and shock front (bottom). Ref1600 and Ref5000 indicate reference solutions from the spatial resolutions of $N=1600$ and $N=5000$ each.}
\end{figure}

The next test problem is the Shu-Osher wave \cite{Shu1989}. This problem is intended to observe the details of the structure after a shock passing through a sinusoidally perturbed density field. The initial conditions are given by
\begin{align}
  &(\rho,\, u_x,\, RT) \notag \\
  &=
  \begin{cases}
    \left(3.857,\,2.629,\,10.333/\rho\right), & 0 \le x < 1,\\
    \left(1 + 0.257\sin[5(x-5)],\,0,\,1/\rho\right), & 1 \le x < 10.
  \end{cases}
  \label{eq:VIC1}
\end{align}
The calculation has been carried out until $t = 0.18$. Figure~\ref{fig:VIC1} shows the density profiles for spatial resolutions of $N = 400,\,800,\,1600$ and $5000$, compared to the high-resolution reference numerical solution \cite{Ramos2021} for $N = 1600$ and $5000$. The reference solution was generated using a third-order strong stability-preserving Runge-Kutta method (SSP-RK3) combined with a fourth-order Roe/WENO scheme for flux calculations. Even at $N=800$, the downstream perturbation structures are well resolved. However, the leading density spike at the shock front exhibits slightly less accurate behavior compared with the high-resolution reference solution, which is expected to be improved with the combination of the higher-order temporal integration scheme.

\subsection{\label{sec:riemann2d}Two‐Dimensional Riemann Problem}

\begin{figure}
  \centering
  \includegraphics[width=\columnwidth]{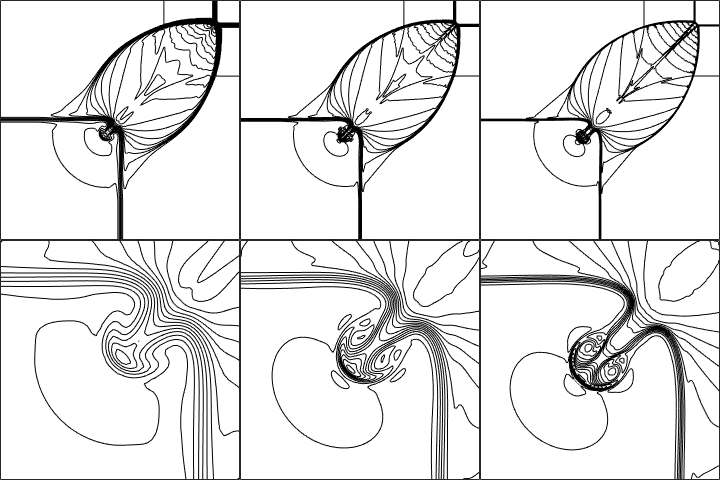}
  \caption{\label{fig:VID1}Density contours at $t=0.25$ for the 2D Riemann problem. Top row: full contours of $256^2$, $512^2$, and $1024^2$ resolutions (left to right). Bottom row: zoom into $(x,y)\in[0.32,0.54]^2$, highlighting small‐scale details.}
\end{figure}

The first validation of the model in 2D setup has been conducted for the canonical two‐dimensional Riemann problem (Configuration~12) \cite{Lax1998}, which has been widely used to evaluate the solver's ability to capture the details determined by the initial conditions. This benchmark consists of four constant initial conditions imposed on a unit square:
\begin{align}
  &(\rho,\, u_x,\, u_y,\, RT) \notag \\
  &\hspace{-0.5em}= 
  \begin{cases}
    \bigl(0.8,\,0,\,0,\,1/\rho\bigr), & 0 \le x < 0.5,\,0 \le y < 0.5,\\
    \bigl(1.0,\,0,\,0.7276,\,1/\rho\bigr), & 0.5 \le x \le 1,\,0 \le y < 0.5,\\
    \bigl(1.0,\,0.7276,\,0,\,1/\rho\bigr), & 0 \le x < 0.5,\,0.5 \le y \le 1,\\
    \bigl(0.5313,\,0,\,0,\,0.4/\rho\bigr), & 0.5 \le x \le 1,\,0.5 \le y \le 1.
  \end{cases} \label{eq:VID1}
\end{align}
The outflow boundary conditions described in Section~\ref{sec:fluidbc} has been applied on all boundaries, and the domain has been discretized in uniform resolutions of $N_x \times N_y=256^2,\,512^2,$ and $1024^2$. Figure~\ref{fig:VID1} shows the density contours at $t=0.25$, using 30 equidistant levels between $\rho_{\min}=0.5107$ and $\rho_{\max}=1.6401$, from the $1024^2$ case. Even without flux limiters or adaptive mesh refinement techniques, the LELBM solution has shown excellent agreement with reference solutions \cite{Lax1998, Kurganov2002}, including small-scale details near the center.

\subsection{\label{sec:double_mach}Double Mach Reflection}

The double Mach reflection \cite{Woodward1984} is a famous benchmark problem for high Mach number compressible flow solvers, featuring self-similar shocks, jet and vortex formation, with high initial condition sensitivity that can produce spurious numerical artifacts under improper configurations \cite{Vevek2019}. Here, a $\mathrm{Ma}=10$ flow impinges on a reflective wall inclined in $30^\circ$ counterclockwise, within the domain of $[0,4]\times[0,1]$. The initial conditions are
\begin{align}
    &(\rho,\,u_x,\,u_y,\,RT) \notag \\
    &=
    \begin{cases}
    \bigl(8,\,4.125\sqrt{3},\,-4.125,\,116.5/\rho\bigr), \\
    &\hspace{-4.7em} x < \tfrac{1}{6}\lor y > \sqrt{3}(x-\tfrac{1}{6}),\\
    \bigl(1.4,\,0,\,0,\,1.0/\rho\bigr), 
    &\hspace{0.5em} \mathrm{otherwise}.
    \end{cases} \label{eq:VIE1}
\end{align}

In the bottom boundary region of $x<1/6$, a fixed post‐shock state of $(\rho,\,u_x,\,u_y,\,RT) =(8,\,4.125\sqrt{3},\,-4.125,$ $116.5/\rho)$ is prescribed, while the left and right boundaries are imposed with inflow Neumann boundary conditions for all moments. At $y=1$, a time‐dependent inflow boundary condition has been imposed to ensure that the shock remains attached. See reference\cite{Vevek2019} for implementation details.

To reduce computational cost, the acoustic scaling of $0.5$ has been applied, as described in Section~\ref{sec:lax}. The scaling lowered the maximum temperature in the initial condition from $14.56$ to $3.64$, while the minimum temperature reached $0.179$, which is the bare minimum before EPR fails. Four uniform discretizations with $N_y=125,\,250,\,500$, and $1000$ are tested to observe the effect of grid resolution. The maximum lattice Courant number reached $16.6$ for this setup.

\begin{figure}
  \centering
  \includegraphics[width=\columnwidth]{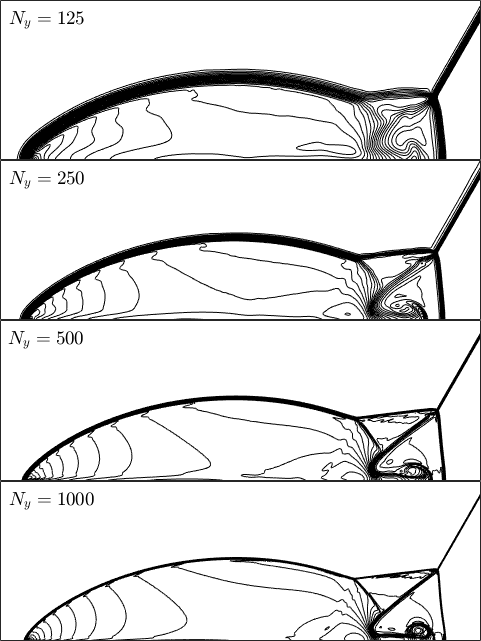}
  \caption{\label{fig:VIE1}Density contours of the Double Mach Reflection problem, over $[0,3]\times[0,1]$, with 43 uniform levels.}
\end{figure}

Figure~\ref{fig:VIE1} shows density contours at $t=0.2$. The primary shock, Mach stem, and slip line are clearly resolved, but the jet instability seen in the higher-order scheme \cite{Vevek2019, Zeng2019} did not appear even at the highest resolution. This is attributed to increased local diffusion at high lattice temperatures, as discussed in Section~\ref{sec:shear_wave}, which dampens small‐scale instabilities from growing. The further analysis of the effect of temperature and resolution on error is provided in Section~\ref{sec:cylinder_shock}.

\subsection{\label{sec:oblique}Oblique Shock}

\begin{figure}
  \centering
  \includegraphics[width=\columnwidth]{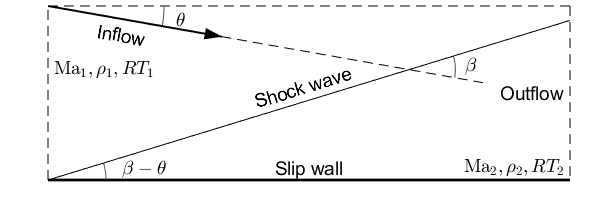}
  \caption{\label{fig:VIF1}Geometry of the oblique shock problem. Subscripts “1” and “2” denote pre– and post–shock states, respectively.}
\end{figure}

To evaluate the physicality of the LELBM model, the oblique shock problem has been tested. This problem is characterized by a supersonic inflow at Mach number $\mathrm{Ma}_1$, defined by the state $(\rho_1,\,RT_1)$, impinging on a slip wall deflected by angle $\theta$ and produces an oblique shock at an angle of $\beta$, as depicted in Figure~\ref{fig:VIF1}. The relationship among $\theta$, $\beta$, and $\mathrm{Ma}_1$ is given by the classical $\theta\mathrm{-}\beta\mathrm{-Ma}$ equation. Once $\beta$ is determined, the downstream Mach number, pressure ratio, density ratio, and temperature ratio can be obtained analytically \cite{John2006}.

\begin{figure}
\centering
\includegraphics[width=0.95\columnwidth]{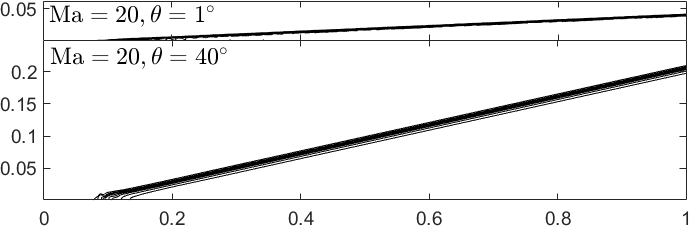}
\caption{\label{fig:VIF2}Density contours with 10 equidistant levels for $\mathrm{Ma} = 20$, at deflection angles of $\theta = 1^\circ$ and $40^\circ$.}
\end{figure}

For fixed reference temperature $RT=0.5$ and $\gamma=1.4$, a comprehensive sweep of the upstream Mach numbers $\mathrm{Ma}_1 = 2,\,2.5,\,3,\,4,\,6,\,8,\,10,\,15,\,20$ and deflection angles $\\ \theta = 1^\circ,\,2^\circ,\,3^\circ, 4^\circ, 5^\circ, 7.5^\circ,\,10^\circ,\,12.5^\circ,\,17.5^\circ,\,20^\circ,\,25^\circ,\,30^\circ,\,40^\circ$ has been performed. The domain has been discretized with a uniform cell size with $\Delta x = \Delta y =1$, using $N_x = 512$ for $\mathrm{Ma}_1 \le 15$ and $N_x=1024$ for $\mathrm{Ma}_1=20$. The shock angle has been numerically extracted by locating the maximum point of the Knudsen number defined in Appendix~\ref{sec:knudsen}, between $x=0.75N_x$ and $x=N_x$. Similarly to the near-inlet boundary treatment described in Section~\ref{sec:double_mach}, a non-reflective boundary condition, is assigned at the lower boundary for $x < 0.1$. Figure~\ref{fig:VIF2} presents the density contours for $\mathrm{Ma}_1 = 20$, with deflection angles of $\theta = 1^\circ$ and $40^\circ$. The maximum lattice Courant number reached $32.7$ for $\mathrm{Ma}_1=20$, $\theta=40^\circ$ case.

\begin{figure}
  \centering
  \includegraphics[width=0.97\columnwidth]{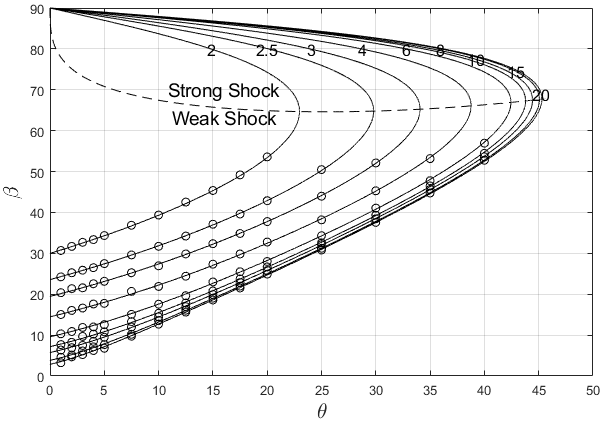}
  \caption{\label{fig:VIF3}Shock angle $\beta$ versus deflection $\theta$ for various $\mathrm{Ma}_1$. Solid lines and markers indicate analytical and numerical solutions each.}
\end{figure}

\begin{figure}
  \centering
  \includegraphics[width=\columnwidth]{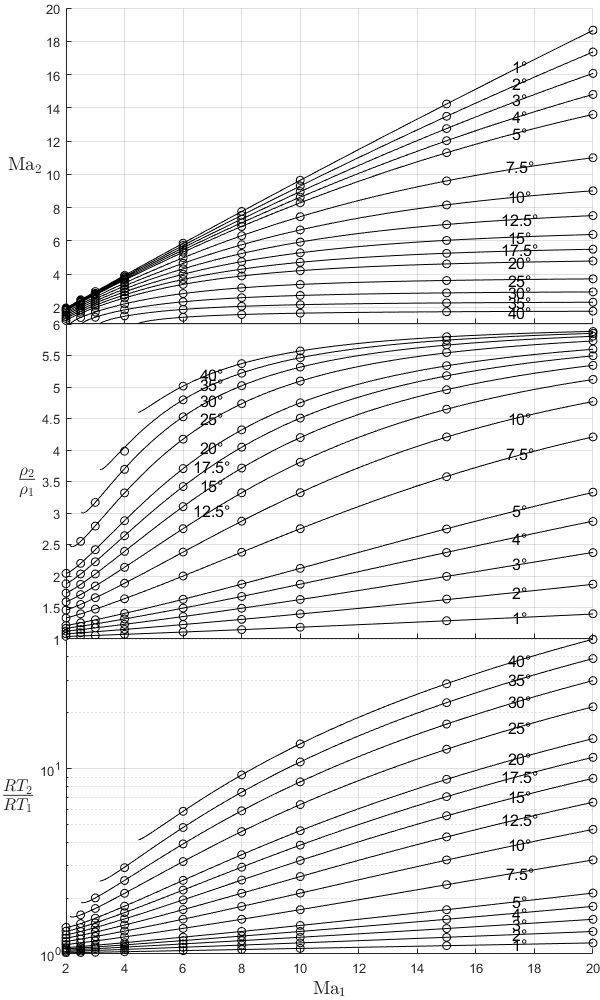}
  \caption{\label{fig:VIF4}Post‑shock properties: $\mathrm{Ma}_2$ (top), $\rho_2/\rho_1$ (middle), $RT_2/RT_1$ (bottom). Solid lines indicate analytical solutions and markers indicate numerical solutions.}
\end{figure}

Figure~\ref{fig:VIF3} shows the comparison of the calculated shock angles with the analytical $\theta\mathrm{-}\beta\mathrm{-Ma}$ curves, while Fig.~\ref{fig:VIF4} is the comparison of the numerical and analytical post‑shock Mach number $\mathrm{Ma}_2$, the density ratio $\rho_2/\rho_1$, and the temperature ratio $RT_2/RT_1$ for each Mach number and the deflection angles. All figures show excellent agreement with analytical solutions from the ideal gas dynamics, even in the extreme case of $\mathrm{Ma} = 20$.

To the author’s knowledge, there exists only one study that has applied the LBM framework to the oblique shock problem and provided sets of comparable results\cite{Qiu2017}. Their work employed non‑oscillatory non‑free‑parameter dissipation (NND) scheme, coupled with a second‑order upwind scheme for flux calculation, and applied gradual refinement near the wall. Table~\ref{tab:oblique_compare} shows the comparison of analytical solutions with the results provided from the NND-LBM framework and the results from the present LELBM model. Despite not incorporating advanced numerical schemes or mesh refinements, the present work achieved near‑perfect agreement with theory, substantially outperforming the previous study. The slight deviations in the measured shock angle arise from the uniform grid resolution, for which the shock has been presumed to be located at the cell center. The evaluation of the accurate shock angle is expected to be available by the application of the local mesh refinement technique, which is the topic of future work.

\subsection{\label{sec:cylinder_shock}Supersonic Flow Past a Circular Cylinder}

\begin{figure}
  \centering
  \includegraphics[width=0.8\columnwidth]{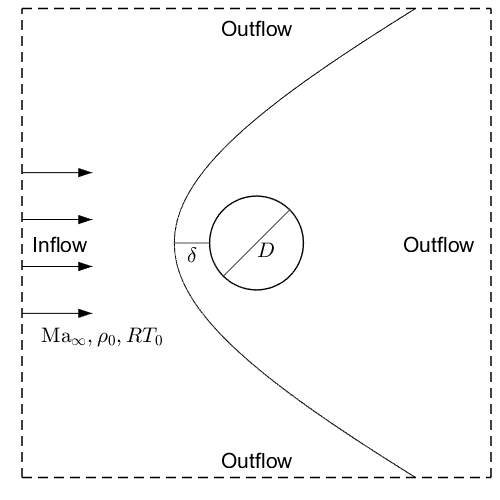}
  \caption{\label{fig:VIG1}Diagram of supersonic flow past a cylinder.}
\end{figure}

Supersonic flow past a circular cylinder is a classic compressible flow problem, which has been widely studied over the past several decades \cite{Alperin1950, Kaattari1961, Kim1956, Sinclair2017}. This problem consists of uniform upstream flow of $\mathrm{Ma}_\infty$ impinging on a cylinder with a diameter $D$, forming a detached bow shock at a distance of $\delta$, as described by Figure~\ref{fig:VIG1}. This section compares the semianalytic prediction of $\delta/D$ \cite{Sinclair2017} with experimental measurements \cite{Alperin1950, Kaattari1961, Kim1956}, continuum FVM \cite{Sinclair2017}, and precedent supersonic LBM results \cite{Tran2022}.

\begin{table}
  \caption{\label{tab:oblique_compare}Comparison of analytical, reference \cite{Qiu2017}, and LELBM results for oblique shock problem under various configurations.}
  \begin{ruledtabular}
  \begin{tabular}{lccccccc}
    & $\mathrm{Ma}_1$ & $\theta$ &      $\beta$ & & $\mathrm{Ma}_2$ & $\rho_2/\rho_1$ & $T_2/T_1$ \\
    \hline
    Analytical & 2 & $2.0^\circ$ & $31.646^\circ$ & & 1.928 & 1.083 & 1.032 \\
    Present    &   &             & $31.697^\circ$ & & 1.928 & 1.083 & 1.032 \\
    Reference  &   &             & $31.990^\circ$ & & 1.959 & 1.083 & 1.032 \\
    \hline
    Analytical & 2 & $5.0^\circ$ & $34.302^\circ$ & & 1.821 & 1.216 & 1.082 \\
    Present    &   &             & $34.358^\circ$ & & 1.821 & 1.216 & 1.082 \\
    Reference  &   &             & $34.291^\circ$ & & 1.894 & 1.152 & 1.059 \\
    \hline
    Analytical & 4 & $2.0^\circ$ & $15.813^\circ$ & & 3.852 & 1.152 & 1.059 \\
    Present    &   &             & $16.036^\circ$ & & 3.852 & 1.152 & 1.059 \\
    Reference  &   &             & $16.257^\circ$ & & 3.963 & 1.151 & 1.059 \\
    \hline
    Analytical & 4 & $5.0^\circ$ & $18.021^\circ$ & & 3.638 & 1.407 & 1.152 \\
    Present    &   &             & $17.766^\circ$ & & 3.639 & 1.407 & 1.151 \\
    Reference  &   &             & $18.371^\circ$ & & 3.904 & 1.406 & 1.152 \\
  \end{tabular}
  \end{ruledtabular}
\end{table}

\begin{table}
  \caption{\label{tab:cylinder_setup}Freestream temperature $RT_\infty$ and resolution $N_D$ for each setup.}
  \begin{ruledtabular}
  \begin{tabular}{lcc}
    Setup & $RT_\infty$ & $N_D$ \\
    \hline
    1     & 0.25          & 128   \\
    2     & 0.50          & 128   \\
    3     & 0.25          & 64    \\
    4     & 0.50          & 64    \\
  \end{tabular}
  \end{ruledtabular}
\end{table}

To evaluate the effect of lattice temperature $RT$ and resolution $N_D$, which is the number of discretizations per diameter $D$, the problem has been calculated for four different setups across the range of Mach number of $\mathrm{Ma}_\infty=1.6,\,1.8,\,2,\,3,\,4,\,5,\,6,\,7,\,8,\,9,$ and $10$. The shear viscosity and thermal conductivity are set to be zero, with a slip-wall boundary condition imposed on the cylinder surface.

\begin{figure}
  \centering
  \includegraphics[width=0.85\columnwidth]{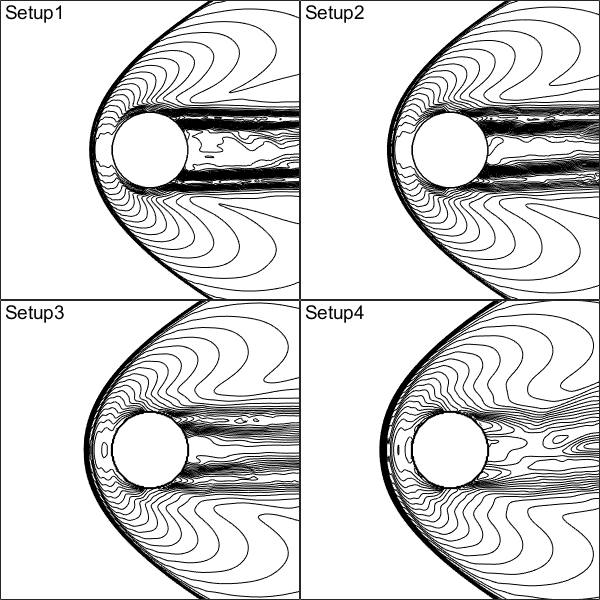}
  \caption{\label{fig:VIG2}Temperature contours for $\mathrm{Ma}_\infty=10$ with 30 equidistant levels.}
\end{figure}

Figure~\ref{fig:VIG2} shows that as the lattice temperature $RT$ increases, or the resolution $N_D$ decreases, the shock interface becomes more diffusive, leading to an overestimation of $\delta$. The characteristic wake‐tail behind the cylinder can be found in the high-Mach flow \cite{deTullio2009, Xu2018} remained unresolved, due to the limited domain size, bounded by the hardware limit. Since this section focuses on the detachment distance ratio, the backstream dynamics has been intentionally omitted to reduce the computational cost. The maximum lattice Courant number for $\mathrm{Ma}_\infty=10$, with $RT_\infty=0.5$ setup reached 16.7 locally.

\begin{figure}
  \centering
  \includegraphics[width=0.95\columnwidth]{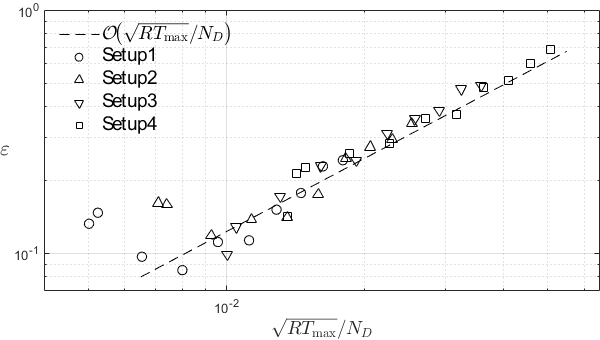}
  \caption{\label{fig:VIG3}Error $\varepsilon$ versus the error parameter $\sqrt{RT_{\max}}/N_D$.}
\end{figure}

Based on the formulation of Lagrangian acoustic stencils, the maximum collisionless streaming distance per timestep scales as $\Delta x_{\max}\sim\sqrt{RT_{\max}}\Delta t$, while the error is also expected to be inversely proportional to the resolution $\varepsilon \sim 1/N_D$. Therefore, we can expect the error per timestep to scale as $\varepsilon\sim\sqrt{RT_{\max}}/N_D$. Therefore, the maximum temperature can be approximated as the stagnation temperature by $RT_{\max}\approx \left(1+\tfrac{\gamma-1}{2}\mathrm{Ma}_\infty^2\right)RT_\infty$ for high Mach numbers. The timestep size $\Delta t$ is omitted, since this is equivalent to the acoustic scaling as in Eq.~\eqref{eq:VIB2}. Figure~\ref{fig:VIG3} confirms this proportionality.

\begin{figure}
  \centering
  \includegraphics[width=0.95\columnwidth]{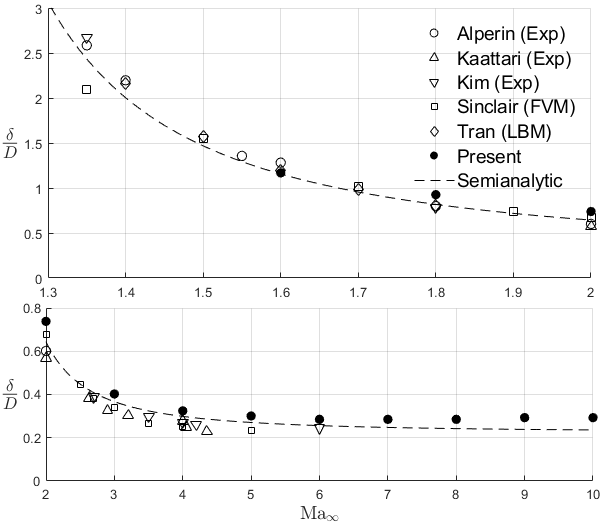}
  \caption{\label{fig:VIG4}$\delta/D$ versus $\mathrm{Ma}_\infty$ for Setup~1, compared with reference data \cite{Alperin1950, Kaattari1961, Kim1956, Sinclair2017, Tran2022} (markers) and semianalytical solution \cite{Sinclair2017} (dashed line). Top: $\mathrm{Ma}_\infty\in[1.3,\,2]$; Bottom: $\mathrm{Ma}_\infty\in[2,\,10]$.}
\end{figure}

Finally, Figure~\ref{fig:VIG4} presents the overall result of $\mathrm{Ma}_\infty$ versus $\delta/D$ from Setup~1, including the reference points \cite{Alperin1950, Kaattari1961, Kim1956, Sinclair2017, Tran2022}. Although the results coincide well with the reference solution, the remaining deviations arise from the factors that the current work did not implement a higher-order temporal integration scheme and mesh refinements, while also the cylinder geometry has been approximated by the "legoland" approximation. It is obvious to expect a reduction in error with the implementation of advanced techniques, but these are the topic for future research.

\subsection{\label{sec:naca0012}Supersonic Flow Past a NACA0012 Airfoil}
Supersonic flow past a NACA0012 airfoil is a well‑known benchmark problem for the evaluation of the accuracy of compressible flow solvers, as it features primary and secondary shock formation and a von K\'arm\'an vortex sheet at the trailing edge of the airfoil \cite{Hafez2007}. Here, the inflow conditions are $\mathrm{Ma}=1.5$ at zero angle of attack, with fluid properties of $\mathrm{Re}=10^4$, and $\mathrm{Pr}=0.7$. The upstream temperature has been set to $RT_\infty=0.5$ in all cases. Simulations were carried out in the domain of $[x,y]=[-1,7]\times[-4,4]$, with the leading edge of the airfoil located at $(0,0)$ and the chord length of $C=1$. Uniform cartesian grids with $N_C=100,\, 200,$ and $400$ were used to assess the influence of spatial resolution, where $N_C$ is the number of discretizations along the chord. Following the method in Section~\ref{sec:cylinder_shock}, the curved geometry is approximated using the “legoland” approach. The airfoil surface is designated with an adiabatic, no‑slip solid boundary conditions described in Section~\ref{sec:solidbc}. 

\begin{figure}
  \centering
  \includegraphics[width=\columnwidth]{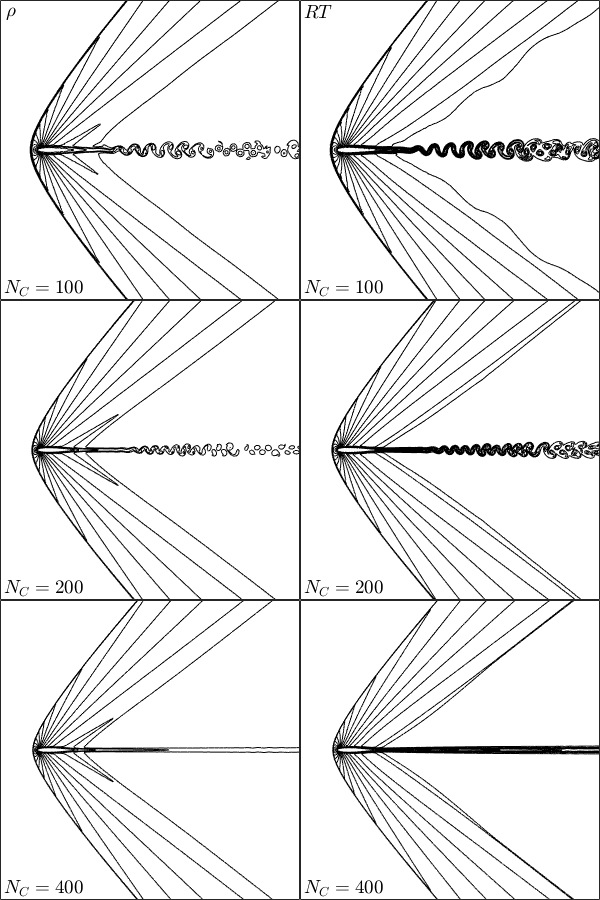}
  \caption{\label{fig:VIH1}Density (left) and temperature (right) contours for supersonic flow past a NACA0012 airfoil at different resolutions of $N_C=100$ (top), $200$ (middle), and $400$ (bottom), with 20 uniform equidistant contour levels.}
\end{figure}

Figure~\ref{fig:VIH1} presents the density and temperature contours for three chord resolutions of $N_C=100, 200,$ and $400$. Although the primary shock is well captured in all cases, the secondary shock in $N_C=100$ case exhibits unsteady behavior caused by a von K\'arm\'an vortex sheet forming too close to the trailing edge. Interestingly, these vortex sheets are most pronounced at the coarsest resolution and diminishes with increasing resolution, leaving only a minor perturbations at $N_C=400$ case. Considering that high‑fidelity simulations with adaptive mesh refinement reported the notable development of vortex‐sheet at further downstream beyond $x>7C$ under the same flow conditions \cite{Thorimbert2022}, this suggests that the development of vortex sheets in low-resolution cases is actually the numerical artifacts of underresolved near‑wall shear stress, producing spurious, downstream vortex structures that disappear as the resolution increases. Similar vortex structures would likely emerge under extended downstream simulation, but hardware limitations restricted further analysis.

\begin{figure}
  \centering
  \includegraphics[width=\columnwidth]{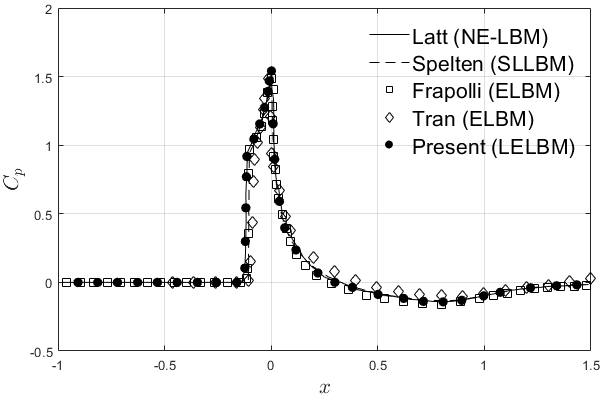}
  \caption{\label{fig:VIH2}Pressure coefficient $C_p$ profile, compared against reference data \cite{Frapolli2017, Latt2020, Tran2022, Spelten2024}.}
\end{figure}

To assess the accuracy of the present model, the pressure coefficient along the surface of the airfoil has been investigated. The pressure coefficient is defined by
\begin{align}
  C_p = \frac{p - p_\infty}{\tfrac{1}{2}\rho_\infty U_\infty^2},
  \label{eq:VIH1}
\end{align}
where $p = \rho RT$, following the ideal gas law and the subscript $\infty$ denotes the upstream conditions. Since the curved geometry has been approximated as the “legoland”, $C_p$ is calculated from the cell immediately adjacent to the wall, rather than from the true surface. Figure~\ref{fig:VIH2} compares the pressure coefficient profile obtained from the present LELBM formulation against various supersonic LBM variants \cite{Frapolli2017, Latt2020, Tran2022, Spelten2024} in $N_C=400$ resolution. Despite the geometric approximation, the resulting profile shows excellent agreement with the reference data. Especially, a noticeable alignment with NE‑LBM result \cite{Latt2020} can be observed. Considering that the EPR proposed in this paper is the generalized extension of NE‑LBM, it is obvious to observe a high level of correspondence under moderate conditions.

\section{\label{sec:conclusion}Conclusions}
This work introduces the first fully unified LBM framework that enables stable, efficient, explicit, Courant‐free simulations at high Mach numbers. The regularized central moment collision combined with the Knudsen limiter stabilized the shock without smearing fine features. Lagrangian acoustic stencils together with entropic population reconstruction guarantee positivity and moment-conserving populations for arbitrary local conditions. The final moment streaming algorithm reduced the memory demand from polynomial order in temperature to a constant overhead, enabling the feasible practical application of LELBM.

While the conventional method required quasi-population for the adjustment of thermoviscous properties, the proposed method achieved the precise control of shear and bulk viscosities, and thermal conductivity by only conducting regularized collision on second- and third-order central moments. In particular, the targeted damping of bulk moments, which was traditionally the source of spurious oscillations, enabled the stabilization of the numerical scheme in underresolved meshes. As demonstrated through shear, thermal, and acoustic plane wave problems, this framework achieves excellent accuracy around $RT=0.5$, with errors growing predictably at higher lattice temperatures, which is equivalent to a larger timestep size.

In conventional numerical schemes, achieving Courant numbers greater than unity is required to solve a globally coupled system, whose computation cost grows superlinearly with problem size, and still can become fragile under challenging local flow conditions. By contrast, the present method has shown excellent performance even in arbitrarily large Courant numbers while retaining an explicit, localized algorithm, with stability, by fully utilizing the Lagrangian nature of the streaming process. As demonstrated by the “hot-box” problem, the wall‐clock time of the LELBM solver scales linearly with the problem size, in excellent agreement with the theoretical expectation. This linear scaling law shows a significant advantage over traditional implicit timestepping methods.

Extensive validation across canonical benchmarks, most notably the oblique shock problem over a broad range of angles and Mach numbers, confirms the robustness and physicality of the LELBM formulation, even for extreme Courant numbers. Simulations of supersonic flow past a cylinder and a NACA0012 airfoil under uniform discretization with the “legoland” approximation also exhibit excellent agreement with theory and experimental data, highlighting the practical accuracy of the present model.

With the construction of the core foundational framework, future enhancements will include higher-order timestepping and flux control schemes to extend accuracy and explore low‑temperature regimes, while the advanced spatial discretization method is also demanded to recover fine structures such as shock interface or vortex structures. Furthermore, extending the LELBM framework to unstructured meshes and developing a generalized boundary condition are required for the general application of the present model. Finally, integration with the chemical reaction model is expected to enable the application of the lattice Boltzmann method to tackle practical industrial multiphysics problems.

\section*{Acknowledgments}
This work was supported by the National Research Foundation of Korea (NRF) grant funded by the Korean government (MSIP 2022R1A2C2005538).

\section*{Data Availability Statement}
The data that support the findings of this study are available from the corresponding author upon reasonable request.

\appendix
\section{\label{sec:polyatomic}Polyatomic Formulation of Monatomic Distribution Function}

Consider the equilibrium distribution of monatomic particles in a $D + K$ dimensional space, where the particles exist in a $D$ dimensional physical space and a virtual $K$ dimensional space with symmetry. Then, the $D + K$-dimensional equilibrium distribution function $\hat{f}^{\mathrm{eq}}$ can be written as
\begin{align}
    &\hat{f}^{\mathrm{eq}}(\rho,\, \bm{\xi},\, \bm{\eta},\, \bm{u},\, RT) \notag \\
    &= \frac{\rho}{(2\pi RT)^{(D + K)/2}} 
    \exp\left( -\frac{|\bm{\xi} - \bm{u}|^2 + |\bm{\eta}|^2}{2RT} \right). \label{eq:A1}
\end{align}
where $\hat{f}$ is the multidimensional distribution function that describes the state of particles existing in $D + K$ dimensions, where $\bm{\xi}$ and $\bm{u}$ are the particle velocity and bulk velocity in the $D$-dimensional physical space, respectively, and $\bm{\eta}$ is the virtual particle velocity in the $K$-dimensional space. The moment of the distribution function can be calculated by multiplying the moment basis function with the distribution function and integrating, as shown below.
\begin{align}
    M_N = \iiint_{-\infty}^{\infty} T_N(\bm{\xi},\, \bm{\eta}) \hat{f} d\bm{\xi} d\bm{\eta}. \label{eq:A2}
\end{align}
Then the relevant equilibrium moments can analytically be expressed as
\begin{subequations}
\begin{align}
     M_0^{\mathrm{eq}} &= \iiint_{-\infty}^{\infty} \hat{f}^{\mathrm{eq}} d\bm{\xi} d\bm{\eta} = \rho, \label{eq:A3a} \\
    M_\alpha^{\mathrm{eq}} &= \iiint_{-\infty}^{\infty} \xi_\alpha \hat{f}^{\mathrm{eq}} d\bm{\xi} d\bm{\eta} = \rho u_\alpha, \label{eq:A3b} \\
    M_{\alpha \beta}^{\mathrm{eq}} &= \iiint_{-\infty}^{\infty} \xi_\alpha \xi_\beta \hat{f}^{\mathrm{eq}} d\bm{\xi} d\bm{\eta} = \rho \left( u_\alpha u_\beta + RT \delta_{\alpha \beta} \right), \label{eq:A3c} \\
    M_{\alpha \beta \gamma}^{\mathrm{eq}} &= \iiint_{-\infty}^{\infty} \xi_\alpha \xi_\beta \xi_\gamma \hat{f}^{\mathrm{eq}} d\bm{\xi} d\bm{\eta}  \notag \\
    &= \rho \left( u_\alpha u_\beta u_\gamma + RT u_\alpha \delta_{\beta \gamma} \chi_{\alpha \beta \gamma} \right), \label{eq:A3d} \\
    M_{\alpha \beta \gamma \delta}^{\mathrm{eq}} 
    &= \iiint_{-\infty}^{\infty} \xi_\alpha \xi_\beta \xi_\gamma \xi_\delta 
    \hat{f}^{\mathrm{eq}} d\bm{\xi} d\bm{\eta} \notag \\
    &\hspace{-1em}= \rho \left[ u_\alpha u_\beta u_\gamma u_\delta + \left( u_\alpha u_\beta + RT \delta_{\alpha \beta} \right) 
    \delta_{\gamma \delta} \chi_{\alpha \beta \gamma \delta} \right], \label{eq:A3e}
\end{align}
\begin{align}
    M_D^{\mathrm{eq}} &= \iiint_{-\infty}^{\infty} |\bm{\xi}|^2 \hat{f}^{\mathrm{eq}} d\bm{\xi} d\bm{\eta} = \rho \left( |\bm{u}|^2 + D RT \right), \label{eq:A3f} \\
    M_{D \gamma}^{\mathrm{eq}} &= \iiint_{-\infty}^{\infty} |\bm{\xi}|^2 \xi_\gamma \hat{f}^{\mathrm{eq}} d\bm{\xi} d\bm{\eta} = \rho \left( |\bm{u}|^2 + (D + 2) RT \right) u_\gamma. \label{eq:A3g}
\end{align}
\end{subequations}

The mixed and purely internal DoF moments can analytically be expressed as follows:
\begin{subequations}
\begin{align}
    M_\zeta^{\mathrm{eq}} &= \iiint_{-\infty}^{\infty} \eta_\zeta \hat{f}^{\mathrm{eq}} d\bm{\xi} d\bm{\eta} = 0, \label{eq:A4a} \\
    M_{\zeta \beta}^{\mathrm{eq}} &= \iiint_{-\infty}^{\infty} \eta_\zeta \xi_\beta \hat{f}^{\mathrm{eq}} d\bm{\xi} d\bm{\eta} = 0, \label{eq:A4b} \\
    M_{\zeta \theta}^{\mathrm{eq}} &= \iiint_{-\infty}^{\infty} \eta_\zeta \eta_\theta \hat{f}^{\mathrm{eq}} d\bm{\xi} d\bm{\eta} = \rho RT \delta_{\zeta \theta}, \label{eq:A4c} \\
    M_{\zeta \beta \gamma}^{\mathrm{eq}} &= \iiint_{-\infty}^{\infty} \eta_\zeta \xi_\beta \xi_\gamma \hat{f}^{\mathrm{eq}} d\bm{\xi} d\bm{\eta} = 0, \label{eq:A4d} \\
    M_{\zeta \theta \gamma}^{\mathrm{eq}} &= \iiint_{-\infty}^{\infty} \eta_\zeta \eta_\theta \xi_\gamma \hat{f}^{\mathrm{eq}} d\bm{\xi} d\bm{\eta} = \rho RT u_\gamma \delta_{\zeta \theta}, \label{eq:A4e} \\
    M_{\zeta \theta \phi}^{\mathrm{eq}} &= \iiint_{-\infty}^{\infty} \eta_\zeta \eta_\theta \eta_\phi \hat{f}^{\mathrm{eq}} d\bm{\xi} d\bm{\eta} = 0, \label{eq:A4f} \\
    M_{\zeta \beta \gamma \delta}^{\mathrm{eq}} &= \iiint_{-\infty}^{\infty} \eta_\zeta \xi_\beta \xi_\gamma \xi_\delta \hat{f}^{\mathrm{eq}} d\bm{\xi} d\bm{\eta} = 0, \label{eq:A4g} \\
    M_{\zeta \theta \gamma \delta}^{\mathrm{eq}} &= \iiint_{-\infty}^{\infty} \eta_\zeta \eta_\theta \xi_\gamma \xi_\delta \hat{f}^{\mathrm{eq}} d\bm{\xi} d\bm{\eta} \notag \\
    &= \rho \left( u_\gamma u_\delta + RT \delta_{\gamma \delta} \right) RT \delta_{\zeta \theta}, \label{eq:A4h} \\
    M_{\zeta \theta \phi \delta}^{\mathrm{eq}} &= \iiint_{-\infty}^{\infty} \eta_\zeta \eta_\theta \eta_\phi \xi_\delta \hat{f}^{\mathrm{eq}} d\bm{\xi} d\bm{\eta} = 0, \label{eq:A4i} \\
    M_{\zeta \theta \phi \psi}^{\mathrm{eq}} &= \iiint_{-\infty}^{\infty} \eta_\zeta \eta_\theta \eta_\phi \eta_\psi \hat{f}^{\mathrm{eq}} d\bm{\xi} d\bm{\eta} \notag \\
    &= \rho RT^2 \delta_{\zeta \theta} \delta_{\phi \psi} \chi_{\zeta \theta \phi \psi}, \label{eq:A4j} \\
    M_K^{\mathrm{eq}} &= \iiint_{-\infty}^{\infty} |\bm{\eta}|^2 \hat{f}^{\mathrm{eq}} d\bm{\xi} d\bm{\eta} = \rho K RT, \label{eq:A4k} \\
    M_{K \alpha}^{\mathrm{eq}} &= \iiint_{-\infty}^{\infty} |\bm{\eta}|^2 \xi_\alpha \hat{f}^{\mathrm{eq}} d\bm{\xi} d\bm{\eta} = \rho K RT u_\alpha, \label{eq:A4l} \\
    M_{K \zeta}^{\mathrm{eq}} &= \iiint_{-\infty}^{\infty} |\bm{\eta}|^2 \eta_\zeta \hat{f}^{\mathrm{eq}} d\bm{\xi} d\bm{\eta} = 0, \label{eq:A4m} \\
    M_{K \alpha \beta}^{\mathrm{eq}} &= \iiint_{-\infty}^{\infty} |\bm{\eta}|^2 \xi_\alpha \xi_\beta \hat{f}^{\mathrm{\mathrm{eq}}} d\bm{\xi} d\bm{\eta} \notag \\
    &= \rho K RT \left( u_\alpha u_\beta + RT \delta_{\alpha \beta} \right), \label{eq:A4n} \\
    M_{K \zeta \beta}^{\mathrm{eq}} &= \iiint_{-\infty}^{\infty} |\bm{\eta}|^2 \eta_\zeta \xi_\beta \hat{f}^{\mathrm{\mathrm{eq}}} d\bm{\xi} d\bm{\eta}= 0, \label{eq:A4o} \\
    M_{K \zeta \theta}^{\mathrm{eq}} &= \iiint_{-\infty}^{\infty} |\bm{\eta}|^2 \eta_\zeta \eta_\theta \hat{f}^{\mathrm{\mathrm{eq}}} d\bm{\xi} d\bm{\eta} = \rho K RT^2 \delta_{\zeta \theta}. \label{eq:A4p}
\end{align}
\end{subequations}
The subscripts $\alpha,\,\beta,\,\gamma$ indicate the translational degrees of freedom (DoF), and $\zeta,\,\theta,\,\psi$ indicate the internal DoF, which do not overlap. Additionally, $\chi$ appearing from third-order moments denotes the summation over all possible cyclic permutations of the subscript indices. Furthermore, the subscript $D$ denotes trace moments for translational DoF, and $K$ for internal DoF trace moments. As expected, since the distribution is symmetric in the $K$-dimensional space, odd-order moments vanish. Non-equilibrium moments can be derived using the Chapman-Enskog expansion (explained in Appendix~\ref{sec:chapman}).
\begin{subequations}
\begin{align}
    M_{\alpha \beta}^{\mathrm{neq}} &= -\rho RT  \tau_{\alpha \beta} \left(\partial_\alpha u_\beta + \partial_\beta u_\alpha - \tfrac{2}{D+K} \partial_\gamma u_\gamma \delta_{\alpha \beta} \right) \label{eq:A5a} \\
    M_{\alpha \zeta}^{\mathrm{neq}} &= 0 \label{eq:A5b} \\
    M_{\zeta \theta}^{\mathrm{neq}} &= \rho RT \tau_{\zeta \theta} \tfrac{2}{D+K} \partial_\gamma u_\gamma \delta_{\zeta \theta} \label{eq:A5c} \\
    M_{\alpha \beta \gamma}^{\mathrm{neq}} &= \left( u_\gamma M_{\alpha \beta}^{\mathrm{neq}} - \rho RT \tau_{\alpha \beta \gamma} \, \partial_\gamma RT \, \delta_{\alpha \beta} \right) \chi_{\alpha \beta \gamma} \label{eq:A5d} \\
    M_{\zeta \beta \gamma}^{\mathrm{neq}} &= 0 \label{eq:A5e} \\
    M_{\zeta \theta \gamma}^{\mathrm{neq}} &= \rho RT \tau_{\zeta \theta \gamma} \left( \tfrac{2}{D+K} u_\gamma \partial_\delta u_\delta - \partial_\gamma RT \right) \delta_{\zeta \theta} \label{eq:A5f} \\
    M_{\zeta \theta \psi}^{\mathrm{neq}} &= 0 \label{eq:A5g}
\end{align}
\end{subequations}

We can observe that each component of the moments in a purely $K$-dimensional space is uncorrelated. This is consistent with the equipartition theorem in statistical mechanics. Therefore, instead of explicitly constructing a $D+K$-dimensional distribution, the problem can be simplified into two distribution functions defined in the $D$-dimensional space.
\begin{subequations}
\begin{align}
    f(\rho,\, \bm{\xi},\, \bm{u},\, RT) 
    &= \iiint_{-\infty}^{\infty} \hat{f}(\rho,\, \bm{\xi},\, \bm{\eta},\, \bm{u},\, RT) d\bm{\eta}, \label{eq:A6a} \\
    g(\rho,\, \bm{\xi},\, \bm{u},\, RT) 
    &= \iiint_{-\infty}^{\infty} |\bm{\eta}|^2 \hat{f}(\rho,\, \bm{\xi},\, \bm{\eta},\, \bm{u},\, RT) d\bm{\eta}. \label{eq:A6b}
\end{align}
\end{subequations}
In many literatures, $f$ and $g$ are referred to as fluons and phonons. Each distribution must satisfy the same moment constraints as $\hat{f}$. Let $M$ denote the fluon moments and $G$ the phonon moments. The equilibrium moments for fluons are identical to Eqs. \eqref{eq:A3a}--\eqref{eq:A3g}, and the newly defined phonon moments correspond to Eqs. \eqref{eq:A4k}--\eqref{eq:A4p}, which can now be expressed as
\begin{subequations}
\begin{align}
    G_0^{\mathrm{eq}} &= \iiint_{-\infty}^{\infty} g^{\mathrm{eq}} d\bm{\xi} = \rho K RT, \label{eq:A7a} \\
    G_\alpha^{\mathrm{eq}} &= \iiint_{-\infty}^{\infty} \xi_\alpha g^{\mathrm{eq}} d\bm{\xi} = \rho K RT u_\alpha, \label{eq:A7b} \\
    G_{\alpha \beta}^{\mathrm{eq}} &= \iiint_{-\infty}^{\infty} \xi_\alpha \xi_\beta g^{\mathrm{eq}} d\bm{\xi} = \rho K RT \left( u_\alpha u_\beta + RT \delta_{\alpha \beta} \right). \label{eq:A7c}
\end{align}
\end{subequations}

\section{\label{sec:chapman}Boltzmann Equation to Navier-Stokes-Fourier Equation}

From the double-distribution function formulation of the Boltzmann equation,
\begin{align}
    \partial_t \{ f,\, g \} + \xi_\alpha \, \partial_\alpha \{ f,\, g \} = -\frac{1}{\tau} \{ f^{\mathrm{neq}},\, g^{\mathrm{neq}} \}. \label{eq:B1}
\end{align}
Boltzmann moment equation can be constructed by integrating both sides with a moment basis function
\begin{align}
    & \partial_t \iiint_{-\infty}^{\infty} T_N(\bm{\xi}) \{ f,\, g \} d\bm{\xi} + \partial_\alpha \iiint_{-\infty}^{\infty} T_{N+\alpha}(\bm{\xi}) \{ f,\, g \} d\bm{\xi} \notag \\
    & = -\frac{1}{\tau_N} \iiint_{-\infty}^{\infty} T_N(\bm{\xi}) \{ f^{\mathrm{neq}},\, g^{\mathrm{neq}} \} d\bm{\xi}. \label{eq:B2}
\end{align}
$T_N$ is the arbitrary $N$th-order moment basis function and $\tau_N$ is the relaxation time corresponding to the moment of $N$th-order for fluon and phonon distribution functions. This relaxation time applies not only to moments of the same order but also to each components in different value. The generated Boltzmann moment equation is
\begin{align}
    &\partial_t \left( M_N + G_{N-2} \right) + \partial_\alpha \left( M_{N+\alpha} + G_{N+\alpha - 2} \right) \notag \\
    &= -\frac{1}{\tau_N} \left( M_N^{\mathrm{neq}} + G_{N-2}^{\mathrm{neq}} \right). \label{eq:B3}
\end{align}

The moments of $M_N$ and $G_{N-2}$ are of the same order since, from Appendix~\ref{sec:polyatomic}, $g$ is the distribution function of internal DoF energy, with the zeroth-order moment having the same order as the second-order moment of fluons. Before performing the Chapman-Enskog expansion, several assumptions are required. First, the total moment can be separated into equilibrium and non-equilibrium moments by
\begin{align}
    \{ M_N,\, G_N \} = \{ M_N^{\mathrm{eq}} + M_N^{\mathrm{neq}},\, G_N^{\mathrm{eq}} + G_N^{\mathrm{neq}} \}. \label{eq:B4}
\end{align}
In addition, the conservation of mass, momentum, and total energy must hold.
\begin{subequations}
\begin{align}
    M_0 = M_0^{\mathrm{eq}} &= \rho, \label{eq:B5a} \\
    M_\alpha = M_\alpha^{\mathrm{eq}} &= \rho u_\alpha, \label{eq:B5b} \\
    M_D + G_0 = M_D^{\mathrm{eq}} + G_0^{\mathrm{eq}} &= \rho \left( |u|^2 + (D + K) R T \right). \label{eq:B5c}
\end{align}
\end{subequations}
This implies that the non-equilibrium moments of Eqs. \eqref{eq:B5a}--\eqref{eq:B5c} are zero.
\begin{subequations}
\begin{align}
    M_0^{\mathrm{neq}} &= 0, \label{eq:B6a} \\
    M_\alpha^{\mathrm{neq}} &= 0, \label{eq:B6b} \\
    M_D^{\mathrm{neq}} + G_0^{\mathrm{neq}} &= 0. \label{eq:B6c}
\end{align}
\end{subequations}
Now, we conduct the Chapman-Enskog expansion.
\begin{subequations}
\begin{align}
    \{ M_N,\, G_N \} &= \{ M_N^{(0)} + \epsilon M_N^{(1)},\, G_N^{(0)} + \epsilon G_N^{(1)} \}, \label{eq:B7a} \\
    \{ M_N^{\mathrm{eq}},\, G_N^{\mathrm{eq}} \} &= \{ M_N^{(0)},\, G_N^{(0)} \}, \label{eq:B7b} \\
    \{ M_N^{\mathrm{neq}},\, G_N^{\mathrm{neq}} \} &= \{ \epsilon M_N^{(1)},\, \epsilon G_N^{(1)} \}, \label{eq:B7c} \\
    \partial_t &= \epsilon \partial_t^{(1)}, \label{eq:B7d} \\
    \partial_\alpha &= \epsilon \partial_\alpha^{(1)}. \label{eq:B7e}
\end{align}
\end{subequations}
Applying Eqs. \eqref{eq:B7a}--\eqref{eq:B7e} to Eq. \eqref{eq:B3} and arranging in each order of $\epsilon$ makes
\begin{subequations}
\begin{align}
    \epsilon^1:\hspace{0.5em} & \partial_t^{(1)} \bigl( M_N^{(0)} + G_{N-2}^{(0)} \bigr) + \partial_\alpha^{(1)} \bigl( M_{N+\alpha}^{(0)} + G_{N+\alpha-2}^{(0)} \bigr) \notag \\
    &= -\frac{1}{\tau_N} \left( M_N^{(1)} + G_{N-2}^{(1)} \right), \label{eq:B8a} \\
    \epsilon^2:\hspace{0.5em} & \partial_t^{(1)} \bigl( M_N^{(1)} + G_{N-2}^{(1)} \bigr) + \partial_\alpha^{(1)} \bigl( M_{N+\alpha}^{(1)} + G_{N+\alpha-2}^{(1)} \bigr) = 0. \label{eq:B8b}
\end{align}
\end{subequations}
Applying Eqs. \eqref{eq:A3a}--\eqref{eq:A3g}, \eqref{eq:A7a}--\eqref{eq:A7c}, and \eqref{eq:B7a}--\eqref{eq:B7e} to Eqs. \eqref{eq:B8a}--\eqref{eq:B8b}, each order of the Boltzmann moment equations can be constructed. For the zeroth-order moment, the mass
\begin{subequations}
\begin{align}
    \epsilon^1:\hspace{0.5em} & \partial_t^{(1)} \rho + \partial_\alpha^{(1)} \rho u_\alpha = 0, \label{eq:B9a} \\
    \epsilon^2:\hspace{0.5em} & 0 + 0 = 0. \label{eq:B9b}
\end{align}
\end{subequations}
First-order, the momentum
\begin{subequations}
\begin{align}
    \epsilon^1:\hspace{0.5em} & \partial_t^{(1)} \rho u_\alpha + \partial_\beta^{(1)} \rho (u_\alpha u_\beta + R T \delta_{\alpha \beta}) = 0, \label{eq:B10a} \\
    \epsilon^2:\hspace{0.5em} & \partial_\beta^{(1)} M_{\alpha \beta}^{(1)} = 0. \label{eq:B10b}
\end{align}
\end{subequations}
Second-order, the energy
\begin{subequations}
\begin{align}
    \epsilon^1:\hspace{0.5em} & \partial_t^{(1)}  \rho \bigl( |u|^2 + (D + K) R T \bigr) \notag \\
    &+ \partial_\gamma^{(1)} \rho \bigl( |u|^2 + (D + K + 2) R T \bigr) u_\gamma = 0, \label{eq:B11a} \\
    \epsilon^2:\hspace{0.5em} & \partial_\gamma^{(1)} M_{D\gamma}^{(1)} + \partial_\gamma^{(1)} G_\gamma^{(1)} = 0. \label{eq:B11b}
\end{align}
\end{subequations}
Second-order, the stress
\begin{subequations}
\begin{align}
    \epsilon^1:\hspace{0.5em} & \partial_t^{(1)} \rho (u_\alpha u_\beta + R T \delta_{\alpha \beta})+ \partial_\gamma^{(1)} \rho (u_\alpha u_\beta u_\gamma \notag \\
    &+ R T u_\gamma \delta_{\alpha \beta} \chi_{\alpha \beta \gamma}) = -\frac{1}{\tau_{\alpha \beta}} M_{\alpha \beta}^{(1)}, \label{eq:B12a} \\
    \epsilon^1:\hspace{0.5em} & \partial_t^{(1)} \rho K R T + \partial_\gamma^{(1)} \rho K R T u_\gamma = -\frac{1}{\tau_{\alpha \beta}} G_0^{(1)}. \label{eq:B12b}
\end{align}
\end{subequations}
Third-order, the heat flux moment equations are constructed.
\begin{subequations}
\begin{align}
    \epsilon^1:\hspace{0.5em} & \partial_t^{(1)}  \rho \left( u_\alpha u_\beta u_\gamma + R T u_\alpha \delta_{\beta \gamma} \chi_{\alpha \beta \gamma} \right) + \partial_\delta^{(1)}  \rho \bigl( u_\alpha u_\beta u_\gamma u_\delta \notag\\
    &+ R T u_\alpha u_\beta \delta_{\gamma \delta} \chi_{\alpha \beta \gamma \delta} + RT^2 \delta_{\alpha \beta} \delta_{\gamma \delta} \chi_{\alpha \beta \gamma \delta} \bigr) \notag \\
    &= -\frac{1}{\tau_{\alpha \beta \gamma}} M_{\alpha \beta \gamma}^{(1)}, \label{eq:B13a} \\
    \epsilon^1:\hspace{0.5em} & \partial_t^{(1)} \rho K R T u_\gamma + \partial_\delta^{(1)}  \rho K R T \left( u_\gamma u_\delta + R T \delta_{\gamma \delta} \right) \notag \\
    &= -\frac{1}{\tau_{\alpha \beta \gamma}} G_\gamma^{(1)}. \label{eq:B13b}
\end{align}
\end{subequations}
Eqs. \eqref{eq:B9a}--\eqref{eq:B9b} is obviously the mass equation even without expansion. After some heavy algebra on Eqs. \eqref{eq:B9a}--\eqref{eq:B13b}, we can obtain the analytical expression of non-equilibrium moments.
\begin{subequations}
\begin{align}
    M_{\alpha \beta}^{(1)} &= -\rho R T \tau_{\alpha \beta} \Bigl[ \left( \partial_\alpha^{(1)} u_\beta + \partial_\beta^{(1)} u_\alpha - \tfrac{2}{D} \partial_\gamma^{(1)} u_\gamma \delta_{\alpha \beta} \right) \notag \\
    &+ \left( \tfrac{2}{D} - \tfrac{2}{D + K} \right) \partial_\gamma^{(1)} u_\gamma \delta_{\alpha \beta} \Bigr], \label{eq:B14a} \\
    M_{\alpha \beta \gamma}^{(1)} &= -\rho R T \tau_{\alpha \beta \gamma} \Bigl\{
    u_\gamma \Bigl[ \left( \partial_\alpha^{(1)} u_\beta + \partial_\beta^{(1)} u_\alpha - \tfrac{2}{D} \partial_\delta^{(1)} u_\delta \delta_{\alpha \beta} \right)\notag \\
    &\hspace{-0.5em}+ \left( \tfrac{2}{D} - \tfrac{2}{D + K} \right) \partial_\delta^{(1)} u_\delta \delta_{\alpha \beta} \Bigr]+ \partial_\gamma^{(1)} RT \delta_{\alpha \beta}\Bigr\} \chi_{\alpha \beta \gamma}, \label{eq:B14b} \\
    G_0^{(1)} &= \rho R T \tau_{\alpha \beta} \tfrac{2K}{D + K} \partial_\gamma^{(1)} u_\gamma, \label{eq:B14c} \\
    G_\gamma^{(1)} &= \rho R T \tau_{\alpha \beta \gamma} \left( u_\gamma \tfrac{2K}{D + K} \partial_\delta^{(1)} u_\delta - K\partial_\gamma^{(1)} RT\right). \label{eq:B14d}
\end{align}
\end{subequations}

Now, suppose that Eq. \eqref{eq:B14a} can be decomposed into the shear and bulk stress term and collide with the corresponding relaxation time of $\tau_f$ and $\tau_b$. Then we can formulate as
\begin{subequations}
\begin{align}
    M_{\alpha \beta}^{(1)} &= M_{\alpha \beta}^{s(1)} + M_{\alpha \beta}^{b(1)}, \label{eq:B15a} \\
    M_{\alpha \beta}^{s(1)} &= -\rho R T \tau_f \left(\partial_\alpha^{(1)} u_\beta + \partial_\beta^{(1)} u_\alpha - \tfrac{2}{D} \partial_\gamma^{(1)} u_\gamma \delta_{\alpha \beta} \right), \label{eq:B15b} \\
    M_{\alpha \beta}^{b(1)} &= -\rho R T \tau_b \left( \tfrac{2}{D} - \tfrac{2}{D + K} \right) \partial_\gamma^{(1)} u_\gamma \delta_{\alpha \beta}. \label{eq:B15c}
\end{align}
\end{subequations}

Similarly for Eqs. \eqref{eq:B14b} and \eqref{eq:B14d}, we separate the shear and bulk viscous dissipation term and the heat conduction term, then collide each with the relaxation time of $\tau_f$, $\tau_b$, and $\tau_h$.
\begin{subequations}
\begin{align}
    M_{\alpha \beta \gamma}^{(1)} &= M_{\alpha \beta \gamma}^{s(1)} + M_{\alpha \beta \gamma}^{b(1)} + M_{\alpha \beta \gamma}^{h(1)}, \label{eq:B16a} \\
    M_{\alpha \beta \gamma}^{s(1)} &= -u_\gamma \rho RT \tau_f \left( \partial_\alpha^{(1)} u_\beta + \partial_\beta^{(1)} u_\alpha - \tfrac{2}{D} \partial_\delta^{(1)} u_\delta \delta_{\alpha \beta} \right) \chi_{\alpha \beta \gamma} \notag \\
    &= u_\gamma M_{\alpha \beta}^{s(1)} \chi_{\alpha \beta \gamma}, \label{eq:B16b} \\
    M_{\alpha \beta \gamma}^{b(1)} &= -u_\gamma \rho RT \tau_b \left( \tfrac{2}{D} - \tfrac{2}{D + K} \right) \partial_\delta^{(1)} u_\delta \delta_{\alpha \beta} \chi_{\alpha \beta \gamma} \notag \\
    &= u_\gamma M_{\alpha \beta}^{b(1)} \chi_{\alpha \beta \gamma}, \label{eq:B16c} \\
    M_{\alpha \beta \gamma}^{h(1)} &= -\rho RT \tau_h \partial_\gamma^{(1)} RT \delta_{\alpha \beta} \chi_{\alpha \beta \gamma}, \label{eq:B16d} \\
    G_\gamma^{(1)} &= G_\gamma^{b(1)} + G_\gamma^{h(1)}, \label{eq:B16e} \\
    G_\gamma^{b(1)} &= u_\gamma \rho RT \tau_b \tfrac{2K}{D + K} \partial_\delta^{(1)} u_\delta, \label{eq:B16f} \\
    G_\gamma^{h(1)} &= -\rho KRT \tau_h \partial_\gamma^{(1)} RT. \label{eq:B16g}
\end{align}
\end{subequations}

Especially, the third-order trace of the non-equilibrium fluon moments $M_{D \gamma}^{(1)}$ can be expressed as
\begin{subequations}
\begin{align}
    M_{D \gamma}^{(1)} &= u_\gamma M_D^{(1)} + 2 u_\alpha M_{\alpha \gamma}^{(1)} 
    - \rho (D + 2) RT \tau_h \partial_\gamma^{(1)} RT, \label{eq:B17a} \\
    M_D^{(1)} &= -\rho RT \tau_b \tfrac{2K}{D + K} \partial_\delta^{(1)} u_\delta. \label{eq:B17b}
\end{align}
\end{subequations}

Assembling Eqs. \eqref{eq:B7a}--\eqref{eq:B7e}, \eqref{eq:B10a}--\eqref{eq:B10b}, and \eqref{eq:B15a}--\eqref{eq:B15c} reconstructs the full expression of the momentum equation.
\begin{align}
    &\partial_t \rho u_\alpha + \partial_\beta \rho \left(u_\alpha u_\beta + R T \delta_{\alpha \beta}\right) = \partial_\beta \tau_{\alpha\beta}, \label{eq:B18}
\end{align}
$\tau_{\alpha\beta}$ is the viscous stress tensor defined as
\begin{align}
    \tau_{\alpha\beta} = \mu_s \left( \partial_\alpha u_\beta + \partial_\beta u_\alpha - \tfrac{2}{D} \partial_\gamma u_\gamma \delta_{\alpha \beta} \right) + \mu_b \partial_\gamma u_\gamma \delta_{\alpha \beta}. \label{eq:B19}
\end{align}
The shear and bulk viscosity are expressed as
\begin{subequations}
\begin{align}
    \mu_s &= \rho R T \tau_f, \label{eq:B20a} \\
    \mu_b &= \rho R T \tau_b \left( \tfrac{2}{D} - \tfrac{2}{D + K} \right). \label{eq:B20b}
\end{align}
\end{subequations}
\eqref{eq:B20b} is exactly the same expression for the bulk viscosity from the model of Li et al.\cite{Li2008}.

Similarly, the total energy equation can be reconstructed from assembling Eqs. \eqref{eq:B7a}--\eqref{eq:B7e}, \eqref{eq:B11a}--\eqref{eq:B11b}, and \eqref{eq:B16a}--\eqref{eq:B16g}
\begin{align}
    &\partial_t \rho \left(\tfrac{1}{2}|u|^2 + C_v T \right) + \partial_\gamma \rho \left(\tfrac{1}{2}|u|^2 + C_p T \right) u_\gamma \notag \\
    &= \partial_\gamma u_\alpha \tau_{\alpha\gamma} + \partial_\gamma \kappa \partial_\gamma T. \label{eq:B21}
\end{align}
Specific heats and thermal conductivity are expressed as
\begin{subequations}
\begin{align}
    C_v &= \tfrac{D+K}{2} R, \label{eq:B22a} \\
    C_p &= \tfrac{D+K+2}{2} R, \label{eq:B22b} \\
    \kappa &= \rho C_p RT \tau_h. \label{eq:B22c}
\end{align}
\end{subequations}
This is the same expression as the Frapolli's \cite{Frapolli2017} derivation, using the population-based Chapman-Enskog expansion.

\section{\label{sec:implicit}Discrete Boltzmann Moment Equation with Implicit Collision}
From the Boltzmann moment equation of
\begin{align}
    D_t M_N = -\frac{1}{\tau_N} M_N^{\mathrm{neq}}. \label{eq:C1}
\end{align}
We take integration on both sides with respect to time to obtain the post-collision moments.
\begin{align}
    \int_t^{t + \Delta t} D_t M_N dt = -\int_t^{t + \Delta t} \frac{1}{\tau_N} M_N^{\mathrm{neq}} dt. \label{eq:C2}
\end{align}
Generally, integration in RHS is done using the trapezoidal rule.
\begin{align}
    & M_N(\bm{x} + \bm{c}_i \Delta t,\, t + \Delta t) - M_N(\bm{x},\, t) \notag \\
    & = -\frac{\Delta t}{2 \tau_N} \left\{ M_N^{\mathrm{neq}}(\bm{x} + \bm{c}_i \Delta t,\, t + \Delta t) + M_N^{\mathrm{neq}}(\bm{x},\, t) \right\}. \label{eq:C3}
\end{align}
The implicit collision on Eq. \eqref{eq:C3} can be treated explicitly by taking the transformation as Eq. \eqref{eq:C4}.\cite{Ubertini2010}
\begin{align}
    \overline{M}_N = M_N + \frac{\Delta t}{2 \tau_N} M_N^{\mathrm{neq}} \label{eq:C4}
\end{align}
The equilibrium moments are identical for both sides in Eq. \eqref{eq:C4}. Then a non-equilibrium moment can be written as
\begin{align}
    M_N^{\mathrm{neq}} = \frac{1}{1 + \tfrac{\Delta t}{2 \tau_N}} \overline{M}_N^{\mathrm{neq}} \label{eq:C5}
\end{align}
Which Eq. \eqref{eq:C3} becomes
\begin{align}
    &\overline{M}_N(\bm{x} + \bm{c}_i \Delta t,\, t + \Delta t) - \overline{M}_N(\bm{x},\, t) = -\omega_N \overline{M}_N^{\mathrm{neq}}(\bm{x},\, t) \label{eq:C6}
\end{align}
$\omega_N$ is the implicit collision frequency defined as
\begin{align}
    \omega_N = \frac{1}{\tfrac{\tau_N}{\Delta t} + \tfrac{1}{2}}. \label{eq:C7}
\end{align}
This maps $0\leq\tau_N<\infty$ to $0\leq\omega_N\leq2$. Then with the regularized central moment collision method discussed in Section~\ref{sec:collision}, the shear viscosity, bulk viscosity, and thermal conductivity are implemented as
\begin{subequations}
\begin{align}
    \mu_s &= \rho R T \left( \tfrac{1}{\omega_f} - \tfrac{1}{2} \right) \Delta t \label{eq:C8a} \\
    \mu_b &= \rho R T \left( \tfrac{2}{D} - \tfrac{2}{D + K} \right) \left( \tfrac{1}{\omega_b} - \tfrac{1}{2} \right) \Delta t \label{eq:C8b} \\
    \kappa &= \tfrac{1}{2} \rho (D + K + 2) R^2 T \left( \tfrac{1}{\omega_h} - \tfrac{1}{2} \right) \Delta t \label{eq:C8c}
\end{align}
\end{subequations}

Besides, Eq. \eqref{eq:C6} is algorithmically identical to the explicit method but with a second-order implicit collision kernel. With LELBM expression, Eq. \eqref{eq:C6} is written as
\begin{align}
    \overline{M}_N^* = \overline{M}_N^{\mathrm{eq}} + (1 - \omega_N) \overline{M}_N^{\mathrm{neq}}. \label{eq:C9}
\end{align}
This is the final expression for moment-based collision in the lattice Boltzmann method. $\overline{M}_N^*$ is the post-collision moments. The non-equilibrium moment we obtain from LBM is related to analytical moments by
\begin{align}
    \overline{M}_N^{\mathrm{neq}} = \frac{1}{\omega_N} \frac{\Delta t}{\tau_N}  M_N^{\mathrm{neq}}. \label{eq:C10}
\end{align}
This formulation applies to both fluon and phonon moments. For the sake of simplicity, the overlining of implicit populations and moments is neglected throughout this paper.

\section{\label{sec:knudsen}Knudsen Limiter}

At the strong shock interface, the EPR (Section~\ref{sec:epr}) often fails. This failure stems from excessively high second- and third-order non-equilibrium moments, which correspond to the covariance and skewness of the distribution function. While refining the stencils in regions of steep gradients can solve this problem, present work instead applied an artificial relaxation to these non-equilibrium moments for the sake of simplicity. This approach is analogous to the artificial viscosity in continuum CFD but is guided by the Knudsen number.

Based on the Chapman-Enskog expansion, the particle distribution function can be expressed as
\begin{align}
    f = f^{\mathrm{eq}} + \sum_{n=1}^{\infty} \epsilon^{n} f^{(n)}, \label{eq:D1}
\end{align}
where $\epsilon$ denotes a smallness parameter, generally of the same order as the Knudsen number \cite{Higuera1990}. If we target the NSF continuum regime, it can be presumed that the Knudsen number is low enough by $\mathrm{Kn} < 0.01$. Assuming that each order term $f^{(n)}$ is of similar magnitude and incorporating statistical descriptions of correlation and skewness, the local Knudsen number can be formulated as
\begin{align}
    \mathrm{Kn} = \max \left(
    \frac{\sum_{\alpha\beta} \left| \widetilde{M}_{\alpha\beta}^{\mathrm{neq} \prime} \right| + \left| G_{0}^{\mathrm{neq} \prime} \right|}{(D+K) R T},\,\frac{\left| \widetilde{M}_{D\gamma}^{\mathrm{neq} \prime} \right| + \left| G_{\gamma}^{\mathrm{neq} \prime} \right|}{(D+K+2) R T^{3/2}} \right). \label{eq:D2}
\end{align}

Then the logarithm of the Knudsen number is used to artificially adjust the collision frequency toward the continuum regime.
\begin{subequations}
\begin{align}
    &\mathrm{Kn}^* = \log_{10} \mathrm{Kn}, \label{eq:D3a} \\
    &\omega^*(\mathrm{Kn}^*,\, \omega) = 
    \begin{cases}
    \omega_H, & \hspace{-3.8em} \mathrm{Kn}^* < \mathrm{Kn}_L^*, \\
    \displaystyle 
    \frac{\omega_H - \omega}{\mathrm{Kn}_H^* - \mathrm{Kn}_L^*} (\mathrm{Kn}^* - \mathrm{Kn}_L^*) + \omega, \\
    & \hspace{-6.7em} \mathrm{Kn}_L^* \leq \mathrm{Kn}^* \leq \mathrm{Kn}_H^*, \\
    \omega, & \hspace{-3.8em} \mathrm{Kn}^* > \mathrm{Kn}_H^*.
    \end{cases} \label{eq:D3b}
\end{align}
\end{subequations}

The proposed function Eq. \eqref{eq:D3b} is the Knudsen limiter. In this implementation, the high collision frequency $\omega_H$ is set to unity, and the thresholds for the logarithmic Knudsen number $\mathrm{Kn}_H^*$ and $\mathrm{Kn}_L^*$ are defined as $-1$ and $-2$, respectively. The limiter is activated when $\mathrm{Kn} \geq 0.01$, progressively increasing the collision frequency, and ultimately relaxing moments toward equilibrium when $\mathrm{Kn} \geq 0.1$. This mechanism closely resembles the stabilization technique proposed by Latt et al.\cite{Latt2020, Coreixas2020, Thyagarajan2023}. The Knudsen limiter is activated generally around the shock interface and inactive even in the presence of turbulence, where the Knudsen number still remains in continuum regime. \\

\nocite{*}

{\bf References}

\end{document}